\documentclass[11pt]{article}  

\usepackage{amsmath,amsfonts,amssymb,amsthm,bbm,stmaryrd}
\usepackage[dvips]{graphicx}
\usepackage{latexsym}
\usepackage{psfrag}
\usepackage[latin1]{inputenc}
\usepackage{hyperref}
\usepackage{color}
\usepackage{enumitem}
\numberwithin{equation}{section}
\usepackage{float}
\usepackage{multirow,setspace}

\newcommand{\be}{\begin{equation}}
\newcommand{\ee}{\end{equation}}
\newcommand{\barray}{\begin{array}}
\newcommand{\earray}{\end{array}}
\newcommand{\bea}{\begin{eqnarray}}
\newcommand{\eea}{\end{eqnarray}}
\newcommand{\bs}{\begin{subequations}}
\newcommand{\es}{\end{subequations}}

\newcommand{\bit}{\begin{itemize}}
\newcommand{\eit}{\end{itemize}}
\newcommand{\bd}{\begin{description}}
\newcommand{\ed}{\end{description}}

\def\nn{\nonumber}

\newcommand{\re}{\mathrm{Re}}
\newcommand{\im}{\mathrm{Im}}

\def\la{\langle}
\def\ra{\rangle}

\def\w{\wedge}
\newcommand{\p}{\partial}
\newcommand{\na}{\nabla}

\newcommand{\R}{\mathbb{R}}
\newcommand{\C}{\mathbb{C}}

\newcommand{\f}{\frac}
\newcommand{\tl}{\tilde}

\renewcommand{\a}{\alpha} \renewcommand{\b}{\beta} \newcommand{\g}{\gamma}  
\renewcommand{\d}{\delta}  \newcommand{\eps}{\epsilon} 
 \renewcommand{\th}{\theta}  \newcommand{\vth}{\vartheta} 
    
\let\m=\mu    \let\n=\nu   \let\r=\rho \let\om=\omega
 \newcommand{\s}{\sigma}  \renewcommand{\t}{\tau}    
\let\G=\Gamma \let\D=\Delta    
\let\Si=\Sigma \let\Om=\Omega

\newcommand{\SL}{\mathrm{SL}(2,\mathbb{C})}

\def\cS{{\cal S}}
\def\cF{{\cal F}}
\def\cN{{\cal N}}

\newcommand{\os}[1]{\overset{\circ}{#1}}

\usepackage{slashed}

\newcommand{\eqonS}{\,\smash{\stackrel{\sscr{\scri}}=}\,}
\newcommand{\eqons}{\,\hat{=}\,}

\usepackage{pgf}
\makeatletter
\newcommand*{\pgfunderleftarrow}{%
  \@ifstar
    {\let\ifpgf@depth\iftrue\mathpalette\@pgfunderleftarrow}
    {\let\ifpgf@depth\iffalse\mathpalette\@pgfunderleftarrow}%
}
\newcommand*{\@pgfunderleftarrow}[2]{%
  #2%
  \edef\pgf@math@fam{\the\fam}%
  \pgfpicture
    \pgfsetbaseline{0pt}
    \pgf@relevantforpicturesizefalse      
    \pgfsetroundcap                       
    \pgfsetarrowsend{to}
    \pgfutil@tempdima=0.28pt%
    \advance\pgfutil@tempdima by.8\pgflinewidth%
    \pgfutil@tempdima-4\pgfutil@tempdima
    \sbox\pgfutil@tempboxa{$\m@th\fam\pgf@math@fam#1#2$}%
    \advance\pgfutil@tempdima-\dp\pgfutil@tempboxa
    \pgfutil@tempdimb\wd\pgfutil@tempboxa
    \pgfpathmoveto{\pgfqpoint{0pt}{\pgfutil@tempdima}}%
    \pgfpathlineto{\pgfqpoint{-\pgfutil@tempdimb}{\pgfutil@tempdima}}%
    \pgfusepath{stroke}
    \ifpgf@depth
      \pgf@relevantforpicturesizetrue
      \pgfpathmoveto{\pgfqpoint{0pt}{-\pgfutil@tempdimb}}%
      \pgfusepath{use as bounding box}%
    \fi
  \endpgfpicture
}
\makeatother
\newcommand{\pbi}[1]{\pgfunderleftarrow{#1}}

\newcommand{\sscr}{\scriptscriptstyle\rm}

\usepackage{manfnt}
\reversemarginpar

\newcommand{\Diff}{{\rm{Diff}}}

\DeclareFontFamily{U}{matha}{\hyphenchar\font45}
\DeclareFontShape{U}{matha}{m}{n}{
      <5> <6> <7> <8> <9> <10> gen * matha
      <10.95> matha10 <12> <14.4> <17.28> <20.74> <24.88> matha12
      }{}
\DeclareSymbolFont{matha}{U}{matha}{m}{n}
\DeclareMathSymbol{\oright}       {2}{matha}{"69}

\usepackage{pifont}

\newcommand{\cR}{{\cal R}}
\newcommand{\F}{\cF}
\newcommand{\xiu}{f}
\newcommand{\bac}{\eta}

\usepackage{mathrsfs}
\newcommand{\Dd}{{\mathscr{D}}}
\newcommand{\scri}{{\mathscr{I}}} 

\newcommand{\Dv}{{\os{D}}}
\newcommand{\sv}{{\os{\s}}}

\newcommand{\jb}{\bar\jmath}
\newcommand{\am}{J}
\newcommand{\Qa}{{\rm X}}
\newcommand{\qa}{{\rm x}}
\newcommand{\DDr}{{\mathbbm D}_\r}

\begin{document}

\title{\bf Center-less BMS charge algebra}

\author{\Large{Antoine Rignon-Bret and Simone Speziale}
\smallskip \\ 
\small{\it{Aix Marseille Univ., Univ. de Toulon, CNRS, CPT, UMR 7332, 13288 Marseille, France}} }
\date{July 16, 2024}

\maketitle

\begin{abstract}
\noindent We show that when the Wald-Zoupas prescription is implemented, the resulting charges realize the BMS symmetry algebra 
without any 2-cocycle nor central extension, at any cut of future null infinity. We refine the covariance prescription for application to the charge aspects, and introduce a new aspect for Geroch's super-momentum with better covariance properties.
For the extended BMS symmetry with singular conformal Killing vectors we find that a Wald-Zoupas symplectic potential exists, if one is willing to modify the symplectic structure by a corner term. The resulting algebra of Noether currents between two arbitrary cuts is center-less. The charge algebra at a given cut has a residual field-dependent 2-cocycle, but time-independent and non-radiative. 
More precisely, super-rotation fluxes act covariantly, but super-rotation charges act covariantly only on global translations.
The take home message is that in any situation where 2-cocycles appears in the literature, covariance has likely been lost in the charge prescription, and that the criterium of covariance is a powerful one to reduce ambiguities in the charges, and can be used also for ambiguities in the charge aspects.
\end{abstract}

\tableofcontents
%
\section{Introduction}
%

Boundaries play an important role in general relativity, turning part of the diffeomorphism gauge redundancy into a physically relevant symmetry.
This is particularly useful  for the physics of gravitational waves, to extract observables from the full theory that can be compared with the experiments. The boundary in this context is future null infinity $\scri$, and the symmetry described by the Bondi-van der Burg-Metzner-Sachs (BMS) group.
A unique set of charges and fluxes for this symmetry were identified a long time ago \cite{Geroch:1977jn,Ashtekar:1981bq,Dray:1984rfa}, but only much later they were given an interpretation in terms of Noether charges and canonical generators for a space-like hypersurface intersecting $\scri$ \cite{Iyer:1994ys,Wald:1999wa,Barnich:2001jy}, see also \cite{Barnich:2011mi,Barnich:2011ty,Flanagan:2015pxa,Barnich:2016lyg,Compere:2018ylh,Grant:2021sxk,Odak:2022ndm}. 

A correct definition of charges should include a realization of the symmetry algebra in terms of a phase space bracket. 
For the BMS charges, this property was established in \cite{Barnich:2011mi}, but only for special asymptotic frames corresponding to round spheres, also known as Bondi frames. For arbitrary frames, a field-dependent 2-cocyle appears. 
This is an undesired limitation, because these frames are physically undistinguishable from the Bondi frames, and there is nothing in the fall-off condition nor in the universal structure that prefers round spheres to other frames.
In this paper we show how this issue is resolved. Following carefully the Wald-Zoupas prescription one finds charges that coincide with those of \cite{Barnich:2011mi} for round spheres, but have an extra term on  general frames \cite{Grant:2021sxk,Odak:2022ndm}. This extra term 
guarantees that the key Wald-Zoupas requirements of stationarity and covariance are satisfied on arbitrary frames, and only with this extra term one matches  the charges of \cite{Geroch:1977jn,Ashtekar:1981bq,Dray:1984rfa}. Including this extra term removes the 2-cocycle in every asymptotic frame, 
and the explicit calculation as well as a general argument show that there is no residual central extension either.

We also take this opportunity to \emph{refine} the Wald-Zoupas covariance prescription, and show that it can be used to discuss covariance of the charge \emph{aspects}, and not only the charges as surface integrals. In particular, we propose a new super-momentum aspect alternative to Geroch's, which gives the same charges and conservation laws when integrated on cross-sections, but different transformation properties when \emph{not} integrated. Specifically, it is exactly background-independent, as opposed to up to an exact 2-form.
The analysis is based on the results of \cite{Rignon-Bret:2024wlu} on the relation between Wald-Zoupas covariance and symmetry algebras, of which this paper provides a longer and more detailed version including field-dependent diffeomorphisms and non-trivial corner terms, and which can be applied to any analysis of boundary symmetries.

We also show that having the Barnich-Troessaert bracket realize the algebra without 2-cocycle means in a precise sense background-independence of the charges. 
This provides a notion of covariance that is simple to implement also in radiative spacetimes. 
In this interpretation and in much of the analysis a central role is played by the anomaly operator \cite{Hopfmuller:2018fni,Chandrasekaran:2020wwn,Freidel:2021cjp,Odak:2022ndm}, which we advertise as a very convenient tool to investigate background-independence and covariance in any situation.

\emph{En route} to these results, we clarify a number of issues relating the covariant description of radiation at $\scri$, and the Bondi coordinates language. Among them, the relation between the `connection coordinate' of \cite{Ashtekar:1981hw} and the `covariant shear' of \cite{Compere:2018ylh}. This includes the relation between  the Ashtekar-Streubel phase space \cite{Ashtekar:1981bq} and the super-translation field \cite{Compere:2018ylh} (also known as super-translation Goldstone mode). We point out that the latter can be endowed with the interpretation of a `bad cut', and can be used to enrich the radiative phase space using it as a coordinate for the late time stationary boundary conditions. We review old results explaining  why the news tensor is in general more complicated than the time derivative of the shear, and why restricting to round spheres is possible but not always convenient. 
We clarify the origin of the complicated transformation laws for the shear, mass and angular momentum in Bondi coordinates by relating them to the choice of a Lie dragged auxiliary vector, to Geroch's super-momentum, and to the total divergences on the cross sections that arise when `integrating the fluxes' to obtain the charges. 
These transformation rules are  apparently sometimes misunderstood in the literature, prompting a discussion of ``covariant'' modifications. 
We explain that there is nothing non-covariant about the transformation rules, and the inhomogeneous terms that appear should not be removed but are crucial to ensure that the charges realize the algebra covariantly and without cocycles.

We then turn attention to an extension of the BMS symmetry to non-globally defined 
conformal Killing vectors that was also considered in \cite{Barnich:2011mi}. 
This extended BMS symmetry (henceforth eBMS) was proposed in \cite{Barnich:2009se,Barnich:2010eb} Its additional symmetries are often referred to `super-rotations', and plays an important role in infra-red problems \cite{Pasterski:2015tva}, flat holography \cite{}, and celestial holography \cite{Pasterski:2021raf}. The situation for the eBMS charge algebra found in \cite{Barnich:2011mi} is much worse, with the 2-cocycle being non-zero and field-dependent on \emph{every} frame. For example the algebra of super-rotations charges evaluated on the Kerr solution has a 2-cocycle function of the angular momentum.
We identify the problem in the fact that the generalization of Geroch's tensor to the eBMS symmetry transforms inhomogeneously.
Remarkably, we find that it is possible to remove this 2-cocycle also for eBMS, under the same assumptions of \cite{Barnich:2011mi} that the transformations preserve the background asymptotic metric and that one can integrate by parts on the cross-sections neglecting boundary terms. The key mechanism is the following: the ``offending" field-dependent term in the 2-cocycle contains a triple derivative of a symmetry parameter that fails to vanish in two distinct situations: if the frame is not a round sphere, or if one allows non-globally defined vector fields.
In the first case, covariance is restored by Geroch's tensor. In the latter, one needs a generalization of Geroch's tensor that was not long ago identitified in the stress-energy tensor of a conformal field theory \cite{Barnich:2010eb,Barnich:2016lyg,Compere:2016jwb,Campiglia:2020qvc,Freidel:2021qpz,Nguyen:2022zgs}.

This is not the end of story however. The generalized Geroch tensor is enough to remove the 2-cocycle in the charges and flux algebras, but this is a manifestation of linearized covariance only, and \emph{finite} covariance is still broken. This is because covariance of the symplectic potential is now satisfied only up to a total divergence on the cross sections, as opposed to exactly. This leads to a breaking of finite covariance because the anomaly operator does not commute with derivatives on the cross-sections. 
We then show that by including the super-translation field of \cite{Compere:2018ylh} 
it is possible to find a Wald-Zoupas symplectic potential for eBMS, at the price of modifying the symplectic 2-form by a corner term. Our proposal is consistent 
with expression for the total flux proposed in \cite{Campiglia:2020qvc,Donnay:2022hkf}, and generalizes it by providing a local expression for it valid on any region of $\scri$ and not only on the whole of $\scri$. In spite of this remarkable situation, only the Noether current algebra is covariant. The algebra of eBMS charges we identify still has a residual 2-cocyle, which is however time-independent. In particular, it contains only the super-translation field and generalized Geroch tensor, and no longer the shear.

We use mostly-plus spacetime signature. We denote future null infinity by $\scri$. Greek letters are spacetime indices, 
lower case latin letters $a,b,...$ are $\scri$ indices, and upper case latin letters $A,B,...$ are indices for  2d cross-sections of $\scri$.
In all cases, $(,)$ denotes symmetrization, $\la,\ra$ trace-free symmetrization, and $[,]$ antisymmetrization. An arrow under a $p$-form means pull-back, $\eqons$ means on-shell of the field equations, and $\eqonS$ means an equality valid at $\scri$ only. For the phase space, we use conventions $\om=dp\w dq$ and $\{q,p\}=1$, and define the canonical generator via action of the vector field on the second slot, namely $-I_{\hat F}\om=dF=\{\,\cdot\,,F\}$, so to have $\hat p=\p_q$. With these conventions a Lie symmetry in a standard conservative system is realized as 
\be\nn
I_\xi I_\chi\Om = \{Q_\xi,Q_\chi\} = \d_\chi Q_\xi = Q_{[\xi,\chi]}.
\ee

\section{Radiation at $\scri$}\label{SecRad}

In this Section we review some facts about the description of gravitational radiation at $\scri$ that would be important in the following.
There are excellent reviews in the literature (e.g. \cite{Ashtekar:2014zsa,Grant:2021sxk}), however we believe that some of the properties that we will use may not be well appreciated,  can be scattered across the literature and be hard to find. We refer in particular to: 
the non-trivial relation between the news and the time derivative of the shear, why Geroch tensor is relevant to compute fluxes between arbitrary cross-sections even if one starts from a round sphere frame, the identification between the `connection coordinate' of \cite{Ashtekar:1981hw} (called `relative shear' in \cite{Ashtekar:2024stm}) and the `covariant shear' of \cite{Compere:2018ylh}, and why care is needed when studying the behaviour of the shear under conformal transformations.

We present our results first in covariant language, and then specialized to the asymptotic expansion in Bondi coordinates. 
We hope in this way to be able to communicate to both communities familiar with each language. 
We also use of the Newman-Penrose (NP) formalism, for which we choose the conventions of \cite{Ashtekar:2000hw} where all spin coefficients have opposite signs in order to make up for the mostly-plus signature and preserve the NP equations. 

The covariant language is based on Penrose's conformal completion, whereby $\scri$ is defined as the boundary $\Om=0$ of the auxiliary manifold with conformal (or `unphysical') metric $\hat g_{\m\n}=\Om^2 g_{\m\n}$. See \cite{Geroch:1977jn,Ashtekar:2014zsa} for details. 
While the conformal factor can be chosen arbitrarily, it is very convenient to restrict it so that $\scri$ becomes a non-expanding horizon in the conformal spacetime. This can be done looking at the normal
$n_\m:=\p_\m\Om$ and requiring $\hat\na_\m n_\n\eqonS 0$. Since by the conformal Einstein's equation this condition is equivalent to $\hat\na_\m n^\m\eqonS 0$, where $n^\m=\hat g^{\m\n}\p_\m \Om$ is the tangent null vector field at $\scri$, this choice of conformal frames is referred to as divergence-free. It will be assumed in the rest of the paper, together with completeness of $\scri$ and its topology $\R\times S$, where the cross-sections $S$ (also known as `cuts') are 2-spheres.
Picking a divergence-free conformal compactification has the consequence that $n$ is an affine geodetic vector at $\scri$, that $n^2:=\hat g^{\m\n}n_\m n_\n=O(\Om^2)$, and that $\scri$ is a non-expanding horizon.\footnote{More precisely, a non-expanding horizon endowed with a canonical extremal weakly isolated horizon structure \cite{Ashtekar:2024bpi}.} 
It follows 
that the pull-back of the conformal spacetime connection defines a unique 3d connection, $D_a:=\hat\na_a$. This connection defines the radiative phase space at $\scri$
\cite{Ashtekar:1981hw,Ashtekar:1981bq,Ashtekar:2014zsa}.\footnote{More precisely, the radiative phase space is defined by an equivalence class of connections that removes the dependence on conformal rescalings of the type $\Om'=(1+\Om\m)\Om$ which change the connection but not the background structure $(q_{ab},n^a)$.} 
It also follows that the induced metric $q_{ab}:=\hat g_{ab}$ is time-independent, $\pounds_n q_{ab}=0$, a condition often referred to as `Bondi condition', and numerous manipulations simplify significantly. 
Since this restriction can be done without loss of generality, all asymptotically flat solutions in Penrose's sense\footnote{And also in the weaker sense in which peeling violations in $\psi_1$ are allowed for $l>1$\cite{Bieri:2023cyn}.} share the same universal structure given by
\be\label{us}
(q_{ab},n^a)\sim(\om^2 q_{ab},\om^{-1}n^a), \qquad  q_{ab}n^b=0, \qquad \pounds_n\om=0.
\ee
In other words, there exists a coordinate system in which every solution induces the same metric and normal vector up to a time-independent conformal transformation.
It means that all asymptotic diffeomorphisms that preserve this universal structure
 are symmetries, in the same way as isometries of the background metric are symmetries for physics in Minkowski spacetime. These are the vector fields $\xi$ such that
\be\label{xius}
\pounds_\xi \hat g_{ab} \eqonS 2\a_\xi \hat g_{ab}, \qquad \pounds_\xi n^a \eqonS -\a_\xi n^a.
\ee
 The equations \eqref{xius} can also be understood as the requirement that the unphysical metric and normal to $\scri$ are left invariant by the combined action of a diffeomorphism plus a conformal transformation with infinitesimal conformal factor $1 - \a_\xi$.
The resulting symmetry group is the infinite-dimensional BMS group $\SL\ltimes \R^{S}$ of Lorentz transformations and super-translations.
The  difference is that super-translations can be uniquely identified, as $\xi = f n$ with $\pounds_n f=0$, whereas to identify a Lorentz transformation we need to choose a cross-section of $\scri$, since there is no unique projector `orthogonal' to $n$. This step is analogue to choosing an origin in Minkowski space in order to extract a Lorentz subgroup from the Poincar\'e group, but with the added difficulty that there is a super-translations' worth of cuts to choose from, as opposed to a translations' worth only. The presence of an infinite number of equivalent Poincar\'e subgroups of the BMS group can be made explicit if we pick coordinates $(u,x^A)$ on $\scri$ such that $n\eqonS\p_u$, we can parametrize the solutions to \eqref{xius} as $\xi = f\p_u + Y^A\p_A$, with $f=T+\f u2\Dd_A Y^A$, where 
$T=T(x^A)$ and $Y^A=Y^A(x^B)$ are the symmetry parameters corresponding respectively to super-translations and conformal Killing vectors (CKV) of the cross-sections, whose covariant derivative is $\Dd_A$, and
which span the (double cover of the proper orthocronous) Lorentz group $\SL$. However while $T$ is uniquely defined, $Y^A$ refer explicitly to the leaves of the $u$-foliation. We also recall that the group of super-translations contains a subgroup of global translations which is also uniquely defined, however its `orthogonal' complement is not, because there is no natural metric in this space. Hence the notion of a `pure super-translation', namely a super-translation without any global translation component, is also not unique but foliation and frame dependent. 
Super-translations and rotations of any Lorentz subgroup preserve the conformal frame, whereas boosts change it. Translations and rotations preserve any given foliation, whereas super-translations and boosts do not.

It is  common in the literature to further restrict the conformal freedom and choose $\om$ so that the induced metric on cross-sections is a unit round sphere. These special conformal completions are called Bondi frames (not to be confused with the Bondi condition above). This can always be done and would not change the symmetry group nor the physics in any way, but it may not be very convenient in practice, because checking conformal invariance of the physical expressions becomes more complicated: one cannot do arbitrary conformal transformations but has to take into account the non-trivial functional dependence that conformal transformations relating round spheres must have (namely, correspond to Lorentz boosts on the celestial sphere).

While the symmetry group is defined \emph{uniquely} in terms of the available intrinsic structure at $\scri$, the covariant phase space requires an embedding of $\scri$ in the conformal spacetime. It is always possible to choose coordinates $(u,\Om,x^A)$ of the embedding so that $n\eqonS\p_u$. Thanks to the Bondi conditions, these coordinates are affine, in the sense that $u$ is an affine parameter for the null geodesics, and $x^A$ are Lie dragged by $n$.
This means that the whole metric at $\scri$ is universal, and not just its pull-back:
\be\label{dhatg0}
\d \hat g_{\m\n}=0.
\ee
As a consequence, the first-order extension of the symmetry vector fields is fixed, and the arbitrariness of their bulk extension starts at $O(\Om^2)$ \cite{Grant:2021sxk}.

In the covariant description, the radiative content of the gravitational field is encoded in  $\hat S_{ab}$, the pull-back to $\scri$ of the unphysical Schouten tensor $\hat S_{\m\n}$. However, this tensor depends  on the  conformal completion chosen: changing it via $\Om\to\om\Om$ does not affect the physics but changes $\hat S_{ab}$. To extract the information on the physical radiation one has to get rid of this dependence.
The problem was solved by Geroch \cite{Geroch:1977jn}, who found that there exists a unique kinematical tensor $\r_{ab}$ whose behaviour under conformal transformations matches exactly the one of the pull-back $\hat S_{ab}$.
The fact that it is unique and kinematical means that it can be subtracted off $\hat S_{ab}$ without affecting the physical content.
The resulting quantity is the news tensor
\be\label{News}
N_{ab}:=\hat S_{ab} -  \r_{ab}.
\ee
It is conformally invariant, traceless, and describes the gravitational radiation in an unambiguous way. 

Geroch's tensor is defined on $\scri$ by four conditions,
\be\label{rhoCond}
\r_{[ab]}=0, \qquad \r_{ab}n^b= 0, \qquad D_{[a}\r_{b]c}= 0, \qquad q^{ab}\r_{ab}=\cR.
\ee
Any tensor like $\rho_{ab}$ whose contraction with $n^a$ gives zero (the second condition) is called `transverse', or `horizontal'.
In the last equation, $\cR$ is the 2d Ricci scalar, and the fact that $\r_{ab}$ is transverse means that one can use any `inverse' in the equivalence class $q^{ab} \sim q^{ab}+n^a X^b$.
These equations imply
\be\label{rgradR}
D^b\r_{ab}= \p_a \cR, \qquad D^b\r_{\la ab \ra}= \f12 \p_a \cR. 
\ee
Furthermore $\pounds_n\r_{ab}=0$ from the Bondi condition.
From the behaviour of $\cR$ under conformal transformations \eqref{us}, 
it follows that 
\be\label{rhoconf}
\r_{ab}'= \r_{ab} -2\om^{-1}D_a D_b \om +4\om^{-2} D_a\om D_b\om -\om^{-2} g_{ab} D^c\om D_c\om. 
\ee
We will denote $\r'-\r=\D_\om\r$, whose linearization for $\om=1+W$ is 
\be\label{rhoconflin}
\D_W\r_{ab}= -2D_a D_b W +O(W^2).
\ee
This conformal transformation matches precisely the one of $\hat S_{ab}$, hence \eqref{News} is conformally invariant.
To prove that a solution to \eqref{rhoCond} exists, it is enough to choose a Bondi frame, which we denote by $q_{AB}=\os{q}_{AB}$ with $\cR=\os{\cR}=2$,
because then 
\be\label{rhoround}
\r_{ab}=\f{\os{\cR}}2\os{q}_{ab}
\ee
is manifestly a solution. To prove that is unique is a bit more elaborate, and crucially relies on the spherical topology of the cross-sections \cite{Geroch:1977jn}. 
Once this is established, the solution in an arbitrary frame is obtained from \eqref{rhoconf}. Since $\r_{ab}$ is uniquely determined by the background metric $q_{ab}$ and the latter is universal, it  is also universal. This may look surprising at first, because BMS boosts induce a conformal transformation of the frame, and $\r_{ab}$ is not conformally invariant. However, the same uniqueness arguments based on the topology of the sphere lead to  \cite{Geroch:1977jn}
\be\label{lierho}
\pounds_\xi\r_{ab}= -2D_a D_b \a_\xi.
\ee
The key point is that this coincides with a linearized conformal transformation \eqref{rhoconflin} with $\a_\xi=W$. Hence combining 
\eqref{lierho} with \eqref{rhoconflin} so to keep $q_{ab}$ invariant as by definition of the BMS symmetry group, $\r_{ab}$ remains also invariant. In other words $\d_\xi \r_{ab}=0$, consistently with being universal.

Geroch's tensor plays a crucial role in turning many frame-dependent statements into conformally-invariant ones. For instance, super-translations contain a unique subgroup of global translations, which on a Bondi frame can be identified as the $l=0,1$ modes of $T$, namely as the solutions to $D_{\la a}D_{b\ra}T=0$.
This equation is however not conformally invariant, and it is only valid on Bondi frames. In arbitrary frames, it is replaced by
\be\label{Tideal}
\left(D_{\la a}D_{b\ra}+\f12\r_{\la ab\ra}\right)T = 0.
\ee
Its conformal invariance can easily be checked using \eqref{rhoconf} and the fact that $T$ has conformal weight 1.

In the literature the news is often presented in terms of 
(the time variation of) an asymptotic shear, in order to give it a more intuitive geometric meaning. This relation however 
requires introducing additional structure, because while the news is a unique covariant tensor, an asymptotic shear refers to a foliation of $\scri$. 
In Bondi coordinates there is a natural foliation given by the level sets of the coordinate $u$, and we will come back to it. 
An alternative way to talk about shear without fixing a specific foliation is to introduce an auxiliary null vector $l$ such that $l\cdot n \eqonS -1$ (also known as `rigging' vector). 
We require $l$ to be hypersurface orthogonal on $\scri$, so that it is equivalent to a choice of foliation. It then 
defines a projector on the space-like cross-sections (`cuts') of $\scri$, which we denote $\g_{\m\n}:=\hat g_{\m\n}+2n_{(\m}l_{\n)}\equiv 2m_{(\m}\bar m_{\n)}$. Notice that $\g_{ab}=q_{ab}$ and that $\g^{\Om\m}=0=\g^\Om_\m$ hence $\g^{\m\n}$ has the same content as $\g^{ab}$, and provides a choice of `inverse' for the induced metric that annihilates the auxiliary null form $l$. 
The shear and expansion of this arbitrary foliation associated with $l$ are 
\be\label{sdef}
\s_{\m\n}:=\g_{\la\m}^\r \g_{\n\ra}^\s \hat\na_\r l_\s, \qquad \th:=\g^{\m\n}\hat\na_\m l_\n.
\ee
They are related to (the pull-back of) the gradient of $l$ by
\be\label{Dtoshear}
D_{a}l_b = \g_{a}^c \g_b^d \hat\na_c l_d - l_a \t_b = \s_{ab}  +\f\th2 q_{ab} - l_a\t_b,
\ee
where 
\be
\t_a:= \pounds_{n} l_a, \qquad \t\cdot n=\t\cdot l=0.
\ee
The time-dependence of the connection can be computed using the fact that the conformal metric's Weyl tensor vanishes at $\scri$, giving
\be\label{Dtonews}
[\pounds_n, D_a]l_b = \hat R_{a\r\s b} n^\r l^\s \eqons \f12 \hat S_{ab} = \f12 (N_{ab}+\r_{ab}).
\ee
Then using the relation between the Schouten tensor and the normal $n$ provided by the Einstein's equation, one can prove that
\begin{align}\label{Nshear}
N_{ab} = 2\pounds_{ n} \s_{ab} - 2(D_{\la a}+\t_{\la a})\t_{b\ra}-2 l_{(a}\pounds_{ n} \t_{b)} -\r_{\la ab\ra},
\end{align}
which is the general relation between the news and the shear.

The general formula \eqref{Nshear} is not very common in the literature, because $\r_{\la ab\ra}$ and $\t_a$ can be set to zero choosing specific background structures, without affecting the physics nor the symmetries. It is however instructive to appreciate the role of the various extra terms, as well as the logic that goes behind the specific restrictions one may choose. The key point is that the shear depends on two background structures: the conformal factor, and the choice of $l$.
The freedom to change $l$ is an internal Lorentz transformation belonging to $n$'s little group, which we'll refer to as class-II  following \cite{Chandra}:
\be\label{classII}
l\to l+\bar a m+a\bar m+|a|^2n, \qquad m\to m+an, \qquad s\in\C.
\ee
This is a 2-real-parameter family a priori, but we restrict it requiring $l$ to remain hypersurface orthogonal. Changing $l$ within this class we change the foliation to which $\s_{ab}$ makes reference, so the first term in \eqref{Nshear} is not class-II invariant. But the $\t_a$ terms in \eqref{Nshear} are also not invariant under \eqref{classII}, and their transformation compensates the transformation of $\s_{ab}$, so that the whole expression is class-II invariant. 
Concerning the conformal factor, under \eqref{us} we have $l\to\om l$, so $\pounds_n\s_{ab}$ is invariant. However $\t_a\to\t_a+\p_a\ln\om$, hence the 
 $\t_a$ terms are not invariant: their transformation compensates the transformation of $\r_{\la ab\ra}$, so that the whole expression is conformally invariant.
 
Having clarified this, let's see what happens when these background terms are simplified. 
As mentioned above, one could limit the conformal frames to be round spheres only, namely `Bondi frames'. Then $\r_{\la ab\ra}=0$, as \eqref{rhoround} shows.
In other words, if we restrict the conformal transformations to those that preserves round spheres, $\hat S_{\la ab\ra}$ is conformally invariant, and Geroch's tensor is only needed to remove the trace part.
Alternatively, one could choose to work with Lie-dragged auxiliary vectors only, then $\t_a=0$. To see what this means, let's fix coordinates so that $n\eqonS \p_u$. Then the class of  hypersurface-orthogonal auxiliary vectors Lie-dragged by this $n$ describes all foliations that differ from the level sets of $u$ by a super-translation only. 
With this choice, \eqref{Nshear} reduces to \cite{Dray:1984rfa} 
 \be\label{NDS}
 N_{ab}=2\pounds_n \s_{ab}-\r_{\la ab\ra},
 \ee
 or in terms of NP scalars, $N=-\dot{\bar\s}-\f 12b$, where  $\s:=-m^a m^b\s_{ab}$ and 
$b:=\bar m^a \bar m^b \r_{ab}$ is the inscrutable notation used in \cite{Dray:1984rfa} for the spin-2 weighted projection of Geroch's tensor.\footnote{Possibly $b$ for Bob?} To check conformal invariance of this expression, one has to be careful, because transforming $l\to\om l$ does not preserve $\t_a=0$. The solution is to add a a class-II transformation with $a=-\om^{-1}\pounds_m \tl u$, where $\tl u=\om u$:
\be\label{inol}
l  \to l'= \om l -\p_A\tl u \, dx^A + \f1{2\om} \p_A \tl u \p^A \tl u\, n.
\ee
This rule for the conformal transformation preserves $\t_a=0$. Using it in the shear, we get 
\be\label{sconf}
\s_{ab}\to \s'_{ab} = \om \s_{ab}
- u \big(D_{\la a} D_{b\ra} \om - 2\om D_{\la a} \ln\om D_{b\ra} \ln\om\big).
\ee
The inhonomogeneous terms can be recognized as $\f 12u\om \D_\om\r_{\la ab\ra}$, hence 
Geroch's tensor in \eqref{NDS} makes the expression conformally invariant in the subset of Lie-dragged $l$'s.
Finally if one chooses both Bondi frames and Lie dragged $l$, then $N_{ab}=2\pounds_n \s_{ab}$, or in terms of NP scalars, $N=-\dot{\bar\s}$.
Conformal invariance of this expression requires one to transform $l$ homogeneously, hence a non-trivial $\t_a$ must be included if the conformal factor does not preserve round spheres. This is for instance the set up used in the review \cite{Grant:2021sxk}. We prefer to use a set up in which we fix $\t_a=0$, because it provides a simplification of many formulas that can be done without any loss of generality, at the small price that the news is given by \eqref{NDS} and not just the time derivative of the shear. It is furthermore the set up that arises naturally when working in Bondi coordinates, as we will review in the next Section.

As discussed above, restricting to $\t_a=0$ means considering only shears adapted to foliations related by a super-translation, on a fixed conformal frame. This has an important consequence for the flux-balance laws, because some of the charge aspects depend on the shear, and thus require a choice of cross section in order to be defined. 
If the initial and final cross section considered belong to the same $u$ foliation, then we can use the same Lie-dragged $l$ to describe them. But if they don't, namely they differ by a super-translation, then the foliation linking them is described by a non-Lie dragged $l$. This problem can be dealt with in two different ways. The first is to stick with the non-Lie dragged $l$, and explicitly map the symmetry parameters and charge aspects of the final cross-section to those of the initial cross-section, which can be done using a BMS transformations. This is for example what is done in \cite{Flanagan:2015pxa,Chen:2022fbu}. But there is a more elegant alternative, which is due to Dray \cite{Dray:1984gz}: One can change frame so that the two cross sections belong now to the same $u$ foliation. With this trick, the symmetry parameters and charge aspects \emph{are the same} on both cross sections, but one is in general no longer working on a round sphere. 
See Appendix~\ref{AppDray} for details. 
In summary, we have  seen two convenient reasons to not limit the conformal compactifications to be only round spheres: first, checking conformal invariance is simpler;
second, it is possible to write the flux between two arbitrary cross-sections using charges described by a Lie dragged $l$. 

The transformation \eqref{inol} can be easily generalized to include an arbitrary super-translation of the foliation that $l$ is orthogonal to. This is done replacing $\tl u=\om u$ with 
\be\label{tlu2}
\tl u=\om(u+T), 
\ee
where $T(x^A)$ is the super-translation. The ensuing transformation of the shear is
\be\label{gens'}
\s_{ab}\to \s'_{ab} = \om \s_{ab}
- (u+T) \big(D_{\la a} D_{b\ra} \om - 2\om D_{\la a} \ln\om D_{b\ra} \ln\om\big) - \omega D_{\la a} D_{b\ra} T.
\ee
The same transformation rule is also studied in \cite{Barnich:2016lyg}, using the Newman-Penrose formalism.
For later purposes, we note here the linearization of \eqref{gens'}, with $\om=1+W$ and $T$ assumed small and same order of $W$,
\be\label{gens'lin}
\s'_{ab} = \s_{ab} +W \s_{ab} - D_{\la a} D_{b\ra} (T+u W).
\ee

The formulas \eqref{Dtoshear} and \eqref{Dtonews} make it clear that the connection $D_a=\hat\na_a$ describes both the news and the shear.
To elaborate further on this relation, one can use the auxiliary rigging vector to define a Newman-Penrose basis at $\scri$ (and there only, we do not require the vectors $(l, n)$ to be null everywhere). Taking $l$ hypersurface-orthogonal implies that the spin coefficient $\r$ is real,  the Bondi condition implies that the real part of the spin coefficient $\g$ vanishes and that the spin coefficient $\t$ describes the non-Lie dragging $\t_a$. All the NP quantities refer to the conformal metric,
but to simplify the notation we don't mark them with hats, and furthermore we will remove the traditional $^\circ$ that stands for leading order terms at $\scri$, with the understanding that all NP symbols used here refer to the leading order asymptotic quantities.
In the Newman-Penrose basis 
\be\label{psi34}
\psi_3  =\f12 \bar m^aD^bN_{ab} =\eth N, 
\qquad \psi_4=-\ddot{\bar\s} =\dot N.
\ee
Here $N=\f12\bar m^a\bar m^b N_{ab}$ is the spin-weighted projection of the news tensor, and $\eth$ (`eth') is the 2d covariant derivative on spin-weighted NP scalars. It appears because $N_{ab}n^b=0$ hence the divergence effectively reduces to a 2d covariant derivative on the cross-sections. `$\eth$-calculus' is very convenient for many manipulations, but can be freely traded for a tensorial notation via
\be
\eth (m^a\bar m^b \psi_{ab}) = m^a\bar m^bm^c\Dd_c\psi_{ab}, 
\ee
where 
the 2d covariant derivative is
\be
\Dd_a \psi_{b}:=\g_a^c\g_b^dD_c\psi_d=(D_a + l_a\pounds_n)\psi_b, \qquad \psi_an^a=0.
\ee
This equation also shows that the 3d derivative $D_a$ acts universally on transverse (or `horizontal') fields, namely in a way independent of the radiation.
The two derivatives coincide on time-independent fields, as well as when pull-backs on cross-sections are involved.
This is what happens in the first equality of \eqref{psi34}, where the divergence is taken with respect to $\g^{ab}$.\footnote{\label{doublederivative}This is only true for first derivatives, for instance for second derivatives we have
\be \nn
\Dd_a \Dd_b f
= D_a D_b f + (D_a l_b+ 2l_{(a}D_{b)}+ l_al_b\pounds_n)\pounds_n f
\ee
and
\be\nn
\Dd_a \Dd_b \s^{ab} =  D_a D_b \s^{ab} + \s_{ab}\dot\s^{ab}.
\ee
}
Since there is no constant tensor on $\scri$, \eqref{psi34} don't have zero modes and we conclude that the news tensor is equivalent to knowing $\psi_3$ and $\psi_4$. 

Concerning the shear, its two components can be split into an `electric' and a `magnetic' part, with the latter super-translation invariant. The magnetic part is related to the news and $\im(\psi_2)$ via  \cite{Newman:1968uj}
\be
\im(\psi_2)\eqons \im\left(\bar\eth^2\s- \s \dot{\bar\s}\right) = \im\left(\Big(\bar\eth^2+\f b2\Big)\s+\s N\right)
= -\f14\eps^{ab}\Big( (\Dd_a \Dd_c +\f12\r_{ac})\s^c{}_b+\f12N_{ac}\s^c{}_b\Big),
\ee
where in the second equality we used \eqref{NDS}, and in the third equality $\eps^{ab}:=-2im^{[a}\bar m^{b]} $.
It is customary to  strengthen the non-radiative conditions requiring $\im(\psi_2)=0$ on top of $N=0$,\footnote{Non-radiative spacetimes so defined possess a unique preferred Poincar\'e subgroup of the BMS group \cite{Ashtekar:2019rpv}.} 
implying
\be\label{magnshear}
\eps^{ab}\left(\Dd_a \Dd_c +\f12\r_{ac}\right)\s^c{}_b=0,
\ee
namely a `purely electric' shear. 
The connections associated with non-radiative spacetimes are called vacuum solutions in the radiative phase space.
We conclude that in any given conformal frame, the (equivalence class of the) connection determines the news and the shear, or equivalently $\im(\psi_2), \psi_3, \psi_4$ and the electric part of the shear, and that a vacuum connection depends only on the electric part of the shear.

It is useful to make this dependence more explicit.
If we specialize \eqref{Dtonews} to a vacuum connection $\Dv$ we find $[\pounds_n, \Dv_a]l_b \eqons \f12\r_{ab}$, and since $\pounds_n\r_{ab}=0$, we conclude that for a Lie-dragged $l$,
\be\label{vacuumshear}
\Dv_{\la a} l_{b\ra} = \sv_{ab} = \f12u\r_{\la ab\ra} - c_{ab}.
\ee
To determine the time-independent field $c_{ab}$, we impose the vacuum condition \eqref{magnshear}.
The differential operator  annihilates Geroch's tensor (it is `purely electric'), and the general solution is
\be\label{elecshear}
c_{ab}= \left(\Dd_{\la a} \Dd_{b\ra} +\f12\r_{\la ab\ra}\right)u_0, \qquad \pounds_n u_0=0,
\ee
or $m^am^bc_{ab}=(\eth^2+\f12b)u_0$ in NP language.
This is the same operator that appears in \eqref{Tideal}, since $T$ is time independent.
It has a four-dimensional kernel,
given on round spheres by the $l=0,1$ spherical harmonics.
Since $c_{ab}$ is entirely determined by a free function on the sphere, it can be always set to zero with a super-translations. 
Once it is set to zero, it remains so for the 4-parameter family of zero modes, namely the global translations. This is the 4-parameter family of shear-free cross-sections, namely the famous `good cuts'.
The solution \eqref{vacuumshear} with $c_{ab}$ given by \eqref{elecshear} determines any vacuum shear as a function of a choice of origin in the radiative phase space, namely $\s=0$ for the chosen $l$, and a choice of `bad cut' $u_0=u_0(x^A)$. 

More in general,
the split into electric and magnetic parts of an arbitrary shear can be parametrized in terms of two  time-dependent functions $\Phi$ and $\Psi$ of conformal weight 1, via
\be\label{emgen}
\sv_{ab} - \f u2\r_{\la ab\ra} = (\Dd_{\la a} \Dd_{b\ra} +\f12\r_{\la ab\ra})\Phi + \eps_{\la a}{}^{c}(\Dd_{b\ra} \Dd_c +\f12\r_{b\ra c})\Psi.
\ee
This formula reduces to the standard Helmholtz decomposition on round spheres (but which is not conformally invariant, hence the need of Geroch's tensor in the general formula). For vanishing news the functions are time-independent.
Then the purely-electric part $\Phi$ is the one that can be set to zero with a super-translation, and the purely-magnetic one $\Psi$ is set to zero adding $\im(\psi_2)=0$ to the definition of non-radiative.

Given a general connection and a vacuum connection parametrized by $u_0$, we define the \emph{relative shear}
\be\label{cS}
{\cal S}_{ab}:=(D_{\la a}-\Dv_{\la a})l_{b\ra} = \s_{ab}-\sv_{ab} = \s_{ab}  - \f u2\r_{ab} + \left(\Dd_{\la a}\Dd_{b\ra}+\f12\r_{\la ab\ra}\right) u_0.
\ee
Applying \eqref{gens'} we see that it is invariant under super-translations, and it
 transforms homogeneously with weight 1 under conformal transformations. Namely 
\be\label{calS'}
{\cal S}_{ab}\to{\cal S}_{ab}'=\om {\cal S}_{ab}.
\ee
The relative shear provides a potential for the news, since
\be\label{NcalS}
N_{ab}=2\pounds_n {\cal S}_{ab}.
\ee
The quantity ${\cal S}_{ab}$ is precisely the `covariant shear' of \cite{Compere:2018ylh}, there obtained from a gBMS coordinate transformation of Minkowski in Cartesian coordinates, here derived in a coordinate independent way from the connection description of $\scri$ \cite{Ashtekar:1981hw}.
It is closely related to the `connection coordinate' of \cite{Ashtekar:1981hw} (denoted relative shear in \cite{Ashtekar:2024stm}), where however $\sv$ is taken as a choice of origin fixed once and for all, as opposed to a variable choice of vacuum.

The relative shear is  convenient to encode temporal boundary conditions on the radiation. We require stationarity in the far future, which we impose asking that the connection goes to a vacuum state:
\be\label{ASbc}
\lim_{u\to\infty} N_{ab} = \f1{u^{1+\varepsilon}}, \qquad \lim_{u\to\infty} \im(\psi_2) = \f1{u^{\varepsilon}}, \qquad \varepsilon>0.
\ee
If we now pick a specific vacuum state $\sv$, and we use it in the definition of the relative shear, we can rewrite the boundary conditions as
\be\label{S0}
\lim_{u\to\infty} \cS_{ab}=0.
\ee
In other words, the relative shear refines the description of the radiative phase space parametrizing the shear into a term $\cS_{ab}$ that vanishes in the far future plus the corner datum 
$u_0$, which represents all possible late times vacuum boundary conditions. In this way, the variations $\d u_0$ (equivalently $\d \sv$) parametrize the directions corresponding to the different boundary conditions. The importance of this decomposition, and of treating $u_0$ as a dynamical quantity, was first pointed out in \cite{He:2014laa}.

\subsection{Bondi asymptotic expansion and anomalies}\label{SecBondi}

Let us now specialize the above covariant formulas to the Bondi expansion. One advantage of it is that it makes computing the action of the BMS transformations on the asymptotic fields completely straightforward. We denote the Bondi coordinates $(u,r,x^A)$ with $r$ the area radius, and assume standard BMS fall-off conditions. We then have
\begin{subequations}\label{gexp}\begin{align}
g_{uu} &= -\f\cR 2+\f{2M}r +  O(r^{-2}), \qquad 
g_{ur} = -1-\f{2\b}{r^2}+O(r^{-3}), \qquad \b:=-\f1{32}C^{AB}C_{AB}, \\
g_{uA} &=-U_A+\f2{3r}(\am_A+\p_A\b-\f12 C_{AB}U^B)+O(r^{-2}), \label{guAexp} \qquad U_A:=-\f12 \Dd^BC_{AB}, \\\label{gABexp}
g_{AB} &= r^2 q_{AB}+r C_{AB}+ O(1).
\end{align}\end{subequations}
We take $\Om=1/r$ as conformal factor, then the only non-vanishing components of the unphysical metric $\hat g_{\m\n}:=\Om^{2}g_{\m\n}$  at $\scri$ are $\hat g_{\Om u} = 1$ and $\hat g_{AB}=q_{AB}$, 
namely 
\be \label{backgrounddecomp}
\hat{g}_{\m \n} dx^\m dx^\n \eqonS 2dud\Om+q_{AB}dx^A dx^B.
\ee
The background 2d metric $q_{AB}$ is universal, $\d q_{AB}=0$,
 we denote $\Dd_A$ its covariant derivative, and $\cR_{AB}=\f12q_{AB}\cR$ its Ricci tensor.
The coordinates $(u,x^A)$ on $\scri$ define a foliation associated with retarded time, and $n\eqonS\p_u$.  
The volume form is $\eps_{\scri}=du\w\eps_S$ where $\eps_S = i_n \eps_{\scri}$ is the area 2-form of the cross-sections.
From the Bondi condition $\pounds_n q_{ab}=0$, hence also $\pounds_n \eps_\scri = 0$ and $d\eps_S=0$.
The embedding makes $u$ an affine parameter for $\scri$, and we have $n^2= \frac{1}{2}\cR\Om^2 +O(\Om^3)$. 

The dynamical fields are $M,\am_A$ and $C_{AB}$. The first two are related to the mass and angular momentum aspects, see below.
They are determined by the asymptotic Einstein's equations via
\begin{align}\label{Mdot}
& \dot M = -\f18 \dot C_{AB}\dot C^{AB}+\f14 \Dd_A\Dd_B \dot C^{AB}+ \f18\Dd^2\cR, \\\label{MEE}
& \dot \am_A = \Dd_AM+\f12 \Dd^B\Dd_{[A}\Dd_C C_{B]}{}^C +\f14C^{AB}\Dd_B\cR + \f12\Dd^B( \dot C_{[B}{}^C C_{A]C})-\f14\dot C_{BC}\Dd_A C^{BC}.
\end{align}
The definition of $\am_A$  
corresponds to the choice $(1,1)$ in the parametrization of \cite{Compere:2019gft}, and it is related to \cite{Barnich:2011mi} and \cite{Flanagan:2015pxa}
respectively by
\be
\am_A = N^{\sscr BT}_A - \p_A\b= N_A^{\sscr FN}+2\p_A\b+\f12C_{AB}U^B,
\ee
or equivalently
\be
g_{uA}=-U_A+\f2{3r}(N^{\sscr BT}_A-\f12 C_{AB}U^B) =-U_A+\f2{3r}(N^{\sscr FN}_A+3\p_A\b).
\ee
The field $C_{AB}$ is related to the shear of the $u$ foliation by
\be
\s_{ab}=-\f12 C_{AB} \, \d^A_a\d^B_b,
\ee
and in terms of \eqref{sdef} it
corresponds to $l=-du$, which is manifestly hypersurface-orthogonal and Lie dragged by $n$. 
An explicit calculation of the unphysical Schouten tensor gives
\be\label{SchoutenBondi}
\hat S_{ab} = -\d_a^A\d_b^B \dot C_{AB} + \f{\cal R}2q_{ab}.
\ee
Recalling the properties of the Geroch tensor listed earlier, the only non-vanishing components of the news tensor are 
\be\label{NAB}
N_{AB}=-\dot C_{AB}-\r_{\la AB\ra},
\ee
in agreement with the general formula \eqref{Nshear}.

The solutions to \eqref{xius} in the Bondi coordinates $(u,x^A)$ read $\xi = f\p_u + Y^A\p_A$, where as before $f=T+\f u2\Dd_A Y^A$, and
$T=T(x^A)$ and $Y^A=Y^A(x^B)$ are the symmetry parameters corresponding respectively to super-translations and conformal Killing vectors of the 2-sphere
associated with the $u$ foliation. 
$T$ and $u$ have conformal weight 1, and $Y$ has conformal weight 0.
Since we have fixed the coordinate gauge freedom in the bulk, we can also fix the bulk extension of the symmetry vector fields, asking that they preserve the Bondi coordinates. This includes preserving the affine embedding, hence \eqref{dhatg0} is satisfied.
The result is
\be\label{xi}
\xi = f\p_u + Y^A\p_A +\Om(\dot f\p_\Om -\p^A f\p_A) - \f12\Om^2(\Dd^2 f \p_\Om - C^{AB} \p_B f \p_A) + O(\Om^3).
\ee
Notice that $\xi$ is field-dependent starting at second order. 
It satisfies
\be\label{TW}
2l^\n\na_{(\m}\xi_{\n)} = \na_\n\xi^\n \, l_\m,
\ee
where $l_\m:=-\p_\m u$. It can be recognized as the Tamburino-Winicour condition for the extension \cite{Tamburino:1966zz}.

To write the action of the symmetries on the dynamical fields, we use the covariant phase space. 
We follow the notation of \cite{Freidel:2021cjp} where $\d$ is the exterior derivative, $I_V$ the internal product with a vector field $V$, and $\d_V = I_V \d+\d I_V$ the field-space Lie derivative. 
Together with their spacetime counterparts $(d,i_v,\pounds_v)$, they define a bi-variational complex with $[d,\d]=0$ (the opposite sign convention is used in \cite{Barnich:2001jy}). 
The field-space vector field corresponding to a diffeomorphism is $V_\xi=\int d^4x \pounds_\xi \phi \f{\d}{\d\phi}$ and we use $\d_{V_\xi}=\d_\xi$ for short.
We also use the anomaly operator $\D_\xi:=\d_\xi-\pounds_\xi-I_{\d\xi}$  \cite{Hopfmuller:2018fni,Chandrasekaran:2020wwn,Freidel:2021cjp,Odak:2022ndm}.  
It measures the breaking of covariance, namely 
discrepancies between $\d_\xi$ and the spacetime Lie derivative $\pounds_\xi$ that can be introduced in the presence of background structures, gauge-fixing, and field-dependent diffeomorphisms.

The action of a BMS transformation in the covariant phase space then corresponds to a transformation $\d_\xi$ where $\xi$ is a symmetry vector field.
To compute it, we have to take into account the presence of two background fields that are used in the asymptotic expansion: the conformal factor $\Om$, and the foliation of $\scri$ provided by the Bondi time $u$, and which is used to define the shear. 
If we see $\Om$ and $u$ as part of a coordinate system, also the remaining $x^A$ coordinates are part of the background, but they are not needed to be included in the list of background fields because none of the quantities used makes reference to a specific choice for them.
Let us denote the background fields collectively with $\bac$. They are universal, hence $\d\bac=0$ and $\d_\xi\bac=0$. For the dynamical fields, here just the metric, we have by definition $\d_\xi g_{\m\n}=\pounds_\xi g_{\m\n}$. 
It follows that for a generic scalar functional $F(g_{\m\n},\bac)$ that depends on both dynamical and background fields, like $M,\am,C,q$ above, there is a discrepancy between the field space and spacetime Lie derivatives:
\be\label{anodef}
\d_\xi F := F(g_{\m\n}+\pounds_\xi g_{\m\n},\bac) - F(g_{\m\n},\bac) = 
\f{\p F}{\p g_{\m\n}}\pounds_\xi g_{\m\n}=\pounds_\xi F+\D_\xi F, 
\ee
where
\be\label{defDxi}
\D_\xi F=(\d_\xi-\pounds_\xi)F = -\f{\p F}{\p\bac}\pounds_\xi\bac
\ee
is the anomaly.
It coincides with the  definition in the previous paragraph because we are only acting on field-space scalars hence $I_{\d\xi}$ is trivial regardless of whether $\d\xi=0$ or not.
From the definition \eqref{anodef}, we see that the action of $\d_\xi$ can be computed writing 
$\pounds_\xi g_{AB} = r^2\d_\xi q_{AB} + r\d_\xi C_{AB}+O(1)$ etc., and using \eqref{gexp} and \eqref{xi} one finds
(see e.g. \cite{Compere:2018ylh,Freidel:2021yqe,Odak:2022ndm})
\begin{subequations}\label{BMStransf}\begin{align}\label{dxiq}
& \d_{\xi} \,{q}_{A B}= (f\p_u+\pounds_Y-2 \dot{f}) {q}_{A B}=0, \\\label{dxiC}
& \d_{\xi} \,C_{A B}= (f\p_u+\pounds_Y-\dot{f}) C_{A B}-2 {\Dd}_{\langle A} \Dd_{B\rangle} f\\\nn
&\qquad\quad=-f N_{AB}+(\pounds_Y-\dot{f}) C_{A B}-2 ({\Dd}_{\langle A} \Dd_{B\rangle}+\f12\r_{\la AB\ra}) f,  \\
& \d_{\xi} \,M= (f\p_u+\pounds_Y + 3 \dot{f}) M - \f{1}{2} {\Dd}_A N^{AB} \Dd_B f +\f{1}{4} \p_u(C^{A B} {\Dd}_A \Dd_B f), \label{dxiM}\\
& \d_{\xi} \,\am_A=  (\t\p_u + \pounds_Y+2 \dot{ f}) \am_A 
+3 M \partial_A  f + \frac{1}{8} N_{BC} C^{B C} \Dd_A  f - \frac{1}{2}C_A^C N_{BC}\Dd^B  f \label{dxiJ} \\\nn
&\qquad\qquad+\frac{3}{2}\Dd_{[A} \Dd^C C_{B]C}\Dd^B  f+\frac{1}{4} \Dd_A\left(C^{B C} \Dd_B \Dd_C  f\right) 
+\frac{1}{2} \Dd_{\langle A} \Dd_{B\rangle}  f \Dd_C C^{B C}
\\\nn&\qquad\qquad + \frac{1}{4}C_{A B} \Dd^B \Dd^2  f
+ \f{1}{8} \r_{BC} C^{B C} \Dd_A  f - \f{1}{2} C_A^C \r_{\la BC\ra} \Dd^B  f .
\end{align}\end{subequations}

The second equality in \eqref{dxiq} follows from the Bondi condition and the restriction of the $Y$'s to be CKVs, hence
\be\label{CKV}
\Dd_{\la A}Y_{B\ra}=0
\ee
for any 2d metric. 
It shows that the symmetry can be understood as the requirement that the unphysical metric is left invariant by the combined action of a diffeomorphism plus a compensating conformal transformation. Taking two derivatives of \eqref{CKV} we find $(\Dd^2+\cR)\Dd Y=-\pounds_Y\cR$. If $q_{AB}$ is a round sphere, this equation reduces to $(\Dd^2+2)\Dd Y=0$ implying that $\Dd Y$ has $l=1$ modes only. This in turns implies that 
\be\label{DDDY}
{\Dd}_{\langle A} \Dd_{B\rangle} \Dd Y=0
\ee
on round spheres.
Switching back to arbitrary frames, 
we conclude that
\be\label{DDrDY}
({\Dd}_{\langle A} \Dd_{B\rangle} +\f12\r_{\la AB\ra})u\Dd_CY^C =0
\ee
for a globally defined CKV. Notice that $u$ is needed here to make the equation conformally invariant. 
This is the operator $\DDr$ in Bondi coordinates, and we have thus seen that it annihilates both global translations and boosts.
The second line of \eqref{dxiC} uses \eqref{NAB}, and allows us to see that only (non-global) super-translations induce a inhomogeneous transformation on the shear. This also implies that a BMS transformation cannot induce a magnetic shear, but only an electric one.

Let us now talk about the anomalies and their meaning. The background fields are $\Om$ and $u$, namely the choice of conformal compactification and of foliation of $\scri$. The anomaly $\D_\xi\Om = -\dot f\Om$ measures the conformal weight 1 of $\Om$, and $\D_\xi u=-T-\dot f u$ measures its conformal weight 1 as well as its 
``super-translation weight" 1, namely the fact that the foliation is not invariant under super-translations.  Similarly for the conformal metric, the anomaly
is $\D_\xi \hat g_{ab} = -2\dot f \hat g_{ab}$ and  picks up its conformal weight 2. The list of anomalies for the purely background fields is 
\begin{align}\label{BMSano1}
& \D_\xi\Om = -\dot f\Om, \qquad \D_\xi u=-T-\dot f u, \\\nn& \D_\xi n_\m = -\dot f n_\m,
\qquad \D_\xi \eps_S = -2\dot f\eps_S, \qquad \D_\xi \eps_\scri= -3\dot f\eps_\scri.
\end{align}
The minus signs in these expressions are conventional, and follow from the definition \eqref{defDxi}.
What we learn from this analysis is that the symmetry group is large enough to probe the background, making the anomaly operator an effective tester of background independence. More precisely, boosts change the conformal frame hence test conformal invariance, and super-translations change the foliation hence test foliation independence. The limitation of testing background independence in this way is that the transformations of the background are limited to those generated by a symmetry, as opposed to arbitrary change of conformal factor and of foliation.
[The anomaly operator can be used to compute the dependence of the fields on background structures in a convenient way. The restriction is that this is done not by arbitrary changes of the background, but by those changes which are allowed by the symmetry vector fields. It is therefore the presence of respectively boosts and super-translations in the symmetry that allows the anomaly operator to be sensitive to the conformal and super-translation weights. ]
This difference shows up if we look at the anomaly of the symmetry vector fields, which is not zero even though from its definition \eqref{xius}, we see that $\xi$ is manifestly conformally invariant and foliation independent. To compute its anomaly, we first observe that it is a purely background field at $\scri$, hence $\d_\chi\xi\eqonS 0$. Then
\be\label{anoxi}
\D_\chi \xi = -\pounds_\chi\xi+O(\Om) = -[\chi,\xi]+O(\Om).
\ee
The anomaly of the vector fields is nothing but their Lie algebra. This fact will play an important role below, making the anomaly operator a convenient tool to study covariance of charges and fluxes. It is also useful for later purposes to single out two sub-cases of \eqref{anoxi}. The general expression for the commutator in a given affine foliation is
\be
[\xi,\chi]=(T_\xi\dot f_\chi +Y_\xi[f_\chi]-(\xi\leftrightarrow \chi))\p_u + [Y_\xi,Y_\chi]^A\p_A.
\ee
For $\xi=\xi_T:=T\p_u$ a pure super-translation, 
\be\label{anoT}
\D_\chi \xi_T = [\xi_T,\chi] = \xi_{T'}, \qquad T'=\dot f_\chi T -Y^A_\chi\p_AT, 
\ee
which we can interpret as the conformal and super-translation weights of the vertical component of the symmetry vector field (namely of $T$ if seen as a vector component and not a scalar, otherwise only the second term would be present). In particular, two super-translations commute. For $\xi=\xi_Y:=u\dot f\p_u+Y^A\p_A$ a (cross-section-dependent) Lorentz transformation, 
\be\label{anoY}
\D_\chi \xi_Y = [\xi_Y,\chi] = \xi_{Y'} + \xi_{f'}, \qquad Y'=[Y,Y_\chi], \quad f'=Y^A\p_Af_\chi -\dot f T_\chi  - u Y_\chi^A\p_A \dot f.
\ee
In particular for $\chi=\chi_T$ a pure super-translation
\be\label{anoY}
\D_{\chi_{T}} \xi_Y = [\xi_Y,\chi_T] = \xi_{T'}, \qquad  T'= Y^A\p_AT -\dot f T,
\ee
which makes it manifest why the notion of Lorentz subgroup of the Lorentz group is cross-section dependent: acting with a super-translation changes the cross-section and the Lorentz symmetry vector is shifted by a super-momentum contribution. For the angular momentum piece $\dot f=0$ and the shift is by $Y^A\p_A T$ only.

To extract the anomaly contribution in \eqref{BMStransf}, we first observe that $q_{AB}$ and $C_{AB}$ can be seen as the only non-zero components of transverse tensors $q_{ab}$ and $C_{ab}$ on $\scri$. This also explains why these functionals do not depend on a specific choice of $x^A$ coordinates, and the only relevant background fields are $\Om$ and $u$.
For transverse tensors, the Lie derivative reduces to $\pounds_\xi = f\p_u+\pounds_Y$  in Bondi coordinates. 
Therefore from the definition \eqref{defDxi} we have
\begin{subequations}\label{BMSanomalies}\begin{align}\label{Dxiq}
& \D_{\xi} \,{q}_{A B}= -2 \dot{f} {q}_{A B}, \\
& \D_{\xi} \,C_{A B}= -\dot{f} C_{A B}-2 {\Dd}_{\langle A} \Dd_{B\rangle} f.  \label{DxiC}
\end{align}\end{subequations}
The first is again the anomaly of the conformal metric that we already know. More interesting is the anomaly of the shear, which recalling that $\s_{AB}=-\f12C_{AB}$, can be rewritten as
\be
\D_{\xi} \,\s_{A B}= -\dot{f} \s_{A B} + {\Dd}_{\langle A} \Dd_{B\rangle} f  \label{Dxisigma}.
\ee
Comparing \eqref{Dxisigma} to \eqref{gens'lin} we see that the anomaly of $C_{AB}$ computes its behaviour under super-translations and conformal transformations, 
including both its conformal weight and the inhomogeneous term, with the specification that conformal transformations are to act on $l$ as in \eqref{inol} and not homogeneously. The reason for this is that $C_{ab}$ is not any shear, but specifically the shear of the $u$-foliation, hence its anomaly follows from the behaviour of $u\to u'$ under conformal transformations and changes of foliation, which is such that $du\to du'$ is still Lie-dragged.
In other words, both $n^a$ and $l_a$ are background fields, hence $\D_\xi\t_a=-\pounds_\xi\t_a$ preserves a vanishing $\t_a$.
We also notice for later purposes that
\be\label{DxiCinv}
\D_{\xi} \,C^{A B}= 3\dot{f} C^{A B}-2 {\Dd}^{\langle A} \Dd^{B\rangle} f, 
\qquad \D_{\xi} (C^{A B}\eps_\scri)= -2 {\Dd}^{\langle A} \Dd^{B\rangle} f\eps_\scri.
\ee

Let us recover also the behaviour \eqref{calS'}, because it will be instructive about the transformation properties of the super-translation/bad-cut field $u_0$.
The transformation of a vacuum shear $\os{C}_{AB}$ can be deduced from \eqref{dxiC} setting the news to zero, 
\be
\d_\xi \os{C}_{AB} = (\pounds_Y-\dot{f}) \os{C}_{A B}-2 ({\Dd}_{\langle A} \Dd_{B\rangle}+\f12\r_{\la AB\ra}) f.
\ee
Using then \eqref{vacuumshear}, \eqref{elecshear} and the universality of both $q_{AB}$ and $\r_{AB}$, we have that 
\be\label{dxiCvac}
\d_\xi \os{C}_{AB} = 2 ({\Dd}_{\langle A} \Dd_{B\rangle}+\f12\r_{\la AB\ra})\d_\xi u_0.
\ee
Comparing the two equations above we conclude that
\be\label{dxiu0}
\d_\xi u_0 =  \pounds_Y u_0 - T - u_0 \dot{f} = \pounds_Y u_0 -f|_{u_0}.
\ee
Notice that it implies $\D_\xi u_0= -f|_{u_0}$ in agreement with its conformal and super-translation weights.
Hence for the relative shear \eqref{cS} we have
\be
\d_\xi {\cal S}_{ab}= \pounds_\xi {\cal S}_{ab} - \dot{f} {\cal S}_{ab} , \qquad \D_\xi {\cal S}_{ab}=  - \dot{f} {\cal S}_{ab},
\ee
consistently with the geometric analysis of the previous section. The relative shear can also be written as
\be\label{calC}
{\cal C}_{ab}:= C_{ab} + (u - u_0) \rho_{\la ab\ra} -2 \Dd_{<a} \Dd_{b>} u_0 = -2{\cal S}_{ab},
\ee
to match the notation of \cite{Compere:2018ylh}.

The transformation \eqref{dxiu0} was posited in \cite{Compere:2018ylh}, in order to obtain the homogeneous transformation of \eqref{cS}.
Our derivation clarifies that \eqref{dxiu0} does not need to be posited, 
but follows from the fact that $u_0$ parametrizes a vacuum shear, and that a vacuum shear is not a new degree of freedom, its transformation follows from the symplectic structure on the radiative phase space of \cite{Ashtekar:1981hw}. It thus also clarifies that the super-translation/bad-cut field $u_0$ is not a new degree of freedom but rather part of the initial (or final) conditions for the gravitational field.

We insisted that the transformation rule \eqref{dxiC} is the appropriate one for a shear associated with an affine foliation. In the covariant description recalled earlier one can use a general shear associated with an arbitrary $l$. In this case the transformation law is \cite{Ashtekar:1981hw,Ashtekar:2024stm}
\be\label{dxiAbhay}
\d_\xi \s_{ab}= [\pounds_\xi,D_{\la a}]l_{b\ra}+2l_{\la a}D_{b\ra}\dot f.
\ee
It is instructive to see how it reduces to \eqref{dxiC} when $l$ is Lie dragged. This was shown in \cite{Grant:2021sxk}, and we report a slightly streamlined version of the proof in Appendix~\ref{AppdxiS}.

Finally let's look at the news tensor. First, from $[\p_u,\d_\xi]= 0$ and  $[\p_u,\D_\xi]=-\dot f\p_u$,
we have
\begin{align}\label{DxiCdot}
\d_{\xi} \,\dot C_{A B}= (f\p_u+\pounds_Y) \dot C_{A B} -2 {\Dd}_{\langle A} \Dd_{B\rangle} \dot f,
\qquad \D_{\xi} \,{\dot C}_{A B}= -2 {\Dd}_{\langle A} \Dd_{B\rangle} \dot{f}.
\end{align}
To understand the meaning of the anomaly of $\dot C_{AB}$, 
observe that it vanishes for a globally defined CKV on round spheres
but not otherwise.
This means that 
identifying non-radiative spacetimes as constant shear, $\dot\s=0$, is not a conformally invariant notion
in general, but only if the conformal transformations are restricted to preserve round spheres. 
This can be confirmed looking at \eqref{sconf}: the trace-less part of Geroch tensor remains zero if the conformal transformation preserves round spheres.

Let us see how the anomaly of $\dot C$ on arbitrary frames is compensated by Geroch's tensor. 
Since $\r_{ab}$ is transverse and time independent, it can be obtained as pull-back of a 2d tensor on a given cross-section. In particular, in Bondi coordinates, the only non-vanishing components of $\rho_{ab}$ are $\r_{AB}$, and using the explicit parametrization \eqref{xi} we obtain $\a_\xi = \dot f$ and  \eqref{lierho} 
becomes
\be\label{LierhoAB}
\pounds_\xi \r_{AB} = -2\Dd_A \Dd_B \dot f.
\ee
It is also universal, namely $\d\rho_{ab}=0$ and therefore $\d_\xi\r_{ab}=0$. It follows that 
\be\label{Dxirho}
\D_{\xi} \,{\r}_{A B} = 2 {\Dd}_{\langle A} \Dd_{B\rangle} \dot{f}.
\ee
Therefore
\be\label{DxiN}
\D_\xi N_{AB}= -\D_\xi (\dot C_{AB}+\r_{\la AB\ra})=0.
\ee
The news tensor is anomaly free, which is nothing but the statement that it is foliation independent and conformally invariant, as we already know.
Yet one should appreciate the facility with which $\D_\xi$ allows us to deduce these properties in a fixed coordinate system and in a fixed conformal frame.
It remains to discuss the meaning of the anomalies of $M$ and $J_A$. We postpone this discussion to Section~\ref{MJano} below, after we have explained their relation to the charges.

To summarize, the anomalies computed by $\D_\xi$ in this Section
measure the loss of covariance caused by the background dependence on foliations or on the conformal factor, as induced by a diffeomorphism thanks to the fact that we identified these background structures with coordinates.
Lack of foliation-independence and/or conformal invariance
can of course be studied independently of $\D_\xi$, but we would like to advertise the anomaly operator as a very convenient tool to do it.
First, it systematizes and generalizes the analysis, making it an algebraic and straightforward operation, and equally adaptable to whatever the background fields are,
see e.g. the different (albeit related) case of arbitrary null surfaces \cite{Odak:2023pga}.
Second, \emph{the analysis of whether something is foliation independent and conformal invariant can be done in a fixed coordinate system.} This should be quite a convenient advantage for that large part of the community that prefers to do calculations in explicit coordinate systems, as opposed to using only covariant and geometric quantities, and we will see it explicit examples of it in the next Sections.
There is also a third advantage.
Notice that the structure of the anomalies is the same for BMS, eBMS, gBMS and BMSW: only the numerical value changes, given respectively by $\dot f=\f12\Dd Y$ with $Y^A$ a CKV globally defined, non-globally defined, an arbitrary vector, and finally $\dot f=W(x^A)$ an arbitrary function on the sphere. 
Therefore \emph{quantities that anomaly-free under the BMSW group are foliation-independent and invariant under arbitrary conformal transformations} respecting the Bondi condition. This offers a very convenient technical tool: instead of imposing BMS covariance only, which needs to be supplemented by an independent test of conformal invariance if one is not restricting attention to Bondi frames, one can get both BMS covariance and general conformal invariance at once using the anomaly operator for the BMSW. This does not mean changing the symmetry group: we keep the same universal structure, and the symmetry group is still BMS. The way we are using BMSW transformations is not as symmetries but as canonical transformations, in order to test covariance and conformal invariance in one go. 
In other words we study background independence under BMSW transformations. 
A similar approach to BMSW transformations was considered also in \cite{Barnich:2016lyg}, in the broader context of also relaxing the Bondi condition.

\subsection{Finite BMSW transformations and finite covariance}

To  operator $\D_\xi$ measures the linearized anomalies. The finite version is obtained performing finite conformal transformations and finite changes of foliation
acting only on the background fields and not on the physical metric, in agreement with \eqref{defDxi}. More precisely, \emph{inverse} transformations, because of the sign convention used in \eqref{defDxi}. 
So for instance the finite anomaly of the conformal metric is simply
\be
\hat g_{\m\n}'=\om^{-2}\hat g_{\m\n}.
\ee
Because we have identified the background fields with coordinates $\eta=(\Om=1/r,u)$, 
changing them in a way compatible with the universal structure can be done computing a finite symmetry transformation on the coordinates. As explained at the end of the previous Section, we can  allow for arbitrary (time-independent) conformal transformations if we use finite BMSW transformations, instead of BMS ones alone. The subset of BMSW transformations made of arbitrary super-translations and conformal transformations is 
\begin{align}\label{finiteBMSW}
	\Om\rightarrow \Om'=\omega \Om, \qquad
	u \rightarrow u'=\tl u - \f\Omega{2\om} \p_{A} \tl u \p^{A} \tl u, \qquad
	x^A \rightarrow x'^{A}=x^A - \f\Om\om q^{AB}\p_B \tl u,
\end{align}
where
\be
\tl u:=\om(u + T), \qquad \pounds_n\om=0.
\ee
The remaining part of the BMSW group is arbitrary $\Diff(S)$ coordinate transformations on the cross sections, and it is not needed since the functionals considered only depend on specific choices of $\Om$ and $u$ but not of $x^A$. For more details on finite BMSW transformations, see \cite{Barnich:2019vzx,Flanagan:2023jio}.
The $O(\Om)$ in \eqref{finiteBMSW} is fixed requiring preservation of the affine embedding, namely
\be\label{ds'}
2du d\Om + \Om{}^{2}g_{AB} dx{}^A dx{}^B  \to 2du' d\Om' + \Om'{}^{2}g_{AB} dx'{}^A dx'{}^B = \om^2(2du d\Om + q_{AB} dx{}^A dx{}^B),
\ee
so that the full conformal metric is rescaled under a finite anomaly transformation, and not just its induced part. The $O(\Om)$ term is not needed to compute the anomalies of fields on $\scri$, but it is useful 
if one looks at spacetime embeddings, like \eqref{ds'}. Another example where it is useful is the transformation of $l$ \eqref{inol}, which can be obtained starting from $l=-du$ and acting with \eqref{finiteBMSW}. The part proportional to $n$ can only be seen embedding $l$ in spacetime (which is done requiring it to be null), and arises from the $O(\Om)$ terms of \eqref{finiteBMSW}. It is nice to include it because it allows us to understand \eqref{inol} as a class-II transformation of the tetrad.
It is however irrelevant to compute the finite anomaly of the shear, which is well defined without embedding. The transformation of the shear of the $u$ foliation under \eqref{finiteBMSW} reproduces \eqref{gens'}, or  equivalently in terms of $C_{ab}$, 
\be\label{finiteC}
C'_{ab} =  \om \Big(C_{ab}
+2 (u+T) (\om^{-1}D_{\la a} D_{b\ra} \om - 2 D_{\la a} \ln\om D_{b\ra} \ln\om) +2 D_{\la a}D_{b\ra}T\Big).
\ee
The finite anomaly is obtained with the inverse transformation, namely switching $(\om,T)\to(\om^{-1},-T)$. We can go back to the linearized anomaly taking $\om=1+W$ with $W\ll 1$ and $T\ll 1$, and identifying $W=\dot f$ we obtain
\be
\D_\xi C_{ab} = -\dot f C_{ab} - 2 \Dd_{\la a}\Dd_{b\ra} f,
\ee
namely \eqref{DxiC} in arbitrary coordinates on $\scri$. We have thus completed the proof that the anomaly \eqref{DxiC} measures the dependence of $C_{AB}$ on the background fields $\Om$ and $u$, by computing its change under a change of $u$ foliation and conformal factors as generated by a BMSW transformations, namely change of foliations by a super-translation and arbitrary conformal transformations. We could have restricted this analysis to BMS transformations only, it would have given the same class of foliation changes, but a smaller class of conformal transformations, restricted to preserving round spheres.
With the same calculation one can show that \eqref{calC} has finite anomaly  ${\cal C}_{ab} \rightarrow \omega {\cal C}_{ab}$.
It then follows from \eqref{NcalS} that the news has vanishing finite anomaly, in agreement with being conformally invariant and super-translation invariant.

\section{BMS flux and charge algebra}

We start by recalling the results of \cite{Iyer:1994ys,Wald:1999wa,Barnich:2001jy,Barnich:2011mi}. 
The covariant phase space 
is constructed equipping the solution space of a field theory at given boundary conditions with a symplectic 2-form current $\om=\d\th$, where the symplectic potential current $\th$ is read from the on-shell variation of the Lagrangian 4-form, via $\d L\eqons d\th$. By Noether theorem, $j_\xi:=I_\xi\th-i_\xi L\eqons dq_\xi$ is on-shell exact in a general covariant theory, for any diffeomorphism $\xi$. As a consequence, the Hamiltonian 1-form
is also exact,
\be\label{Ixiom}
-I_\xi\om \eqons d( \d q_\xi - q_{\d\xi} - i_\xi\th). 
\ee
We restrict attention to vacuum general relativity in metric variables. We can take for $\th$ the standard Einstein-Hilbert symplectic potential, in which case $q_\xi$ is the Komar 2-form,
\be\label{thg}
\th = \f1{3!} \th^\m \eps_{\m\n\r\s}~dx^\n\w dx^\r \w dx^\s , \qquad
\th^\m = \f1{8\pi} g^{\r[\s} \d \G^{\m]}_{\r\s}, \qquad
q_\xi = -\frac 1{32\pi} \epsilon_{\m\n\r\s}\na^\m\xi^\nu dx^\r\w dx^\s,
\ee
in units $G=c=1$.
We then have
\begin{align}\label{Ixiom2}
-I_\xi\om &\eqons - \f1{32\pi} \eps_{\m\n\r\s}\big[ (\d\ln{\sqrt{-g}}) \na^\r\xi^\s + \d g^{\r\a}\na_\a\xi^\s 
+ \xi^\r \left(\na_\a \d g^{\a\s} + 2 \na^\s \d\ln\sqrt{-g}\right) \nn\\ &\hspace{3cm} - \xi_\a\na^\r\d g^{\s\a} \big] dx^\m\w dx^\n.
\end{align}
For the application of this formula to BMS symmetries, 
we consider a hyperbolic space-like hypersurface $\Si$ with a single boundary at future null infinity $\scri$, denoted $S$, with the sphere topology. 
Integrating $\Om_\Si:=\int_\Si\om$ endows $\Si$ with a phase space of partial Cauchy data, which include radiation as well as the `Coulombic data', like $M$ and $\am_A$ in Bondi coordinates.
Integrating $\Om_{\cal N}:=\int_{\cN}\om$ on a region $\cN$ of $\scri$ between two partial Cauchy slices endows $\cal N$ with the radiative phase space of connections (namely news and shear) that we reviewed in the previous Section, and does not contain Coulombic data.
The result of  \cite{Barnich:2011mi} in Bondi coordinates is
\be\label{BT11}
-I_\xi\Om_\Si \eqons \d Q^{\sscr BT}_\xi - \F_\xi^{\sscr BT},
\ee
where\footnote{The calculation \cite{Barnich:2011mi} is done using the expression of \cite{Barnich:2001jy} for the Hamiltonian 1-form, which differs from \eqref{Ixiom} and \eqref{Ixiom2} by a term $\eps_{\m\n\r\s} g^{\r\a}\d g_{\a\b} \na^{(\s}\xi^{\b)}$, but this extra term vanishes in the limit.}
\begin{align}\label{qBT}
Q^{\sscr BT}_\xi = \f1{8\pi}\oint_S(2f M+ Y^A\am_A)\eps_S, \qquad {\F}^{\sscr BT}_\xi 
= - \f1 {32\pi} \oint_Sf \dot C_{AB}\d C^{AB}\eps_S,
\end{align}
and we used
\be\label{xieps}
i_\xi \eps_\scri= f\eps_S.
\ee
The notation follows the previous Sec.~\ref{SecBondi}, and coincides with \cite{Barnich:2011mi} except for the Lorentz aspect (which includes angular momentum and center of mass), which is related to the $N^{\sscr BT}_A$ used in \cite{Barnich:2011mi} by $\am_A=N^{\sscr BT}_A+\f1{32}\p_A (C^{BC}C_{BC})$. 
The reason for the change becomes clear once we express the aspect in the Newman-Penrose notation described in the previous Section (see also  \cite{Barnich:2019vzx}). We have in fact
\be\label{DS}
m^A\am_A = -\left( \psi_1 + \s\eth \bar\s+\f12\eth(\s\bar\s)\right),
\ee
which coincides with the integrand of the Dray-Streubel formula \cite{Dray:1984rfa}. 
We also point out that
\be\label{calP}
m^AN^{\sscr FN}_A:=m^A\left(\am_A+\f14C_{AB}D_CC^{BC} +\f1{16}\p_A(C^{BC}C_{BC}) \right) = -\psi_1,
\ee 
which is the angular momentum aspect denoted $N_A$ in \cite{Flanagan:2015pxa} and used in \cite{Hawking:2016msc}.
The Newman-Penrose expression for $M$ is 
\be\label{MNP}
M = -\left( \psi_2+\s\dot{\bar\s} +\f12\left(\eth^2 \bar\s - cc\right)\right) = -\re(\psi_2 + \s \dot{\bar\s}), 
\ee
and coincides with the integrand of Geroch's supermomentum \cite{Geroch:1977jn} (see expression in \cite{Dray:1984rfa}), but  \emph{only on round spheres}. This discrepancy will be crucial below to understand the origin of the 2-cocycle. We remark that it concerns only the mass aspect, whereas \eqref{DS} matches \cite{Dray:1984rfa} on any frame. 
For completeness we also report the non-integrable term in coordinate-independent form,
\be
 {\F}^{\sscr BT}_\xi = - \f1 {8\pi} \oint_Sf \pounds_n \s_{ab}\d \s^{ab}\eps_S 
 = - \f1 {4\pi} \oint_Sf \re(\dot\s\d \bar\s) \eps_S.
\ee

The right-hand side of \eqref{BT11} contains a field-space exact (``integrable") piece, and a non-exact (``non-integrable") piece.
This split is clearly arbitrary, as integrable terms can be freely moved to the non-integrable piece. 
Once a split $-I_\xi\Om_\Si \eqons \d Q_\xi - \F_\xi$ is chosen, the integrable piece provides a surface charge $Q_\xi$ that acts as canonical generator on the subset of the phase space where the non-integrable piece $\F_\xi$  vanishes. The split also determines the flux-balance law $d q_\xi\eqons F_\xi$ satisfied by the charges. This makes it clear that a useful requirement for the split is that both $\F_\xi$ and $F_\xi$ vanish around solutions satisfying some notion of stationarity, otherwise it would be hard to relate the generator to physical observables.
For the charges  \eqref{qBT} one finds the following flux,
\be\label{BTflux}
Q^{\sscr BT}_\xi[S_2]-Q^{\sscr BT}_\xi[S_1] 
\eqons F^{\sscr BT}_\xi:=-\f1{32\pi} \int_{S_1}^{S_2}\left( \dot C_{AB}\d_\xi C^{AB} + \p^A \cR \p_A f + C^{AB}\Dd_{A}\Dd_{B}\Dd Y\right)\eps_\scri,
\ee 
with $\d_\xi C^{AB}$ given in \eqref{dxiC}.
Accordingly, the charges are conserved if the time derivative of the shear vanishes and if we restrict attention to Bondi frames, since then $\cR$ is constant and
\eqref{DDDY} holds 
for globally defined CKVs. These conditions are met by all non-radiative asymptotically flat spacetimes in Bondi frames \cite{Geroch:1977jn}.
They are however not met by non-radiative spacetimes in arbitrary frames in which $q_{AB}$ is not a round sphere. In this case none of the three terms vanishes: the news is not the time derivative of the shear, $\p_A\cR\neq 0$, and \eqref{DDDY} does not hold. We thus have a failure of the stationarity condition, namely a non-zero flux in spite of the absence of radiation.

Another important requirement for the split is that 
the prescribed charges should realize the symmetry algebra.
This is a non-trivial property, because $d\om\eqons 0$ guarantees that the symplectic two-form is independent of $\Si$ only in the absence of radiation, and the symmetries moving the corners of $\Si$ don't correspond to Hamiltonian vector fields.
In general, two symmetries $\xi$ and $\chi$ give
\be
I_\xi I_\chi\Om_\Si = \d_\chi Q_\xi - I_\chi \F_\xi \neq \d_\chi Q_\xi. 
\ee
It was then proposed in \cite{Barnich:2011mi} to define a bracket 
with the non-integrable term subtracted off,
\be\label{BTbracket}
\{ Q_{\xi}, Q_\chi \}_* := \d_\chi Q_\xi - I_\xi {\F}_\chi = I_\xi I_\chi \Om_\Si +I_\chi \F_\xi - I_\xi \F_\chi.
\ee
The second equality
shows that $\{, \}_*$ reduces to a Poisson bracket for the subspace with vanishing non-integrable term $\cF_\xi$.
Applying this definition to \eqref{qBT} one finds \cite{Barnich:2011mi}
\be
\{ Q^{\sscr BT}_{\xi}, Q^{\sscr BT}_\chi \}_* \eqons Q^{\sscr BT}_{\llbracket \xi,\chi \rrbracket} + K^{\sscr BT}_{(\xi,\chi)},
\ee
where 
\be\label{KBT}
K^{\sscr BT}_{(\xi,\chi)} = \oint_S k^{\sscr BT}_{(\xi,\chi)}, \qquad k^{\sscr BT}_{(\xi,\chi)}=
\f1{32\pi} \left[\xiu_\xi\left(C^{AB}\Dd_{A}\Dd_{B}\Dd_CY^C_\chi + \p^A\xiu_\chi \p_A \cR\right) - (\xi\leftrightarrow \chi)\right]\eps_S,
\ee
and $\llbracket \xi,\chi \rrbracket :=[\xi,\chi] -\d_\xi\chi+\d_\chi\xi$  is the modified Lie bracket needed to describe the algebra of field-dependent diffeomorphisms \cite{Barnich:2011mi}, here due to the choice of Tamborino-Winicour extension \eqref{xi} (the standard bracket is enough if one restricts attention to the vector fields on $\scri$ only).
The algebra is thus realized, but only up to the 2-cocycle $K^{\sscr BT}$.
It is field-dependent, hence not a central extension. 
This is problematic, because it hinders the interpretation of the charges as canonical generators even when $\cF_\xi$ vanishes, and also makes it hard to find representations for quantization. It motivates the search for a different split, whose charge prescription gives an algebra free of field-dependent cocycles. A partial answer to this question was given in \cite{Compere:2020lrt} in the more general context of the generalised BMS symmetry \cite{Campiglia:2015yka,Compere:2018ylh}, where a split was found so that the cocycle vanishes at least in the limit $u\to -\infty$, but not for arbitrary cross sections of $\scri$. 
We now show that it is possible to remove the cocycle  for arbitrary cross-sections of $\scri$.

As shown in \cite{Chandrasekaran:2020wwn,Freidel:2021cjp,Freidel:2021yqe,Chandrasekaran:2021vyu}, a 2-cocycle signals the presence of non-covariant terms in the charge prescription. 
To understand the origin of this loss of covariance, 
we begin by observing that the cocycle \eqref{KBT} vanishes on Bondi frames, since then $\cR$ is constant and \eqref{DDDY} holds. These are the same conditions that give a vanishing flux 
in non-radiative spacetimes, hence the presence of the cocycle is related to the failure of the stationarity condition. 
The fact that the cocycle vanishes on Bondi frames but not otherwise is a first hint that it is unphysical, because there is nothing that distinguishes these 
frames in the BMS fall-off conditions (one should not confuse the fact that BMS transformations preserve round spheres with \emph{preferring} them). 
A second hint comes from the results of \cite{Freidel:2021cjp}, where it was shown that the lack of covariance, or anomaly, in the choice of symplectic potential contributes to the cocycle of Noether charges,
and of \cite{Freidel:2021yqe}, where \eqref{KBT} was indeed reproduced as a purely anomaly contribution.
We are now going to show that this cocycle is in fact a consequence of having selected a symplectic potential which satisfies the Wald-Zoupas requirements of covariance and stationarity only on round spheres. In other words, the split is not invariant under arbitrary (time-independent) conformal transformations. Replacing it with the correct potential that satisfies these requirements on arbitrary frames produces a modification of the charges whose algebra has no cocycle.
 
 The split \eqref{qBT} can be associated to a specific choice of symplectic potential. To see which one,
we take the limit to $\scri$ of \eqref{thg}. This gives for the pull-back
\be\label{barth}
\pbi{\th} = \th^{\sscr BT} -\d b^{\sscr BT}, 
\ee
where
\be\label{bBT}
\th^{\sscr BT} = -\f1{32\pi} \dot C_{AB}\d C^{AB}\eps_\scri, \qquad b^{\sscr BT}= \f1{16\pi}\Big(2 M-\f12 \Dd_A\Dd_B C^{AB} -\f18 \dot C_{AB}C^{AB}  \Big)\eps_\scri.
\ee
Using \eqref{xieps}, we see that 
\be
{\F}^{\sscr BT}_\xi = \oint_S i_\xi \th^{\sscr BT}.
\ee

The calculation of the charges is slightly more subtle, because of the term $q_{\d\xi}$ \cite{Odak:2022ndm,Ashtekar:2024stm}.
This was absent in the original derivation \cite{Wald:1999wa}, where it was assumed that $\d\xi=0$. 
This assumption is supported by the fact that the BMS algebra is a universal property of asymptotically flat metrics. And indeed, the vector fields on $\scri$ are field-independent. The problem though is that the limit to $\scri$ of the Komar 2-form depends on the second and even third order of the extension, as remarked already in \cite{Geroch:1981ut}. This brings in the field-dependence of the Tamburino-Winicour extension \eqref{xi}, which thanks to its property of preserving bulk Bondi coordinates is the customary choice in a large part of the literature, e.g. \cite{He:2014laa,Campiglia:2015yka,Flanagan:2015pxa,Compere:2018ylh,Compere:2019gft,Campiglia:2020qvc,Freidel:2021yqe,Chandrasekaran:2021vyu,Donnay:2022hkf,Geiller:2022vto,Riva:2023xxm}. The term $q_{\d\xi}$ is thus crucial to remove the  spurious contribution to $\d q_\xi$ introduced by the field dependence of the extension.
Explicitly, the pull-back at $\scri$ of the Komar 2-form gives
\be\label{Komarscri}
q_\xi = \f1{16\pi}\left[\xiu\left(2M +\f14\Dd_A\Dd_BC^{AB} +\f1{8}\dot C_{AB}C^{AB}\right)+2Y^A J_A\right]\eps_S,
\ee
up to a total divergence that vanishes upon integration on the cross-section. The latter includes divergent terms that while not contributing to the 
charges,\footnote{The divergent terms are no longer total divergences for the weaker fall-off conditions relevant for the gBMS \cite{Compere:2018ylh,Campiglia:2020qvc,Chandrasekaran:2021vyu}, BMSW \cite{Freidel:2021yqe} and RBS \cite{AS2} extensions of the asymptotic symmetries, and make renormalization of the symplectic potential necessary.}
make the limit sensitive to the subleading terms of $\xi$. 
Indeed, $q_{\d\xi}$ is non-zero in spite of  $\d\xi=O(\Om^2)$, and given by \cite{Odak:2022ndm} 
\be\label{sxi}
q_{\d\xi} = \d s_\xi, \qquad s_\xi := -\f1{64\pi} C^{AB} \Dd_A\Dd_B \xiu \eps_S,
\ee
up to a total divergence.
 Adding up according to \eqref{Ixiom} we recover the result \eqref{BT11}, and in the process we learn that 
\begin{align}\label{qBT2}
 Q^{\sscr BT}_\xi=\oint_S q^{\sscr BT}_\xi, \qquad
 q^{\sscr BT}_\xi = q_\xi +i_\xi b^{\sscr BT}- s_\xi = \f1{8\pi}(2f M+ Y^A\am_A)\eps_S,
\end{align}
up to a total divergence. 
The spurious contribution \eqref{sxi} has the structure of a soft term, hence missing it would change the behaviour of the charges, for instance spoiling basic properties such as vanishing in Minkowski for any symmetry parameter.
The care required in dealing with the extra term $q_{\d\xi}$ is of course not needed if one starts directly from the expression \eqref{Ixiom2}, where it is already subtracted out.\footnote{It is also not needed 
if one constructs the charges `integrating' the fluxes as opposed to bootstrapping them from the Komar formula, see discussion in \cite{Ashtekar:2024stm}.}
However \eqref{Ixiom2} hides the role of the symplectic potential, and it is important to our considerations below
to have identified that the BT charges are associated to the choice of symplectic potential \eqref{bBT}.

Furthermore, \eqref{qBT2} allows us to derive the flux of the charges using Noether's theorem. That requires though the extra step of identifying \eqref{qBT2} as an improved Noether charge. This can be done observing that 
\be
s_\xi = \D_\xi c, \qquad c=-\f1{8\pi}\b\eps_S,
\ee
from which it follows that 
\be
q^{\sscr BT}_\xi=q_\xi +i_\xi \ell^{\sscr BT}- I_\xi \d c,
\ee
up to a closed 2-form. Here $\ell^{\sscr BT} = b^{\sscr BT}+dc$ and  $s_\xi = i_\xi dc-I_\xi\d c$ \cite{Odak:2022ndm}. Then the Noether current formula \cite{Freidel:2021cjp,Chandrasekaran:2021vyu}
$dq^{\sscr BT}_\xi\eqons I_\xi\th^{\sscr BT}-\D_\xi\ell^{\sscr BT}$ leads to \eqref{BTflux}.
 
\subsection{Wald-Zoupas prescription and covariance of the current algebras} \label{SecWZ}

In this Section we review some aspects of the Wald-Zoupas prescription that we'll need below: first, how ambiguities in the charges are dealt with, and its extension to non-trivial corner shifts. We then present the new results of \cite{Rignon-Bret:2024wlu} relating the Wald-Zoupas covariance and the cocycle, and finally propose a refinement of the Wald-Zoupas procedure to fix residual ambiguities.

Changes in the split can be controlled by shifts of the symplectic potential $\th\to\bar\th = \th+\d \ell-d\vth$. 
The Wald-Zoupas prescription aims at selecting a possibly unique, preferred $\bar\th$ imposing basically two physical requirements.\footnote{On top of more technical requirements such as local and analytical behaviour on the fields, which we take for granted.}
The first is `stationarity', namely $\bar\th=0$ on special solutions. One could take this to mean existence of a translational time-like Killing vector, but general relativity admits solutions without it and with gravitational waves, so this is too restrictive. While it is not known how to identify gravitational radiation in general (meaning in a background independent and gauge invariant way),
the situation simplifies in the presence of physical boundary, where one can posit boundary conditions that allow an unambiguous identification of gravitational radiation. 
In the context of this paper the boundary is $\scri$. The chosen boundary conditions are those that define the universal structure associated with the BMS group, and these allow one to identify 
 non-radiative asymptotically flat spacetimes as solutions with vanishing news, as discussed in the previous Section.\footnote{ Other examples of boundaries at which one can successfully apply the Wald-Zoupas prescription include arbitrary null hypersurfaces \cite{Chandrasekaran:2018aop,Odak:2023pga,Chandrasekaran:2023vzb} and non-expanding horizons \cite{Ashtekar:2021kqj}, $\scri$ of asymptotically De Sitter spacetimes \cite{Kolanowski:2021hwo}
 and some extensions of the BMS symmetry \cite{AS2}. See \cite{Odak:2022ndm} for a study of the most general circumstances under which the prescription is applicable.
 
   }
In the following we will often refer to these non-radiative spacetimes as the `stationary' solutions, meant in this general sense and not in the sense of admitting a time translation Killing vector.    
As a consequence, one looks for the preferred $\bar\th$ only for the pull-back at the physical boundary,
\be\label{thbar}
\pbi{\th} = \bar\th - \d \ell + d\vth.
\ee
Choosing the split after pull-back may introduce a dependence on the background structures used to define the boundary conditions,
and this would spoil the physical applications. To avoid this,
the second crucial requirement is covariance, namely $\bar \th$ should be independent of any background structure. 
This can be stated as $\varphi^*\th^{\sscr WZ}[\phi, \d \phi; \bac] = \th^{\sscr WZ}[\varphi^* \phi, \d( \varphi^* \phi); \bac]$ for a diffeomorphism $\varphi$ that corresponds to an asymptotic symmetry, and reduces to
\be\label{covth}
\d_\xi\bar\th=\pounds_\xi\bar\th
\ee
at the linearized level. This property can be equivalently interpreted as stating that the symplectic potential should be invariant when the transformation acts on the background fields only.
In the case of the BMS group this property includes limited conformal transformations that preserve round spheres. 
If one wants to allow for arbitrary 
(time-independent) conformal transformations, it should be added as an additional requirement
on top of  \eqref{covth} (see e.g. \cite{Grant:2021sxk}).
However as we have explained in the previous Section, arbitrary conformal invariance can be studied computing the anomalies associated with the BMSW group. 
In doing so we are not changing the universal structure, and the symmetry group remains the BMS group. We are merely using BMSW as an auxiliary group to test arbitrary conformal transformations.

The original Wald-Zoupas prescription made the additional requirement that no corner term $\vth$ was needed, so that we can actually write
\be\label{cond0}
\pbi{\th} = \bar\th - \d b.
\ee
This guarantees that the symplectic 2-form current defined on $\scri$ by $\bar\th$ is the same as the one defined by $\th$ given by  \eqref{thg}.
It was listed as `condition 0' in \cite{Odak:2022ndm} (with covariance and stationarity being conditions 1 and 2). 
The inclusion of $\vth$ is compatible with the field equations and with the Wald-Zoupas prescription, but introduces various additional subtleties, especially in the way ambiguities are dealt with. Let us first discuss the consequences of covariance assuming \eqref{cond0}.
If a $\bar\th$ satisfying \eqref{covth} is found in the class \eqref{cond0},
one can define charge aspects $\bar q_\xi$ via
\be\label{IxiomWZ0}
\d d \bar q_\xi:\!\!\eqons\! -I_\xi \pbi{\om}+d i_\xi\bar\th =\d I_\xi\bar\th.
\ee
From this it follows that 
\be\label{jN}
d \bar q_\xi \eqons \bar\jmath_\xi:=I_\xi\bar\th,
\ee
up to a field-space constant. This can be removed requiring that the Noether current $\jb_\xi$ vanishes on a reference solution amongst the stationary ones, for instance Minkowski spacetime. The idea is that this be enough for it to vanish on every stationary solution.
Otherwise, the stationary condition satisfied by the symplectic potential would not guarantee charge conservation. 
Then, integrating \eqref{jN} on a region $\D\scri$ delimited by two cross sections $S_1$ and $S_2$, we obtain 
the flux-balance laws 
\be\label{FB1}
\bar Q_\xi[S_2] -\bar Q_\xi[S_1] \eqons F_\xi:=\int_{\D\scri}I_\xi\bar\th.
\ee
Given the right-hand side of \eqref{FB1}, the charges can be explicitly computed `integrating the fluxes' using the Einstein's equations. This is the procedure used for BMS charges in  \cite{Ashtekar:1981bq,Dray:1984gz,Ashtekar:2024stm}.  
The charges so defined are not unique, but ambiguous up to a constant in time,
\be\label{chargeamb}
\bar Q_\xi\to \bar Q_\xi + \Qa_\xi, \qquad \pounds_n \Qa_\xi=0.
\ee
This ambiguity was resolved in \cite{Ashtekar:1981bq,Dray:1984gz,Ashtekar:2024stm} first arguing that the only time-independent 
and background-independent quantities that can be constructed out of the radiative phase space are also universal, and then requiring that all charges vanish on the reference solution.
The Wald-Zoupas paper used a different way to fix the same ambiguity. If condition 0 holds, replacing \eqref{cond0} in \eqref{Ixiom} gives 
\be\label{Ixiom3}
-I_\xi\Om_\Si \eqons \oint_S \d(q_\xi +i_\xi b - s_\xi) - i_\xi\bar\th.
\ee
Here we assumed that all field dependence in $\xi$ comes from the extension of the symmetry vectors fields, as is the case for BMS, so that $i_{\d\xi}b=0$ and $q_{\d\xi}=\d s_\xi$, see \eqref{sxi}. Then  $d\om\eqons 0$ implies
\be\label{chargevar}
\bar Q_\xi[S_2] -\bar Q_\xi[S_1] =\oint^{S_2}_{S_1} q_\xi +i_\xi b - s_\xi , 
\ee
up to a field-space constant. 
The idea is then to fix \eqref{chargeamb} requiring
\be\label{barQWZ}
\bar Q_\xi = \oint_S q_\xi +i_\xi b - s_\xi,
\ee
and the field space constant as before via the reference solution. If all background fields are time independent, \eqref{barQWZ} fixes both ambiguities at once.
The relation \eqref{barQWZ} can be understood as an instance of the improved Noether charge formula \cite{Harlow:2019yfa,Freidel:2020xyx}
\be\label{iN}
	\bar Q_\xi = \oint_S \bar q_\xi = \oint_S q_\xi +i_\xi\ell -I_\xi c,
\ee
where the boundary Lagrangian is $\ell = b + dc$ and $s_\xi = i_\xi dc - I_\xi \d c$ \cite{Odak:2022ndm}. 
In other words, one can use a corner improvement of the type allowed by condition 0 to get rid of the extension dependence of the Komar 2-form. 
From this perspective, satisfying \eqref{jN} requires $\D_\xi\ell=0$.

This procedure shows via \eqref{Ixiom3} that the charges can be also interpreted as canonical generators for the phase space on $\Si$, albeit in the following weak sense: They are proper canonical generators for a symmetry $\xi\in TS$, whereas for a symmetry $\xi$ that moves the corner
it is a canonical generator only (for arbitrary perturbations) around the non-radiative solutions.
On the other hand, the fluxes \eqref{FB1} provide canonical generators for all symmetries on the radiative phase space on $\D\scri$ \cite{Ashtekar:2024stm}.
For BMS, this procedure gives the same unique set of charges that are found with the `integrating the fluxes' procedure. 
A practical convenience of the Wald-Zoupas procedure is that one can determine the charges starting from knowledge of the Komar 2-form and its limit to $\scri$. 

We have discussed the charge ambiguity \eqref{chargeamb} and how it can be fixed. 
There is a considerably larger ambiguity in the charge \emph{aspects}.
As we see from \eqref{jN}, the aspects $\bar q_\xi$ are defined up to the addition of a closed 2-form on $\scri$, 
\be\label{aspectamb}
\bar q_\xi \to \bar q_\xi +\qa_\xi, \qquad \pbi{d\qa_\xi}=0. 
\ee
Since the cohomology is not trivial, $\qa_\xi$ needs not be exact, and this is what gives rise to the charge ambiguity \eqref{chargeamb}, with $\Qa_\xi=\oint_S\qa_\xi$.
More precisely, since the charges are integrals on the cross-sections, is only the non-exact part of $\qa_\xi$ after pull-back on the cross-sections that is relevant to the charge ambiguity. 
Its time independence follows from\footnote{And we assume $\pounds_n \oint_S \qa_\xi=\oint_S\pounds_n \qa_\xi$ which should be guaranteed by smoothness of the fields and $S$.}
\be 
\pounds_n \qa_\xi = d i_n \qa_\xi \qquad \Rightarrow \qquad  \pbi{\pounds_n \qa_\xi} = \Dd_A \qa^A_\xi\eps_S.
\ee 
The larger ambiguity in the aspects includes a non-trivial time dependence, provided it reduces to a total divergence on the cross sections. 
Since the charges are defined as integrals on cross-sections, it is tempting to assume that their aspect 2-forms does not have time components, namely that we can write
\be
q_\xi = q_\xi^{\sscr S}\eps_S,
\ee
for some scalar quantity $q_\xi^{\sscr S}$.
In this case the ambiguity reduces to $\pounds_n \qa_\xi=0$, since $\pounds_n \eps_S=\pounds_n\eps_\scri=0$.
Not all aspects are of this type, however.
It is also important to add that the aspect ambiguity cannot be fixed \`a la Wald-Zoupas using the symplectic structure on $\Si$ and  the `Komar bootstrap'. 
The reason for this is that one needs to \emph{integrate} $d\om\eqons 0$ in order to obtain a relation between the charges. In other words, 
removing the integral in \eqref{barQWZ} gives an equivalence only up to an arbitrary total divergence on the cross sections.
It follows that 
even if the charges are uniquely fixed at the end of the procedure, there is still freedom to add time-dependent closed 2-forms to the aspects, provided their time variation is an exact 2-form.

At the end of the day, stationarity and covariance of the symplectic potential have secured two important properties for the charges: they are conserved on non-radiative spacetimes, and are related to canonical generators (in the weak sense for the phase space on $\Si$, and their fluxes in the general sense for the phase space on $\D\scri$).
Notice also that this prescription for the charges corresponds to a `split' in which the same quantity $\bar\th$ determines both the non-integrable term and the charge flux,
a property which is not true for a generic split.
A comparison of properties for a generic split and a Wald-Zoupas one is summarized in Table~\ref{tableWZ}.

A new result relevant for us is that the covariance \eqref{covth} of the preferred symplectic potential also guarantees that the Noether currents \eqref{jN} realize the symmetry algebra  
free of any field-dependent 2-cocycle
\cite{Rignon-Bret:2024wlu}. We now review it, providing more details than it was possible in the letter.
The first step to prove this is to use the commutator $[\d_\chi -\pounds_\chi, I_\xi ] = I_{\llbracket \xi,\chi\rrbracket}$, which 
together with \eqref{covth} immediately implies
\be\label{jbcov}
(\d_\chi-\pounds_\chi)\bar\jmath_\xi = \bar\jmath_{\llbracket \xi, \chi\rrbracket}. 
\ee
This equation is interesting in its own right:
It means that the only background-dependent part of the 
current comes from the symmetry vector fields, see \eqref{anoxi}, hence it has an intuitive meaning of covariance. 

The second step is to define a \emph{current bracket} similar to the Barnich-Troessaert  bracket \eqref{BTbracket}, 
\be\label{currentalgebra}
\{\jb_\xi,\jb_\chi\}_*:=I_\xi I_\chi \bar\om +d(i_\xi I_\chi{\bar\th} - i_\chi I_\xi{\bar\th}) \eqons (\d_\chi -\pounds_\chi) \jb_\xi = \jb_{\llbracket\xi, \chi\rrbracket}.
\ee
The last equality follows from \eqref{jbcov}, and shows that the algebra is  realized covariantly and without any field-dependent 2-cocycle.
There is no central extension either, but this is (obviously) not a consequence of covariance, but rather of the fact that we assumed that the Noether currents satisfy the stationarity condition. If this is violated, namely we admit a non-vanishing field-independent term  $-\bar a_\xi$ on the right-hand side of \eqref{jN}, it would result in a central extension $-\bar a_{\llbracket \chi,\xi\rrbracket}$ on the right-hand side of \eqref{currentalgebra}.
It follows from \eqref{currentalgebra} that the algebra of fluxes between any two cross sections is also covariant, namely free of 2-cocycles, and furthermore center-less,
\be
\{F_\xi,F_\chi\}_* = \d_\chi F_\xi  -\oint^{S_2}_{S_1}i_\chi \jb_\xi = F_{\llbracket\xi, \chi\rrbracket}.
\ee
This equation also shows that the flux algebra is sensitive to the dissipation at the initial and final cuts.

There are also strong implications for the charge algebra. Integrating \eqref{currentalgebra} on $\D\scri$ we obtain the difference of two Barnich-Troessaert  brackets \eqref{BTbracket} associated with the two cuts, and from the right-hand side we learn that this difference gives a center-less realization of the algebra. It 
follows that the only 2-cocycle allowed by the Wald-Zoupas split is time-independent:
\be\label{BTbracket1}
\{ \bar Q_{\xi}, \bar Q_\chi \}_* := \d_\chi \bar Q_\xi - I_\xi \bar {\F}_\chi = I_\xi I_\chi \bar \Om_\Si +I_\chi \bar \F_\xi - I_\xi \bar \F_\chi
 \eqons \bar Q_{\llbracket \xi,\chi \rrbracket} + \bar K_{(\xi,\chi)},
\qquad \pounds_n \bar K_{(\xi,\chi)}=0.
\ee
In this result we assumed that the charge ambiguity \eqref{chargeamb} has been fixed \`a la Wald-Zoupas matching the canonical generators on $\Si$, so to match the definition of the Barnich-Troessaert bracket.
While this result still allows in principle for a field-dependent 2-cocycle, it is severely constrained. It cannot for instance depend on the shear, unlike \eqref{KBT}.

Whether the residual 2-cocycle can also be removed is not controlled by the covariance of the symplectic potential, but may 
be granted by the boundary conditions. For instance, if all time-independent and background-independent admissible terms are also universal, then the cocycle is reduced to a central extension. And if there are none, it  vanishes. This is what will see below happens for the BMS charges. For the sake of a general discussion, let us suppose that a cocycle is present, and ask whether there are ambiguities left that can be used to try to remove it.
Being time-independent, it can be directly affected by the charge ambiguity \eqref{chargeamb}. If this ambiguity has been fixed as above and the 
Wald-Zoupas potential is unique, this possibility is ruled out. The only option left in this case would be to to relax condition 0 and allow corner improvements in the symplectic 2-form. 
The bottom line is that Wald-Zoupas covariance guarantees a covariant realization of the current algebra, and reduces the allowed cocycle in the charge algebra to be time-independent. Additional conditions are needed to remove the residual charge cocycle. 

We remark that the matching of flux and non-integrable term is crucial for the correct interpretation of the Barnich-Troessaert bracket, because it is only in this case that it correctly reproduces the standard action  on non-radiative spacetimes, namely
\be\label{nonradcov}
\{ \bar Q_{\xi}, \bar Q_\chi \}_* = \d_\chi \bar Q_\xi.
\ee
The  Barnich-Troessaert bracket can also be interpreted as a precise definition of charge anomaly, since
 \be\label{Qcov}
\D_\chi \bar Q_\xi := \oint_S \D_\chi \bar q_{\xi} = \{ \bar Q_{\xi}, \bar Q_\chi \}_*.
\ee
This allows us to say that covariance of the charge algebra, namely $\d\bar K_{(\xi,\chi)}=0$, really means that the only background dependence of the charges is through the symmetry vector fields.
Our analysis shows that this is a meaningful and unambiguous statement also for radiative solutions, hence it generalizes \eqref{nonradcov} to the full phase space. 

In the BMS case, the background fields are $\xi$, $\Om$ and the foliation $u$, or equivalently the Lie-dragged $l$. These fields are affected by a BMS transformation, and the anomaly operator computes this action without touching the physical fields. This is why we can use it to test background-independence. 
On the other hand, notice that in \eqref{Qcov} we are not changing the cross section: $\pounds_\chi$ acts on the integrand only. Accordingly, the choice of cross section is taken to be a physical input, and not part of the background whose dependence is measured by the anomaly operator. It should be indeed clear that two different cross sections $S$ and $S'$ contain different information about the physics, since there can be radiation between $S$ and $S'$. The distinction can be made clearer if we write the full functional dependency of the charges as $\bar Q_\xi[S; g, \Om, l]$. Then the meaning of the covariance \eqref{Qcov} is (linearized) independence from $\Om$ and $l$, but not from $S$ or $\xi$:
\be
\bar Q_{\xi'}[S; g, \Om', l'] \simeq \bar Q_\xi[S;g, \Om, l] + \D_\chi \bar Q_{\xi}[S;g, \Om,l] = \bar Q_\xi[S;g, \Om, l] + \bar Q_{[\xi,\chi]}[S;g, \Om,l].
\ee
Since a decomposition of $\xi$ into super-translations and Lorentz makes necessarily reference to a choice of cross-section, this formula (and more precisely its finite version) allows one to map the charges on a given cross-section to the charge expression one would use for a different cross section.

A second important remark is that knowledge of the charges is \emph{not enough} in order to compute the bracket and check their covariance: one needs to know the Noether current $\jb_\xi\eqons d\bar q_\xi$ as a 3-form. 
In fact, terms in the Noether currents which are total divergences on the cross sections drop out of the charges (and the fluxes), but contribute to their transformation laws because in general 
\be\label{LieDcomm} 
[\pounds_\xi,\Dd_A]\neq 0,
\ee
even if it is true that $[\pounds_\xi, d] = 0$. This is because the pullback operator does not commute with $\pounds_\xi$ if $\xi$ is not tangent to the cross section. We find that such terms typically vanish in the non-radiative case, hence this subtlety is not present in the simpler notion of non-radiative covariance \eqref{nonradcov}.

A related implication is that in general also
\be\label{DDcomm} 
[\D_\xi,\Dd_A]\neq 0.
\ee
This means that it is crucial  to require covariance of the symplectic potential as a 3-form as in \eqref{covth}, and not only up to total divergences on cross-sections.
Otherwise linearized covariance would not imply full covariance, since the successive action of $\D_\xi$ needed to study finite covariance would produce anomalies which are not total divergences. 

Finally, let us discuss the covariance of the charge aspects. 
Covariance of the currents  \eqref{jbcov} implies that
\be \label{AnoAspects}
(\d_\chi-\pounds_\chi) \bar q_\xi = \bar q_{\llbracket \xi,\chi\rrbracket} + \bar c_{(\xi,\chi)}, \qquad \pbi{d\bar c}_{(\xi,\chi)}=0.
\ee
If we integrate over a cross-section
we recover the Barnich-Troessart bracket \eqref{BTbracket1}, and identify the time-independent cocycle with $\bar K_{(\xi,\chi)}=\oint_S \bar c_{(\xi,\chi)}$. 
Notice though that $\bar c_{(\xi,\chi)}$  is not necessarily anti-symmetric: We are only guaranteed that its integral is, thanks to the relation between the bracket and the symplectic 2-form.
From this perspective, the necessary condition for covariant charges is 
that the pull-back of ${\bar c}_{(\xi,\chi)}$ on the cross-sections is a total divergence, up to a field-space constant,\footnote{Therefore, as for the current, we allow the existence of central charges, thus we want a charge satisfying $\d(\d_\chi - \pounds_\chi) Q_\xi = \d Q_{\llbracket\xi, \chi\rrbracket}$. By doing this, the covariance condition is set on one form in the field space, similarly to the conditon $(\d_\xi - \pounds_\xi) \bar{\theta}$ necessary for ensuring the covariance of the currents.}
\be\label{ctotder}
\pbi{\bar c}_{(\xi,\chi)} = \Dd_A \bar c^A\eps_S +C_{(\xi,\chi)},\qquad \d C_{(\xi,\chi)}=0.
\ee

One can then ask whether it is possible to obtain an even stronger property, namely covariance of the charge aspects. It is natural to define covariance of the aspects as the requirement that $\d\bar c_{(\xi,\chi)}=0$, or the even stronger
\be \label{covaspcond}
(\d_\chi-\pounds_\chi) \bar q_\xi = \bar q_{\llbracket \xi,\chi\rrbracket}.
\ee
In situations in which there is a residual ambiguity of adding closed 2-forms to the charge aspects, requiring their covariance offers a finer way to fix it.
In the notation used above, one can use any residual freedom in $\vth^e$ to attempt setting $\bar c_{(\xi,\chi)}$ to zero, up to a field-space constant (Assuming that it is already a total divergences on the cross-sections so that covariance of the charges is obtained, otherwise this would be a pointless exercise).
This goes a step beyond what can be achieved with the standard Ashtekar-Streubel or Wald-Zoupas prescriptions, that only deal with the charges and not the aspects. 
This is a very strong requirement, and there is little guarantee that it can be achieved. 
Already achieving \eqref{Qcov} with vanishing cocycle or field-space constant is to be considered a non-trivial success.
A word of caution is also due. While \eqref{AnoAspects} is a natural definition since it 
captures the intuitive notion of covariance as background-independence, it does not translate to a realization of the algebra in terms of a bracket \`a la Barnich-Troessaert, because it is not antisymmetric.

\subsection{Corner improvements}\label{SecCorner}

Let us see what happens if we relax condition 0, and allow a non-field-space exact corner term $\vth$ as in \eqref{thbar}.
This is motivated for instance by systems in which a $\bar\th$ satisfying stationarity and covariance, or even finiteness, cannot be found otherwise \cite{Compere:2008us,Harlow:2019yfa,Campiglia:2020qvc,Odak:2021axr,Chandrasekaran:2021vyu,Odak:2022ndm,Donnay:2022hkf}. The main technical difference in dealing with this case is that we cannot work with the original symplectic 2-form, since now
$\bar\om=\d\bar\th=\pbi{\om}-\d\pbi{d\vth}$. 
The symplectic current $\bar\om$ is only defined at $\scri$, but we can define a symplectic 2-form associated with $\bar\om$ on any space-like $\Si$ as long as it has a single boundary at $\scri$. To do so, we need first to fix the ambiguity in $\vth$, which follows from its definition \eqref{thbar} and is given by 
\be\label{vthe}
\vth \to \vth+\vth^e, \qquad \pbi{d\vth^e}=0.
\ee
We fix this ambiguity prescribing $\vth^e$. It amounts to 
define the preferred symplectic potential via
\be\label{barth1}
\th \eqonS \bar\th -\d \ell +d\vth.
\ee
That is, \eqref{thbar} \emph{without} the pull-back. Once we have done this, we take an arbitrary extension of $\vth$ in the bulk, and define
\be\label{barOm}
\bar\Om_\Si:=\int_\Si (\om-d\d\vth)=\Om_\Si -\oint_S\d\vth.
\ee
This expression is independent of the choice of extension of $\vth$. On the other hand, it is affected by 
the ambiguity \eqref{vthe}: Changing $\vth^e$ changes $\bar\Om_\Si$ by a constant in time.

Having a new symplectic 2-form affects the 
Wald-Zoupas procedure as follows.
The charge definition   is now 
\be\label{IxiomWZ}
\d d \bar q_\xi:\!\!\eqons\! -I_\xi \pbi{\bar\om} +d i_\xi\bar\th =\d I_\xi\bar\th.
\ee
The Noether currents \eqref{jN} and flux-balance laws \eqref{FB1} are still the same, but the relation to the canonical generators at $\Si$ is changed to
\be\label{Ixiom5}
-I_\xi\bar \Om_\Si = -I_\xi\Om_\Si +\oint_S\d_\xi\vth-\d I_\xi\vth \eqons \oint_S\d(q_\xi + i_\xi\ell - s_{\xi} -I_\xi\vth) -i_\xi\bar\th +(\d_\xi-\pounds_\xi)\vth.
\ee
The corner anomaly that appears here is restricted by the covariance and stationarity to be constant in time. 
To prove this, we use first $d\bar\om\eqons 0$ and \eqref{IxiomWZ} to deduce that
\be
 \oint^{S_2}_{S_!}\d(q_\xi + i_\xi\ell - s_{\xi} -I_\xi\vth) +(\d_\xi-\pounds_\xi)\vth = \d \oint^{S_2}_{S_1}\bar q_\xi = \d\int_{\D\scri}I_\xi\bar\th. 
\ee
Then \eqref{jN} and the improved Noether charge formula \eqref{iN} (applied with $\ell' = \ell+dc$ and $\vth'=\vth+\d c$) 
imply that
\be\label{dDvth}
\pounds_n \oint (\d_\xi-\pounds_\xi)\vth =0 \quad \Rightarrow \quad   \pounds_n (\d_\xi-\pounds_\xi)\vth=\Dd_A C_\xi^A\eps_S,
\ee
for some $C_\xi^A$.

Even if time-independent, such a corner anomaly is problematic. It cannot be field-space exact, otherwise we would still satisfy condition 0. 
Therefore it cannot be reabsorbed in the charge. 
We can still fix the charge ambiguity requiring as before
\be\label{barQWZ1}
\bar Q_\xi = \oint_S q_\xi +i_\xi \ell - s_\xi  -I_\xi\vth, 
\ee
but now the relation of the charges to the canonical generators is spoiled, since
\be\label{Ixiom4}
-I_\xi\bar \Om_\Si \eqons \oint_S\d \bar q_\xi  -i_\xi\bar\th +(\d_\xi-\pounds_\xi)\vth.
\ee
This problem can be avoided if the covariant $\bar\th$ corresponds to a $\vth$ that is also covariant, at least up to a total divergence.
This is where the ambiguity $\vth^e$ can turn out to be useful. 
The discrepancy is in fact guaranteed to be time-independent, see \eqref{dDvth}, therefore it may be possible to remove it using the ambiguity  in the charges, or equivalently in adding $\vth^e$.
Once this is done (if it can be done),
 the only residual ambiguity is that the final covariant $\vth^e$ may not be unique. If this happens, it cannot be fixed using 
\eqref{Ixiom4}, and an independent prescription would be required. One way to do so is to look at the cocycle, which may still be present even if \eqref{dDvth} holds. 
If this is the case, then we have the additional freedom to play with $\vth^e$ ambiguity within the anomaly free class to try to remove it.\footnote{One could also do this within condition 0, namely if the charges related to $I_\xi\Om_\Si$ are not covariant, one can consider modifications adding a $\vth^e$ satisfying $\pbi{d\vth^e}=0$, which would only act on the ambiguity \eqref{chargeamb} while preserving the matching to the canonical generators.}
If there is no cocycle,  it may be possible to restrict this remaining ambiguity requiring covariance of the aspects, and not only of the charges.

The relation \eqref{Qcov} between the charge anomalies and the Barnich-Troessaert bracket requires \eqref{Ixiom4}. If there is an anomalous corner term as in \eqref{Ixiom}, then the bracket has to be modified (in order to be anti-symmetric) to
\be\label{QQnotcangen}
\{ \bar Q_{\xi}, \bar Q_\chi \}_* := I_\xi I_\chi \bar \Om_\Si +I_\chi \bar \F_\xi - I_\xi \bar \F_\chi = \d_\chi \bar Q_\xi - I_\xi \bar {\F}_\chi +\oint_S I_\chi\D_\xi\vth
 \eqons \bar Q_{\llbracket \xi,\chi \rrbracket} + \bar K_{(\xi,\chi)},
\ee
with a cocycle that is still time-independent thanks to the covariance of the symplectic potential. 
We thus see that an anomalous $\vth$ is responsible for both spoiling the relation to the canonical generators and introducing a cocycle.

\begin{table}\begin{center}\begin{spacing}{1.3} 
\begin{tabular}{|l|l|l|}
\hline \emph{Quantity} & \emph{Generic split} & \emph{Wald-Zoupas split} \\\hline
Symplectic flux &  $\th'=\pbi{\th}+\d\ell'-d\vth'$ & $\bar\th=\pbi{\th}+\d\bar\ell-d\bar\vth$  \\
Non-integrable term &  $\cF_\xi= \oint i_\xi\th'+\D_\xi\th'+I_{\d\xi}\th'$
&  $\cF_\xi= \oint i_\xi\bar\th$  \\
Noether current & $j_\xi:=I_\xi\th'-\D_\xi\ell' 
$ & $\bar\jmath_\xi:=I_\xi\bar\th$ \\
Charge flux & $F_\xi:=\int j_\xi$ &  $F_\xi:=\int \jb_\xi$  \\
Current 2-Cocycle & 
arbitrary & absent \\
Charge 2-Cocycle & arbitrary & time-independent 
\\\hline
\end{tabular}\end{spacing}\end{center}\vspace{-1cm}
\caption{\label{tableWZ}\emph{\small{Comparing an arbitrary split with a Wald-Zoupas split, defined both by \eqref{thbar} where $\th$ is the standard EH potential, but the second additionally satisfies \eqref{covth} and the stationary condition. The 2-cocycles refer to the brackets \eqref{currentalgebra} and \eqref{BTbracket}. }}}
\end{table}

We remark that all results of this Section are valid also for field-dependent diffeomorphisms. One should however distinguish two very different situations in which these can arise. First, as field-dependence of the arbitrary bulk extension of a boundary symmetry. This is for instance the case of the BMS and eBMS symmetries, see \eqref{xi}. In this context 
$I_{\d\xi}$ matters for the bulk $\th$, but has trivial action on quantities defined intrinsically at $\scri$ such as $\bar \th$. It may still be useful to keep $\llbracket \xi,\chi\rrbracket$ so to be able to use 4d Lie brackets, but field-dependence can be forgotten altogether if one restricts attention to the boundary, and writes only the 3d boundary Lie bracket. This is what was done in \cite{Rignon-Bret:2024wlu}, to keep the presentation as focused as possible.
A very different situation occurs if there is an actual field-dependence of the symmetry parameters, for instance if the boundary conditions are not universal but field-dependent. The technical difference is that $\d\xi$ is not a symmetry vector field in the first situation, whereas it is in the second.
One may then consider two inequivalent definitions of covariance: $(\d_\xi-\pounds_\xi)\bar\th=0$ as before, or
$\D_\xi\bar\th=(\d_\xi-\pounds_\xi-I_{\d\xi})\bar\th=0$.
This second option is weaker, and does not guarantee that \eqref{IxiomWZ} is exact: that has to be an additional requirement, and its minimal form is $I_{\d\xi}\bar\th\eqons d Z$ for some $Z$, see discussion in \cite{Odak:2023pga}. Then covariance of the symmetry algebra can also be obtained, but one has to define the bracket subtracting two additional terms,
\be
\{\jb_\xi,\jb_\chi\}'_*:= I_\xi I_\chi \bar\om +d(i_\xi I_\chi{\bar\th} - i_\chi I_\xi{\bar\th}) +I_{\d_\chi\xi}\bar\th - I_{\d_\xi\chi}\bar\th
\eqons (\d_\chi -\pounds_\chi) \jb_\xi - \bar\jmath_{\d_{\xi}\chi}
= \jb_{\llbracket\xi,\chi\rrbracket}.
\ee
This definition is however less satisfactory in our opinion, because there is no guarantee that the fluxes are canonical generators in the dense subset of the radiative phase space with vanishing news at the initial and final cross-sections. For these reasons it seems to us that even in the case of field-dependent symmetries (which is not relevant for the rest of this paper), one should insist on \eqref{covth} as the definition of covariance, and not $\D_\xi\bar\th=0$.

The fact that anomalies produce cocycles in the charge algebra was pointed out already in \cite{Chandrasekaran:2020wwn,Freidel:2021cjp,Freidel:2021yqe}. We have sharpened those results identifying the precise meaning of the cocycle as a background-dependence on the conformal factor and foliation, and more importantly showing the implications of the Wald-Zoupas requirements for the flux and charge brackets.

If \eqref{Qcov} is satisfied the resulting charges satisfy all the properties that one may look for: they are conserved on solutions satisfying the stationarity condition, coincide with canonical generators for arbitrary perturbation around the stationary solutions, and provide an anomaly-free realization of the symmetry algebra, namely no cocycle in the  Barnich-Troessaert bracket.
If furthermore one proves that the prescription is unique, then the problem of associating charges to a given spacetime symmetry is completely solved. 
This turns out to be the case for the BMS charges, for which a unique prescription satisfying all these properties exists. 
We now show how this viewpoint improves on the split \eqref{qBT} and leads to a prescription in which both current algebra and charge algebra are free of cocycles.
We then discuss how one can go one step beyond and select charge aspects with covariant properties.

\subsection{Covariant BMS charges}

Let us now go back to  the  Barnich-Troessaert split \eqref{BT11}. 
As we have seen, it corresponds to a symplectic potential $\th^{\sscr BT}$ that is covariant and stationary on round spheres.
Hence the  Barnich-Troessaert charges \eqref{BT11} are valid Wald-Zoupas charges, provided one restricts attention to Bondi frames.
It follows that if we want to remove the cocycle in general, all we need is to do is to pick a symplectic potential that is covariant and stationary on arbitrary frames, and not only on round spheres. As pointed out in \cite{Odak:2022ndm}, this is achieved adding and subtracting the term $\f1{32\pi}\d(\r_{AB} C^{AB})$ in \eqref{barth}, so that $\dot C_{AB}$ is replaced with $N_{AB}$ as in \eqref{NAB}.
Namely, we write
\be\label{barth2}
\pbi{\th} =  \th^{\sscr BT} -\d b^{\sscr BT} = \th^{\sscr BMS} -\d b^{\sscr BMS}, 
\ee
where now
\be\label{thBMS}
\th^{\sscr BMS} :=  \th^{\sscr BT} + \d\ell^{\sscr G}= \f1{32\pi} N_{AB}\d C^{AB}\eps_\scri, \qquad b^{\sscr BMS} := b^{\sscr BT}+\ell^{\sscr G}, \qquad 
 \ell^{\sscr G}:=-\f1{32\pi}\r_{AB}C^{AB}\eps_\scri,
\ee 
and
\be\label{qBMS1}
q_\xi^{\sscr BMS} = q_\xi^{\sscr BT} + i_\xi\ell^{\sscr G}
\ee
up to a closed 2-form.
The new symplectic potential is manifestly stationary for non-radiative spacetimes on arbitrary frames, and its covariance and conformal invariance follow from \eqref{DxiN} and \eqref{DxiCinv}, so that $(\d_\xi-\pounds_\xi)\th^{\sscr BMS}=0$.
It is also easy to check that it matches the one given by Wald and Zoupas \cite{Wald:1999wa}, here specialized to Bondi coordinates and for a shear associated with the $u$-foliation. To see that, recall that in their expression, Wald and Zoupas use Ashtekar's `connection coordinate' \cite{Ashtekar:1981hw}, which as explained in the previous Section is a relative shear with the vacuum fixed once and for all; in other words, with $\d\sv=0$. Therefore $\d C_{AB}=\d {\cal C}_{AB}$, and the equivalence follows. 

The associated charge split is
\begin{align}\label{qWZBMS}
Q^{\sscr BMS}_\xi = \f1{8\pi}\oint_S (2\xiu  M_\r+ Y^A\am_A)\eps_S, \qquad 
\F^{\sscr BMS}_\xi = \f1{32\pi}\oint_S \xiu N_{AB}\d C^{AB}\eps_S,
\end{align}
where now 
\be\label{SM}
M_\r:= M - \f18\r_{AB}C^{AB} = -\re(\psi_2 - \s N)
\ee
coincides with (the pull-back of) Geroch's super-momentum \eqref{SM} on  arbitrary frames and not only on round spheres. 
The Lorentz charge \eqref{DS} picks up a similar shift through the boost dependence in $f$. 
The charges \eqref{qWZBMS} are defined up to a field-space constant, which is fixed uniquely to zero by the requirement that all charges vanish in Minkowski spacetime.\footnote{\label{Qeta0} This is not manifest for the Lorentz part. It relies on integration by parts, and the properties \eqref{magnshear} of a vacuum shear and \eqref{CKV} of the symmetry vector fields.}
The expression in NP language is useful in the sense that it does not make reference to explicit coordinates for $\scri$,
and we can also rewrite in the same way the non-integrable piece, 
\be
\F^{\sscr BMS}_\xi = -\f1{16\pi}\oint_S \xiu N_{ab}\d \s^{ab}\eps_S= \f1{4\pi}\oint_S \xiu \re(N\d \s)\eps_S.
\ee
The flux of the charges can be computed from \eqref{jN}, giving
\be\label{BMSflux}
Q^{\sscr BMS}_\xi[S_2]-Q^{\sscr BMS}_\xi[S_1] \eqons F^{\sscr BMS}_\xi= -\f1{16\pi}\int N_{ab}\d_\xi\s^{ab} \eps_\scri.
\ee
The charge flux $F^{\sscr BMS}$ and the non-integrable term $\cF^{\sscr BMS}$ have the same functional dependence in any frame, in agreement with the Wald-Zoupas prescription described earlier, and unlike for the split \eqref{qBT}. Furthermore $F_\xi^{\sscr BMS}$ with $\d_\xi \s^{ab}$ given by \eqref{dxiC} matches the Ashtekar-Streubel flux \cite{Ashtekar:1981bq} with $\s^{ab}$ the shear of the $u$-foliation.
We conclude that imposing covariance and stationarity on every frame modifies the split used in \cite{Barnich:2011mi} and leads to the result of \cite{Wald:1999wa}, namely charges given by Geroch and Dray-Streubel's expressions, with flux given by the Ashtekar-Streubel expression.

Let us now see how the modification in the split also removes the cocycle.
To that end, we insert the relations \eqref{thBMS} and \eqref{qBMS1} in the Barnich-Troessaert bracket \eqref{BTbracket}, 
\begin{align}\label{BMSbracket}
\{Q^{\sscr BMS}_\xi,Q^{\sscr BMS}_\chi\}_* &:=\d_\chi Q^{\sscr BMS} _\xi - I_\xi \F^{\sscr BMS}_\chi 
= Q^{\sscr BMS}_{\llbracket \xi,\chi\rrbracket} + K^{\sscr BT}_{(\xi,\chi)} +\oint_S i_\xi \D_\chi \ell^{\sscr G} - i_\chi \D_\xi \ell^{\sscr G},
\end{align}
where we used the fact that $i_{\d\xi}\ell^{\sscr G}=0$. The anomaly of $\ell^{\sscr G}$  is given by \cite{Odak:2022ndm}
\be\label{Gano}
\D_\xi \ell^{\sscr G} = -\f1{32\pi}(C^{AB}\Dd_A \Dd_B \Dd_CY^C_\xi +2\r^{AB}\Dd_{\la A}\Dd_{B\ra}\xiu_\xi)\eps_\scri.
\ee
It can be computed with the formulas in Sec.~\ref{SecBondi}, and relies crucially on \eqref{lierho}.
The term with $C^{AB}$ cancels the one coming from the cocycle \eqref{KBT}. The second term in \eqref{Gano} can be integrated by parts, and using $2\Dd^B\r_{\la AB\ra}=\p_A \cR$ we see that it cancels the second term of the cocycle. In other words, the contribution from the anomaly of $\ell^{\sscr G}$ in \eqref{BMSbracket} matches precisely the cocycle \eqref{KBT} up to an integration by parts, and therefore
\begin{align}
\D_\chi Q_\xi^{\sscr BMS} = \{Q^{\sscr BMS}_\xi,Q^{\sscr BMS}_\chi\}_*  
= Q^{\sscr BMS}_{\llbracket \xi,\chi\rrbracket}.  \label{WZbracket}
\end{align}
It implies from the  definition of the Barnich-Troessaert bracket that in non-radiative spacetimes we recover that standard coadjoint orbits $\d_\chi Q^{\sscr BMS}_{\xi}=Q^{\sscr BMS}_{\llbracket \xi,\chi\rrbracket}$.

We conclude that adding the  boundary Lagrangian $\ell^{\sscr G}$ makes the symplectic potential covariant and stationary on arbitrary frames, and removes the 2-cocycle. 
Furthermore, the calculation shows that there is no central extension either, since every term of the original cocycle \eqref{KBT} is removed, including the field-independent ones. The absence of central extensions can also be argued for on more general grounds. In fact the only allowed central extension by the covariance requirement would be a universal quantity that must furthermore be foliation independent and conformal invariant,  a linear and anti-symmetric function of $\xi$ and $\chi$, and not a total divergence. Inspection of the quantities at disposal should convince the reader that there isn't any such term. 
It also follows that the Ashtekar-Streubel flux provides a covariant realization of the algebra between any two cross-sections of $\scri$, in agreement with the general results of the previous section:
\be\label{FFBMS}
\d_\chi F^{\sscr BMS}_\xi = F^{\sscr BMS}_{\llbracket \xi,\chi \rrbracket} +\oint^{S_2}_{S_1}i_\chi I_\xi\th^{\sscr BMS}
\qquad \Leftrightarrow\qquad \{F^{\sscr BMS}_\xi,F^{\sscr BMS}_\chi\}_*= F^{\sscr BMS}_{\llbracket\xi, \chi \rrbracket}.
\ee

In the above calculation we used the anomaly operator $\D_\xi$, which is a very convenient tool for the covariant phase space. But the removal of the cocycle is just a consequence of covariance, and can be proved  without any reference to the anomaly operator. For completeness, we do so in Appendix~\ref{AppProofs} with a calculation along the lines of \cite{Barnich:2011mi}. In the same Appendix we also explain the cocyle's removal from the perspective of the improved Noether charge construction and anomalies of its boundary Lagrangian, along the lines of \cite{Freidel:2021yqe}.

The charges and fluxes can be conveniently split in super-momentum and Lorentz using the (foliation-dependent) parametrization \eqref{xi} of the symmetry vector fields.
Specializing $\xi=\xi_T:=T\p_u$ with obtain the super-momentum charge
\begin{align}\label{QT}
Q^{\sscr BMS}_T = \f1{4\pi}\oint_S T  M_\r \eps_S = - \f1{4\pi}\oint_S T \re(\psi_2 - \s N)\eps_S,
\end{align}
with flux
\be\label{QTflux}
Q^{\sscr BMS}_T[S_2]-Q^{\sscr BMS}_T[S_1] \eqons - \f1{32\pi}\int\left(TN^{AB}N_{AB} + 2N^{AB}(\Dd_A\Dd_B+\f12\r_{AB})T  \right) \eps_\scri,
\ee
and whose transformation under a BMS symmetry $\chi$ can be read from \eqref{WZbracket} to be: 
\begin{align}\label{dxiQT}
\{Q^{\sscr BMS}_T,Q^{\sscr BMS}_\chi \}_*  = Q^{\sscr BMS}_{[\xi_T,\chi]} = Q^{\sscr BMS}_{T'},  \hspace{2.1cm} T'=\dot f_\xi T - Y_\xi[T].
\end{align}
For  $\xi=\xi_Y:=\f u2\Dd Y\p_u+Y^A\p_A$ we have the Lorentz charge
\be \label{Qam}
Q^{\sscr BMS}_Y = \f1{8\pi}\oint_S Y^A(\am_A-u\p_A M_\r)\eps_S = -\f1{4\pi}\oint_S \re\left(\bar Y\left( \psi_1 + \s\eth \bar\s+\f12\eth(\s\bar\s)\right)+u\eth M_\r\right)\eps_S,
\ee
with flux 
\be
Q^{\sscr BMS}_Y[S_2]-Q^{\sscr BMS}_Y[S_1] \eqons - \f1{64\pi}\int \left( u\Dd Y N^{AB}N_{AB}+N^{AB}(\Dd YC_{AB} - 2\pounds_Y C_{AB})\right) \eps_\scri,
\ee
and transformation law
\begin{align}\label{dxiQam}
& \{Q^{\sscr BMS}_Y,Q^{\sscr BMS}_\chi \}_*  = Q^{\sscr BMS}_{[\xi_Y,\chi]} =  Q^{\sscr BMS}_{Y'} + Q^{\sscr BMS}_{f'}, \qquad Y'=[Y,Y_\chi],  \\\nn
&\hspace{7.8cm} f'=Y[f_\chi]-\f12\Dd Y T_\chi-\f u2Y_\chi[\Dd Y].
\end{align}
In writing the Newman-Penrose version of \eqref{Qam} we defined $Y:=m_A Y^A$ and used $X_AY^A=2\re(X\bar Y)$. The Newman-Penrose expressions require a choice of auxiliary vector $l$, which is here $l=-du$.
The fluxes split into `hard' and `soft' contributions, defined respectively as the part quadratic and linear in the news. The super-momentum flux \eqref{QTflux} has the additional property of being purely `hard' for global translations, and in particular strictly negative for the energy ($T=1$) in the presence of radiation, which is Bondi's famous result.
Notice the role played by Geroch's tensor in order to make the last two statements valid in arbitrary frames.

The covariance property guarantees that the charges inherit the specific properties of the BMS algebra consistently. For instance, 
a generic BMS transformation acts homogeneously on the super-momentum as in \eqref{dxiQT}, but inhomogeneously on the Lorentz charges, with a shift by a super-momentum term 
determined  by  the parameter $T_\xi$ of the transformation, as in \eqref{dxiQam}. This is the well-known `super-translation ambiguity' of angular momentum.
It is a direct consequence of the BMS algebra, specifically the fact that there is no preferred Lorentz subgroup of the BMS group in radiative spacetimes, and the
Dray-Streubel charge \eqref{Qam} correctly captures this feature. 
Another feature that stands out is that any two super-translation charges commute. 
This brings to the forefront that the $l$-dependence of the charges leads to different behaviour under super-translations: invariance for the super-momentum, and a shift by a super-momentum for the Lorentz charge.

In concluding this Section, we remark that integrating by parts was crucial to remove the cocycle. This suggests that even though the Wald-Zoupas split \eqref{qWZBMS} has achieved covariance of both Noether currents and charges, the question is still open for the aspects. This is what we would like to address next. To do so, let us first review some properties of the charges, which will also be useful to understand the role of total divergences and to explain the anomalies of $M$ and $J_A$.

\subsection{Understanding the anomalies of $M$ and $J_A$}\label{MJano}

We have identified a covariant Noether current for the BMS symmetries, given by
\be\label{jNBMS}
j_\xi^{\sscr BMS}= -\f1{16\pi} N_{ab}\d_\xi\s^{ab} \eps_\scri\eqons dq^{\sscr BMS}_\xi.
\ee
Let us analyse separately its $T$ and $Y$ components. 
Using \eqref{NAB} and \eqref{dxiC} for a pure super-translation we can rewrite the $T$-current as
\begin{align}
j_T^{\sscr BMS}&= \f1{32\pi}N_{AB} \d_T C^{AB}\eps_\scri = -\f1{32\pi}(\dot{C}^{AB}+\r^{\la AB\ra})( T\dot C_{AB} - 2 \Dd_{A} \Dd_{B} T)\eps_\scri \nn\\
	&= \f1{32\pi}\Big[T (- \dot{C}^{AB} \dot{C}_{AB} + 2 \Dd_{A} \Dd_{B} \dot{C}_{AB} + \Dd^2 \cR) -  T \rho^{AB}\dot{C}_{AB} \nn\\
	&\qquad + 2\Dd_A( \dot{C}^{AB} \p_B T - T D_B \dot{C}^{AB} + \rho^{\la AB\ra} \Dd_B T -\f12 T\Dd^A \cR)\Big]\eps_\scri \nn \\
	&\eqons \f1{4\pi}(T \dot M_\r - \Dd_AV^A_{(N,T)})\eps_\scri =
	 \f1{4\pi}\p_u\left( T M_\r + \Dd_AV^A_{(C+u\r,T)}\right)\eps_\scri.\label{jTBMS}
\end{align}

In the second equality we integrated by parts, and in the third  we used the Einstein equation \eqref{MEE} and the short-hand notation
\be\label{defV}
V^A_{[F,f]}:= \f14 (F^{\la AB\ra}\Dd_B f-f\Dd_B F^{\la AB\ra}).
\ee
This calculation shows explicitly how one can obtaining the charges `integrating the fluxes' as advocated in \cite{Ashtekar:1981bq}, there performed in arbitrary coordinates and here specialized to Bondi coordinates. That is, we are able to rewrite the current as an exact 3-form using the Einstein's equations, and in the process we introduce `Coulombic' degrees of freedom such as $M$ that are not present in the flux. The result can be written as
\be\label{DP}
j^{\sscr BMS}_T \eqons \f1{4\pi}D_aP^a_T\eps_\scri = \f1{4\pi}dP_T,
\ee
where
\be\label{GSM}
P_T^a
= \left(TM_\r, \ \f14 \big(T\Dd_B N^{AB}-  N^{AB}\Dd_B T\big) \right).
\ee
This quantity coincides with Geroch's super-momentum  \cite{Geroch:1977jn}, in Bondi coordinates and with $l=-du$.
Its Hodge dual defines the 2-form $P_T := \f12P^a_T \eps_{\scri abc}dx^b\w dx^c$, whose pull-back on the cross sections gives $P^u_T\eps_S=TM_\r\eps_S$,
confirming what previously stated for \eqref{MNP} and \eqref{SM}.
Accordingly, 
\be
q_T^{\sscr BMS} = P_T+\qa_\xi,
\ee
where the integration constant $\qa_T$ leads to  the ambiguities  \eqref{chargeamb} and  \eqref{aspectamb},
and if we restrict it to be a total divergence it will affect only the aspects.
Let us fix $\qa_T=0$ for now, which is consistent with the BMS charges \eqref{QT},
and come back to this point and the uniqueness of charges and aspects later.

This analysis shows that $M$ is a component of a vector, and this observation allows us to understand the reason for the complicated inhomogeneous terms in its transformation law \eqref{dxiM}. These terms have a structure similar to the angular components of $P_T$, therefore they capture the mixing of $u$ and $A$ components of this vector when we change reference system by a BMS transformation. The statement can be made precise if we compute the anomaly of $P_T$. For the time component we find
\begin{align}
	\Delta_\chi P^u_T &= (\d_\chi - \pounds_\chi) P^u_T = T \d_\chi M_\rho - \chi^a \p_a (TM_\r) + P^a_T \p_a \chi^u \nn \\
	&=  3 \dot{f}_\chi T M_\rho + (T \dot{f}_\chi- Y^A_\chi \p_A T) M_\rho  - \f14 \Dd_A (T N^{AB} \Dd_B f_\chi).
\end{align}
The first term will compensate the anomaly of the volume form, see \eqref{BMSano1}, and the second term can be recognized as the covariant transformation law $P^u_{[\xi_T,\chi]}$, see \eqref{anoT}. 

For the angular components, which we remark match $P_T^A=-V^A_{(N,T)}$, we find
\begin{align}
	\Delta_\chi P^A_T &
	= \f14(\d_\chi - \pounds_\chi) (T\Dd_B N^{AB} - N^{AB} \Dd_B T) 
= 3\dot f_\chi P^A _T + P^A_{[\xi_T,\chi]} + \f14 \p_u(TN^{AB}\Dd_B f_\chi).
\end{align}
Adding up gives
\be\label{DPSM}
\Delta_\chi P_T = P_{[\xi_T,\chi]} + di_v\eps_S, \qquad v:=\f18T N^{AB} \Dd_B f_\chi \p_A.        
\ee

If we integrate on the cross-sections we find the covariant transformation law \eqref{dxiQT}.
Since $P_T$ provides a good aspects for the BMS charges, this is a consistency check of the charge algebra.
It further shows that there is nothing `non-covariant' with the transformation law of $M$, it is precisely what one needs in order for $M$ to be the time component of a \emph{covariant} BMS super-momentum vector. 
Hence there is no need to change the definition of mass aspect, as sometimes considered in the literature.\footnote{For instance the alternative charge prescription of taking $\re(\psi_2)$ alone as mass aspect will fail to be covariant beyond non-radiative spacetimes.}

Geroch's super-momentum satisfies a covariance property that is actually stronger than \eqref{dxiQT}. It transforms covariantly not only when integrated on cross-sections, but on \emph{every} two-dimensional compact region of $\scri^+$, thanks to the fact that the anomaly of its aspect is an exact form. 
Covariance of  Geroch's super-momentum means that its background dependence comes only from the symmetry vector field, and it is conformally invariant and $l$-independent, precisely as proved in  \cite{Geroch:1977jn}. Conformal invariance is actually easy to show simply counting conformal weights, but $l$-independence (namely foliation independence when $l$ is restricted to be hypersurface orthogonal) is not.
Our calculation based on the anomaly operator provides an independent proof of it, albeit in the restricted setting in which $l$ is changed not arbitrarily but as the effect of a BMS(W) transformation, namely within the hypersurface-orthogonal and Lie-dragged class. The exact 2-form anomaly in \eqref{DPSM} means that only the charge is $l$-independent, and not the aspect, again in agreement with \cite{Geroch:1977jn}.

The formula \eqref{DPSM} allows us to comment also on the covariance properties of $P_T$ as an aspect. First, we see that the anomaly is not anti-symmetric, hence we cannot interpret the aspect anomaly as a bracket. Second, it is field-dependent, and vanishes on non-radiative spacetimes. Hence Geroch's aspect satisfies our definition of aspect covariance \eqref{covaspcond} on non-radiative spacetimes. In general spacetimes it is covariant only under rotations, and not under super-translations and boosts. This is because these two transformations change the foliation, hence the Lie dragged $l$ changes as well.

The fact that the super-momentum is independent of $l$  has an important consequence. It makes it possible to capture its covariance \emph{without} making explicit reference to its angular components. The key for this trick is the last equality of \eqref{jTBMS}. The angular components are total time derivatives, hence they can be reabsorbed in the time component. This allows us to define the super-momentum charge aspect
\be\label{qTBMS}
\pbi{q}^{\sscr BMS}_T := \f1{4\pi}( T M_\r + \Dd_AV^A_{(C+u\r,T)} +\qa_T )\eps_S,\qquad \p_u \qa_T=0.
\ee
It gives the same super-momentum charges as  $P_T$. As aspects, they are related by the ambiguity \eqref{aspectamb} with $\Qa_\xi=0$ and 
\be
\qa_T = i_{\pounds_n V}\eps_\scri - \Dd_A V^A\eps_S, \qquad V^A=V^A_{(C+u\r,T)}, \qquad dP_T= 4\pi d q^{\sscr BMS}_T. 
\ee

The formula \eqref{qTBMS} is given for constant $u$ cross-sections, but since \emph{any} cross-section of $\scri$ can be written as constant $u$ in some BMS coordinate system, it leads to covariant charges. 
As an aspect though, it has different covariance properties than $P_T$. Let us compute its anomaly explicitly, since it will also offer a complementary understanding of the anomaly of $M$.
First, from \eqref{dxiM} we have
\be\label{DxiMr}
\D_{\chi} \,M_\r= 3 \dot{f}_\chi M_\r- \f{1}{2} {\Dd}_A N^{AB}  \Dd_B f_\chi - \f{1}{4} N^{AB} {\Dd}_A \Dd_B f_\chi.
\ee
The eye-catching factor of 3 occurs simply because the area 2-form and the symmetry parameter $T$ also have non-trivial conformal weights, see \eqref{BMSano1} and \eqref{anoT}. Including them to obtain (a piece of) the actual aspect, we find
\begin{align}
\D_\chi (TM_\r\eps_S) &= (\D_\chi T)M_\r\eps_S + T\left(\dot{f}_\chi M_\r- \f{1}{2} {\Dd}_A N^{AB}  \Dd_B f_\chi - \f{1}{4} N^{AB} {\Dd}_A \Dd_B f_\chi \right)\eps_S \nn\\
&= [ \xi_T,\chi]^uM_\r \eps_S - T\left(\f12 {\Dd}_A N^{AB}  \Dd_B f_\chi + \f{1}{4} N^{AB} {\Dd}_A \Dd_B f_\chi \right)\eps_S.
\end{align}
These steps are quite straightforward, but we believe it is instructive to see how the pieces add up together towards the expected covariant transformation law. 
The remaining step concerns the inhomogeneous terms. 
Comparing \eqref{DxiMr} to \eqref{dxiM} we see that the shift from $M$ to $M_\r$ eliminates the inhomogeneous terms that would not vanish for non-radiative spacetimes in arbitrary frames.
What remains is still not a total divergence. This may look surprising since the other terms in \eqref{qTBMS} are total divergences. The answer is in the non-commutativity \eqref{LieDcomm} and \eqref{DDcomm}, which means that 
we need non-total divergences in order to be able to remove the anomaly contribution of the total divergences in  \eqref{qTBMS}. 

Explicitly, we have 
\begin{align}
&[\D_\chi,\Dd_A] V = \dot V\Dd_A f_\chi, \\\label{mbappe}
&[\D_\chi,\Dd_A] V^A = \dot V^A\Dd_A f_\chi  - 2 V^A\Dd_A \dot{f}_\chi, \\
&[\D_\chi, \Dd_A] V^{AB} =  \dot{V}^{AB} \Dd_A f_\chi -4 V^{AB}\Dd_A \dot{f}_\chi.
\end{align}
The rest of the  calculation confirms this mechanism, and we obtain
\be\label{Tcov}
\D_\chi q_T^{\sscr BMS} = q^{\sscr BMS}_{\llbracket \xi_T,\chi\rrbracket} + \bar c_{(\xi_T,\chi)}.
\ee

Here
\be\label{cBMS}
c^{\sscr BMS}_{(\xi_T,\chi)}=  \f1{8\pi} \Dd_A \left(T\Dd_B \DDr^{AB} T_\chi -  \DDr^{AB} T_\chi \Dd_B T \right)\eps_S,
\ee
and we used the shorthand notation
\be\label{DDr}
\DDr{}^{ab}:={\Dd}^{\langle a} \Dd^{b\rangle}+\f12\r^{\la ab\ra}.
\ee

Since it is a total divergence, it vanishes upon integration on the cross sections and we recover \eqref{dxiQT}: The charges satisfy a covariant symmetry algebra, without central extension. 
As an aspect on the other hand it is still non covariant, but with a different anomaly than $P_T$. It is field-independent, and vanishes for global translations. 

Coming back to $\d_\xi M$, its inhomogeneous terms guarantee \eqref{Qcov}, which requires computing $i_\chi d\bar q_\xi$ and terms that are total divergences are no longer so after the interior product. 
In the non radiative case the Noether current vanishes, hence this issue is no longer relevant.
Indeed the transformation of the mass boils down to the simple $\d_\chi M_\r = \pounds_{Y_\chi} M_\r + 3\dot f_\chi M_\r$, or equivalently the manifestly covariant $\d_\chi (TM_\r\eps_S) = \pounds_{Y_\chi}(TM_\r\eps_S)+[\xi_T,\chi]^uM_\r\eps_S$.

Conversely, \eqref{qTBMS} provides a definition of `super-translation-covariant' mass aspect, in the sense of \eqref{Tcov}, and as opposed to $M$
or $M_\r$ alone.\footnote{
As a side comment, notice that if we restrict to round spheres and to global translations, 
\[
q_T^{\sscr BMS} = \f1{4\pi}T\left(M-\f14\Dd_A\Dd_B C^{AB}\right)\eps_S.
\]
It was suggested in  \cite{Moreschi:1998mw,Dain:2000lij} to take this expression as super-momentum aspect for any $T$, because it has the interesting feature of a `purely hard' flux. This option would however violate covariance, which as we see from \eqref{qTBMS} requires additional terms. Remarkably, it turns out that if one improves the Ashtekar-Streubel symplectic form by a corner term, then there exists a related expression that \emph{is} covariant, and maintains its `hard-flux' property \cite{newST}.} 
Notice that the required total divergences can only be deduced using the procedure of integrating the fluxes, and not through the `Komar bootstrap'. This is because as explained in Section~\ref{SecWZ}, there is no relation between the total divergence found in \eqref{qTBMS} and the one that is obtained keeping track of 
 total divergences in the Komar 2-form and the shifts $i_\xi\ell$ and $s_\xi$. 

Any hope that the two match through some unforeseen mechanism appears to be ruled out by an explicit calculation, which we do not report here. 

\bigskip

The complicated transformation law \eqref{dxiJ} of $J_A$ can be explained in a similar manner. 
The $Y$ part of the Noether current is 
\begin{align}
j_Y^{\sscr BMS} &=\f1{32\pi}N_{AB} \d_Y C^{AB}\eps_\scri \nn\\&= -\f1{32\pi}(\dot{C}^{AB}+\r^{\la AB\ra})
( u\dot f \dot C_{AB} +(\pounds_Y+3\dot f)C_{AB}- 2u \Dd_{A} \Dd_{B} \dot f)\eps_\scri.
\end{align}
The procedure to obtain the charge integrating the flux is considerably more complicated than the one for the super-momentum. 
One has to first integrate by parts to turn the $\dot C_{AB}\dot C^{AB}$ term into the Einstein equation \eqref{Mdot}  for $\dot M$, just like we did above. Then, one has to integrate by parts a large number of times in order to reconstruct the Einstein equation \eqref{MEE} for $\dot J_A$, and in the process take into account the CKV identity \eqref{CKV} and related identities for higher order derivatives. 
The result will be of the form
\be\label{dqYBMS}
j_Y^{\sscr BMS}\eqons  \f1{8\pi}D_aJ_Y^a\eps_\scri = \f1{8\pi}dJ_Y,
\ee
where
\be\label{JYu}
J_Y^u = Y^A(J_A-u\Dd_A M_\r). 
\ee
The angular components of this covariant vector are the \emph{raison d'etre} for the inhomogeneous terms in the transformation law of $ J_A$.
We did not attempt to compute them, and we content ourselves with the indirect proof obtained from the removal of the Barnich-Troessaert cocycle in \eqref{dxiQam}. 
The long calculation proving \eqref{dqYBMS} was successfully completed in \cite{Dray:1984gz}, using the Newmann-Penrose formalism. Unfortunately total divergences were discarded, and therefore only \eqref{JYu} was obtained. This is sufficient to prove that the fluxes integrated to the Dray-Streubel's charges. To know also the angular components $J_Y^A$,
one has to redo the calculation of \cite{Dray:1984gz} keeping track of all total divergences (As explained above, 
it is not possible to shortcut this calculation by deducing the total divergences from the limit of the Komar 2-form). 
For the (brave!) reader  interested in determining them, let us point out that one should not expect them to be total time derivatives. 
This special property made sense for the super-momentum because it is related to its covariant transformation law \eqref{dxiQT}, and to the $l$-independence of  Geroch's super-momentum. But any Lorentz charge necessarily refers to a cross section since the notion of Lorentz subgroup does. Consistently, the covariant transformation law \eqref{dxiQam} means that when we change cross section, a super-momentum shift is required. The shift would be unnecessary if it were possible to capture the covariant transformation law simply adding a total divergence to the $u$ component.

\subsection{A new covariant super-momentum aspect}

Let us come back to the question of uniqueness of the charges and their aspects, starting with the former. 
The Wald-Zoupas procedure of matching the $\Om_\Si$ calculation gave us directly \eqref{qWZBMS} up to a field-space constant which is removing by the requirement that all charges vanish in Minkowski. Therefore the charges are unique (but not the aspects). 
With the procedure of integrating the fluxes, one can use the fact that all time-independent quantities in the radiative phase space are also universal. Therefore the ambiguity \eqref{chargeamb} can be removed by the same requirement that the charges all vanish in Minkowski, and this singles out \eqref{qWZBMS} again. 
Notice that the property is only valid for the charges and not for the aspects, see footnote~\ref{Qeta0}, which are therefore left ambiguous at this stage. If one does not want to use the reference solution, or if there isn't any obvious one standing out, then it is  possible to address the ambiguity requiring covariance of the charges. In the BMS case at hand this argument gives the same answer \eqref{qWZBMS}. In fact, the only quantities that can be added to it without spoiling covariance are conformal invariant and foliation-independent, and in order to be time independent they can only be built out of the universal fields $(q_{ab}, n^a, \r_{ab})$ (plus $\xi^a$ which can enter via $I_\xi$ since $[\pounds_n,I_\xi]=0$), e.g. $\oint \Dd Y\cR\eps_S$. A moment of reflection shows that this is not possible, hence the charges are unique.

A separate discussion is necessary for the vacuum shear, and more precisely for the bad cut mode $u_0$. This is not well defined in the whole phase space, but only in the non-radiative subset of vacua. Therefore it does not change the above argument about universality in the radiative phase space. 
On the other hand, we have introduced in Section~\ref{SecRad} an enlarged  radiative phase space, adding precisely $u_0$ as corner datum, representing the (late) time boundary condition. The enlarged phase space contains a non-universal and time-independent quantity, hence the above argument no longer applies, and $u_0$ can be used to construct new charges. 
We can now distinguish two different situations, depending on whether we impose condition 0 or not. If we allow for non-trivial corner terms it is  indeed possible to find alternative covariant charges. This investigation will be reported elsewhere \cite{newST}. For this paper we maintain the original Wald-Zoupas prescription at least for BMS, hence condition 0. Then the only allowed ambiguity 
comes from a $\vth$ with $\pbi{d\vth}=0$, hence $\qa_\xi= F\eps_S$ with $\dot F=0$, meaning that $F=F(q_{ab}, n^a, \r_{ab},u_0)$. But $u_0$ here is the only quantity which is not super-translation invariant, hence it necessarily breaks covariance. The only way to use this ambiguity and preserve covariance of the charges is thus to restrict $F$ to be a total divergence. It follows that the charges are untouched, and their uniqueness is preserved also in the enlarged phase space with the Ashtekar-Streubel symplectic structure. On the other hand, we can use the ambiguity in $F$ to change the aspects. Since the aspects were not covariant to begin with, it is interesting to ask whether there exists a $\vth$ that improves their covariance. The answer is affirmative for the super-momentum aspect. 

To that end, we consider the following corner improvement:
\be
\vth= - \frac{1}{8 \pi}\Dd_A [(\Dd_B\DDr^{AB} u_0) \d u_0 - \DDr^{AB} u_0 \Dd_B \d u_0]\eps_S, \qquad \pbi{d\vth}=0.
\ee
with
\be
(\d_\xi-\pounds_\xi) \vth=\frac{1}{8 \pi} \Dd_A \left[\d u_0  \Dd_B \DDr^{AB} T_\xi - \DDr^{AB} T_\xi \Dd_B \d u_0\right]\eps_S.
\ee
It satisfies condition 0, and does not change the charges nor the Noether currents, so their covariance is preserved. It only changes the aspects, which now read
\be
\bar q_T^{\sscr BMS} = q_T^{\sscr BMS} +\f1{4\pi} \qa_T\eps_S= \f1{4\pi}( T M_\r + \Dd_AV^A_{[{\cal C},T]})\eps_S, \qquad
\qa_T=-2\Dd_A V^A_{[\D_\r u_0,T]}\eps_S = -4\pi I_{\xi_T}\vth.
\ee
In other words, the role of the corner improvement is to fix the integration constant in the potential of the news to be given by the bad cut boundary condition, hence turning the shear appearing in the aspect \eqref{qTBMS} into the relative (or `covariant') shear \eqref{calC}. Computing the algebra we find
\be
\Delta_\chi \bar q_\xi^{\sscr BMS} = \bar q_{\llbracket \xi,\chi\rrbracket}^{\sscr BMS},
\ee
for any BMS symmetry $\xi$. We find it quite remarkable that this is possible. We refer to this new aspect as Goldstone-improved super-momentum aspect.

Of course, \emph{l'appetito vien mangiando} and one may wonder whether it is possible to similarly improve the Lorentz aspect. We did not succeed in doing so, and we believe it is not possible because of the usual argument that cross-section dependence of any Lorentz subgroup is an unavoidable property of any covariant charge, hence its aspect needs to carry this background-dependence. 

\section{Extended BMS flux and charge algebra}

We turn now to examine whether it is possible to satisfy the Wald-Zoupas conditions on charges and fluxes for the extended BMS (eBMS) symmetry. 
The fall-off conditions on the metric are the same as in the BMS case,  but one allows non-globally defined CKVs. The non-globally defined CKVs are referred to collectively as `super-rotations', even though they include both `rotation-like' and `boost-like' components.
The presence of singularities in the vector fields changes the topology of the cross sections, allowing punctures on the sphere. This in turns allows one to pick a conformal frame corresponding to cross sections with a \emph{flat} metric, and $\cR=0$ (everywhere except at isolated points corresponding to the punctures). Being of constant curvature, this frame can be thought of as a special case of `Bondi frame'. 
Since the symmetry vector fields are still CKVs, \eqref{CKV} is valid, and we can still assume that $\d q_{ab}=0$ namely that the background metric is universal.
However, 
\eqref{DDDY} does \emph{not} hold for non-globally defined conformal vector fields.
Therefore the charge prescription  \eqref{bBT} fails to satisfy the stationarity and covariance conditions in \emph{every} frame, whether of constant curvature or not, and this manifests itself in the  field-dependent 2-cocyle \eqref{KBT}.

To address these issues, we can still follow Geroch's lead and observe that the Schouten tensor, or equivalently the (time derivative of the) shear of a Lie-dragged $l$, are not conformally invariant, hence we should improve the symplectic potential introducing the news tensor \eqref{News}. 
But recall that
the sphere topology is crucial to prove that $\r_{ab}$ is unique. If the topology is not that of a sphere, it is easy to see that for each given conformal completion there are infinitely many solutions to Geroch's conditions. This is the main novelty of eBMS, see Appendix~\ref{AppSET} for details.
The new degrees of freedom in Geroch's tensor can be interpreted in terms of a conformal field theory for which $\r_{ab}$ is the stress-energy tensor \cite{Barnich:2010eb,Compere:2016jwb,Barnich:2016lyg,Compere:2018ylh,Campiglia:2020qvc,Nguyen:2022zgs,Donnay:2022hkf}.\footnote{In these references, this generalization of Geroch's tensor is sometimes denoted still $\r_{ab}$ as in here, sometimes $N_{ab}^{\sscr vac}$, sometimes $T_{ab}$ in reference to the conformal field theory.} 
As $\r_{ab}$ is now neither unique nor universal, its behaviour under diffeomorphisms and conformal transformations is decoupled, leading to a non-trivial transformation rule under the eBMS symmetry group, 
\be\label{dxirho}
\d_\xi \r_{AB} = \pounds_\xi \r_{AB} + 2\Dd_A \Dd_B\dot\xiu \neq 0.
\ee
This is the expression 
in Bondi coordinates, for an arbitrary $q_{AB}$. If we specialize to the flat metric, whose components can be taken to be $q_{zz}=0$ and $q_{z\bar z}=1$ in complex coordinates, then $Y^z$ is meromorphic and $Y^{\bar z}$ anti-meromorphic, and \eqref{dxirho} reads $\d_\xi \r_{zz} = \pounds_\xi \rho_{zz} + \p_z^3 Y^z$.
The right-hand side vanishes for super-translations, but not for super-rotations. If the latter are restricted to be globally defined CKVs, then we expect it to vanish as well as to recover the BMS result. To see this, we distinguish two cases. If we make the restriction to globally defined CKVs assuming the sphere topology, then $\r_{AB}$ is unique and the result follows from the original Geroch's analysis. If we do it with the punctured sphere topology, then $\r_{AB}$ is no longer unique, but there is a choice that guarantees that the right-hand side vanishes in every frame, and this is given by the special solution $\r_{AB}=0$ on the flat frame, see App.~\ref{AppG1} for a proof. In other words, there is always a preferred $\r_{AB}$ for globally defined CKVs, given by Gerch's tensor on the sphere topology, and by a conformal transformation of the trivial tensor on the punctured sphere.

Concerning the transformation rules of the background and dynamical fields, we can still use \eqref{BMStransf}, with the proviso that we allow for non-globally defined CKVs. 
Remarkably, the anomaly of the non-universal version of Geroch tensor $\r_{AB}$ is still the same expression \eqref{Dxirho} that we had for BMS.\footnote{
There is an analogy to what happens to $q_{AB}$ in going from BMS to more general symmetries such as gBMS or BMSW: the rule $\d_\xi q_{AB}$ changes, but the background dependence is still the same, and so is $\D_\xi q_{AB}$.}
As a consequence, \eqref{DxiN} still holds: \emph{the news tensor is conformally invariant in eBMS just as it was in BMS}, in spite of different transformation rules for $\d_\xi\rho_{ab}$ and $\d_\xi N_{ab}$.
One may also think that  super-rotations create magnetic shear, since (2.52) does not hold. However this is not the case thanks to Geroch's tensor, which makes the decomposition of the shear in electric and magnetic parts conformally invariant and also super-rotation invariant [24], as can be seen using (2.31).

This observation suggests to look for a Wald-Zoupas charge prescription following the same idea we used in the BMS case, namely 
we add and subtract the term $\f1{32\pi}\d(\r_{AB} C^{AB})$ to \eqref{barth} in order to turn $\dot C_{AB}$ into $N_{AB}$.
This leads to he same charges \eqref{qWZBMS}, 
but there is a catch: The symplectic potential picks up an additional contribution due to  the non-universality of $\r_{ab}$,
\be\label{theBMS1}
\th^{\sscr BMS} = \f1{32\pi} (N_{AB}\d C^{AB} - C^{AB}\d\r_{AB})\eps_\scri.
\ee
The new contribution can also be read from the Noether current associated with the same charges. In fact $dq^{\sscr BMS}_\xi$ contains a term $\pounds_\xi\r_{AB}$, which 
goes into reconstructing the flux $N_{AB}\d_\xi C^{AB}$. Except that in the eBMS case in order to do so we have to use \eqref{dxirho} and this leads precisely to the extra piece $C^{AB}\d_\xi\r_{AB}$, in agreement with \eqref{theBMS1}.

The first term of \eqref{theBMS1} is perfectly covariant, but the second one only up to a total divergence on the cross sections:
\be\label{Dxibt}
(\d_\xi-\pounds_\xi) \th^{\sscr BMS} = \f1{16\pi}\Dd_A(\d\r^{AB}\Dd_B f)\eps_\scri.
\ee
This follows from \eqref{Dxirho} and 
\be
\Dd^B\d\r_{AB}=\d\Dd_A\cR=0,
\label{confdivrho}
\ee
which still holds in spite of the non-universality of $\rho_{ab}$. 
The linearized non-covariance \eqref{Dxibt}
vanishes upon integration (One may question the validity of this statement in a context in which singularities in the vector fields are allowed, but it is the assumption taken in \cite{Barnich:2011mi} and which we follow here).
Therefore 
the flux satisfies the linearized covariance condition, but the symplectic potential 3-form and the Noether current do not.
and this is enough to have a cocycle-free flux algebra: \eqref{theBMS1} satisfies \eqref{FFBMS}. In fact, even the charge algebra is center-less:
on can follow the same steps of \eqref{BMSbracket}, and even though \eqref{theBMS1} has an extra piece and $\d_\xi\r_{ab}\neq 0$, the fact that the anomaly is still given by \eqref{Dxirho} is enough to achieve \eqref{WZbracket}. In other words, the BMS charges satisfy a covariant algebra even if $\rho_{ab}$ is not universal!

We are now in a position to clarify the origin of the 2-cocycle found in \cite{Barnich:2011mi} for the eBMS symmetry. It comes from having neglected the contribution of Geroch's tensor, which as in the BMS case, it removes both terms of \eqref{KBT}. One may have thought that choosing a flat metric  and assuming an initial $\r_{ab}=0$, Geroch's contribution would be irrelevant to the charge algebra, similarly to what happens in the BMS case if we stick to Bondi frames. The reason why this is not the case is that eBMS transformation \eqref{dxirho} \emph{is not homogeneous}. Hence keeping $\r_{ab}=0$ is not consistent when computing the algebra, and this is why a breaking of covariance appears, in the form of the 2-cocycle of \cite{Barnich:2011mi}.
Having clarified this and exhibited an eBMS charge algebra free of cocycles, it may come as a disappointment the fact that \eqref{theBMS1} is actually \emph{not} a valid Wald-Zoupas potential for eBMS. The reason is that we have a non-trivial commutator \eqref{DDcomm}, hence the property \eqref{Dxibt} is not sufficient to have covariance under \emph{finite} conformal transformations. This can be explicitly checked using the finite transformation rules
\eqref{finiteC} and \eqref{rhotransformation} for the shear and Geroch tensors.

The non-covariant term $C^{AB}\d\r_{AB}$ also spoils the stationarity condition. However, since it depends on time only through the shear, stationarity can be restored up to a corner term:
\be\label{menez}
-\f1{32\pi}C^{AB}\d\r_{AB} \eps_\scri = -\f1{32\pi} (u-u_0) N_{AB}\d \r^{AB}\eps_\scri +d\vth,
\ee
where
\be\label{vtheBMS}
\vth = -\f1{32\pi} \left( (u-u_0)C^{AB} +\f u2(u-2u_0)\r^{\la AB\ra} \right)\d\r_{AB} \eps_S,
\ee
and $u_0=u_0(x^A)$ is a constant of integration. 
We are thus led to relax condition 0 and  consider the new symplectic potential
\be\label{theBMS}
\th^{\sscr eBMS} := \th^{\sscr BMS} -d \vth
= \frac{1}{32 \pi} (N_{AB} \d C^{AB} - (u - u_0) N_{AB} \d \rho^{AB}) \eps_\scri.
\ee
It is manifestly stationary. How about its covariance? A term like $u N_{AB}\d\r^{AB}$ depends on the foliation, hence it cannot be covariant. 
This is where the integration constant $u_0$  plays a key role. We take it to be charged under conformal transformations and super-translations like $u$, hence $\D_\xi u_0 = -f|_{u_0}$.
This means that it transforms precisely like \eqref{dxiu0}, hence we can identify it as the super-translation field (aka super-translation Goldstone aka bad-cut field) corresponding to the boundary condition defined by the relative shear \eqref{cS}, and $\D_\xi (u-u_0) = -\dot f(u-u_0)$.
One can then easily verify that \eqref{theBMS} is fully covariant, at both linearized and finite level. In other words, allowing for an integration constant in \eqref{menez} and endowing it with the interpretation of a vacuum connection at late times, we have made the expression independent of the foliation.
We have thus been able to identify a Wald-Zoupas potential for eBMS!
Before discussing its uniqueness, let us add a few remarks.

Finding a Wald-Zoupas potential for eBMS has required two new ingredients with respect to the BMS case: $(i)$ enrich the radiative phase space to include a vacuum connection as corner data representing late time boundary conditions, and $(ii)$ relax what we referred to as condition 0 of the Wald-Zoupas paper and allow a change of symplectic structure by a corner term.
The new symplectic current is
\be
\omega^{\sscr eBMS} = \om - d \d \vth = \frac{1}{32 \pi} (\d N_{AB} \d C^{AB} - \d[(u - u_0) N_{AB}] \d \rho^{AB}) \eps_\scri.
\ee 
It is not defined in the spacetime bulk but only at $\scri$, since it depends on the `edge modes' $\rho_{ab}$ and $u_0$. Its integral over $\Si$ is however well-defined. 
Therefore while the symplectic current is not defined on the phase space $\Si$, the symplectic 2-form is.
The fact that covariance may be restored using fields defined only at the boundary is consistent with the Wald-Zoupas philosophy, and very much at the root of the edge modes approach to gauge symmetries in the presence of boundaries \cite{Donnelly:2016auv}.

The corresponding Noether currents are
\be\label{jeBMS}
j^{\sscr eBMS}_\xi = I_\xi \th^{\sscr eBMS} 
= \f1{32\pi} \left(N_{AB}\d_\xi C^{AB} - (u-u_0) N_{AB}\d_\xi \r^{AB}\right)\eps_\scri \eqons dq_\xi^{\sscr eBMS},
\ee
with
\be\label{qeBMS}
q^{\sscr eBMS}_\xi =q^{\sscr BMS}_\xi -I_\xi\vth = q^{\sscr BMS}_\xi + \f1{32\pi} \left((u-u_0)C^{AB} +\f u2(u-2u_0)\r^{\la AB\ra}\right)\d_\xi\r_{AB}\eps_S,
\ee
up to a closed 2-form. This gives the same super-translation charges and fluxes \eqref{QT}, the 
only modification occurs for super-rotations.
 It follows from the results of \cite{Rignon-Bret:2024wlu} recalled above that these Noether currents provide a \emph{center-less realization of the eBMS symmetry algebra} via \eqref{currentalgebra}. It can be verified explicitly with a calculation similar to the one of \eqref{BMSbracket}.

In order to integrate the current over all of $\scri$, one needs the fall-off conditions
\be\label{Nu2}
\lim_{u\to\infty} N_{ab} = \f1{u^{2+\varepsilon}},
\ee
which are stronger  than the Ashtekar-Streubel ones \eqref{ASbc}. In other words, the eBMS symmetry can be made covariant locally or over all of $\scri$, but in the latter case only with respect to a \emph{smaller set of solutions} than the BMS symmetry.
Integrating \eqref{jeBMS} over all of $\scri$ and assuming \eqref{Nu2} gives a result consistent with  \cite{Campiglia:2020qvc,Donnay:2022hkf}.
Our analysis thus strengthens the proposal of these papers, by showing that their total flux satisfies the Wald-Zoupas conditions, and furthermore that it can be obtained from a local 3-form which also satisfies the Wald-Zoupas conditions. As a consequence not only the total flux on $\scri$, but also the flux between two arbitrary cuts of $\scri$ provides a center-less realization of the eBMS algebra.

Finally, let us talk about the uniqueness of  \eqref{theBMS}. The question is whether there exist exact forms in field space or in (the boundary of) spacetime that can be added without spoiling covariance and stationarity requirements. 
These requirements eliminate most terms that one can write down. Some options remain, for instance $\d [(u - u_0) N_{AB} N^{AB} \eps_I]$ or $d[(u - u_0) N_{AB} \d C^{AB} \eps_S]$. These however contain second derivatives in time, hence they are ruled out for the same reason they are ruled out in the standard BMS analysis, namely the analyticity requirement that the equivalence class of symplectic potentials should be compatible with the 
second-order nature of the field equations. The only allowed second-order derivatives on a null boundary are then purely spatial or mixed time-space, but not purely temporal. This argument eliminates the examples above, and barring unforeseen terms, we believe that the eBMS Wald-Zoupas symplectic potential we have found is unique.

\subsection{Uniqueness of the eBMS charges}

Let us now look at the charges \eqref{qeBMS}. Since they come from a covariant flux, the only possible cocyle in their Barnich-Troessaert bracket is time-independent. An explicit calculation (using integration by parts and neglecting boundary terms following \cite{Barnich:2011mi}) gives
\begin{align} \label{KeBMS}
K_{(\xi, \chi)}^{\sscr eBMS} = - \frac{1}{16 \pi} \oint_S \left( u_0 \, \d_{\xi} \rho_{AB} (\Dd^{\la A} \Dd^{B\ra}+\f12\r^{\la AB\ra}) T_\chi \eps_S -  (\xi \leftrightarrow \chi)\right)\eps_S.
\end{align}
It is time-independent and furthermore non-radiative, depending only on the background metric and on the edge modes. Since the edge modes are not universal however, it is not a simple central extension, hence it signals a lack of covariance of the charges. If we look more closely, we see that the cocyle vanishes for globally defined CKV, but also for super-rotations if we consider global translations. We thus have a covariant charge algebra of super-rotations with global translations, but not between arbitrary eBMS transformations. 

The cocycle is proportional to $u_0$, so it vanishes for the boundary condition $u_0=0$. But there is nothing physical nor covariant about this value, which depends  on an arbitrary choice of origin in the vacuum sector of the radiative phase space. 
Since the initial charges had no cocycle, \eqref{KeBMS} comes entirely from the corner improvement, which is in fact anomalous, 
\be
	(\d_\xi - \pounds_\xi) \vartheta = \frac{1}{16\pi} [u \Dd_{\la A} \Dd_{B \ra} T_\xi \d \rho^{AB} + \f12 u^2 \Dd_{\la A} \Dd_{B \ra} \dot{f}_\xi  \d \rho^{AB} - u_0 (\Dd_{\la A} \Dd_{B \ra} + \f12 \rho_{\la AB \ra}) T_\xi  \d \rho^{AB}]  \eps_S.
\ee
Integration on the cross section removes the time dependence, in agreement with the general argument \eqref{dDvth}.
The anomaly spoils the relation to the canonical generator, as discussed in \ref{SecCorner}. Therefore to obtain the cocycle we have to use  \eqref{QQnotcangen} and we obtain 
\be
	K_{(\xi, \chi)}^{\sscr eBMS} = \oint_S I_\xi (\d_\chi - \pounds_\chi) \vartheta - I_\chi (\d_\xi - \pounds_\xi) \vartheta. 
\ee
Next, we ask if the charges are unique, and if not, whether there is an alternative free of cocycle. The ambiguity in the charges is the usual freedom of adding a time-independent function, which can be parametrized by the freedom of adding a time-independent corner term,
\be
\vth\to \vth+\vth^{\sscr e}, \qquad \pbi{d\vth^e}=0, \qquad \bar\Om_\Si\to\bar\Om_\Si - \d\oint_S\vth^e.
\ee
If it were possible to remove the anomaly in this way, the new charge would be consistent with the canonical generator on $\Si$.
For instance, one could try to remove the relative factor 1/2 in \eqref{qeBMS} which prevents replacing the shear $C_{AB}$ with the covariant shear ${\cal C}_{AB}$.
However this factor is necessary for the integration by parts in time \eqref{menez}. A simple example would be adding
\be
\vth^e = -\f1{64\pi} u_0^2 \r^{\la AB\ra} \d\r_{AB} \eps_S,
\ee
leading to the new corner potential
\be\label{vtheBMS1}
\vth_1=\vth+\vth^e = -\f1{32\pi} (u-u_0)\left( C^{AB} +\f12(u-u_0)\r^{\la AB\ra} \right)\d\r_{AB} \eps_S.
\ee
It gives rise to the same Wald-Zoupas symplectic potential \eqref{theBMS}, and the new charges
\be\label{qeBMS1}
q^{\sscr eBMS}_\xi =q^{\sscr BMS}_\xi -I_\xi\vth = q^{\sscr BMS}_\xi + \f1{32\pi} (u-u_0)\left(C_{AB}+\f12(u-u_0)\r_{AB}\right)\d_\xi\r^{AB}\eps_S.
\ee
They may appeal aesthetically more than \eqref{qeBMS}. But the extra shift actually worsen the covariance. The cocycle becomes
\begin{align} \label{chargecocycle}
K_{(\xi, \chi)}^{\sscr eBMS1} = - \frac{1}{16 \pi} \oint_S u_0\left(\d_{\xi} \rho_{AB} ( \Dd^{\la A} \Dd^{B\ra} T_\chi+ \f{u_0}2 \Dd^{\la A} \Dd^{B\ra} \dot f_\chi) -  (\xi \leftrightarrow \chi)\right)\eps_S 
\end{align}
and we have lost the property that global translations act covariantly on super-rotations in every frame. 
We were not able to find a choice of $\vth^e$ that removes the cocycle, and the attempts we made suggest that \eqref{vtheBMS} is uniquely selected in that it minimizes the non-covariance of the charges. 
In spite of its limited success, this example is useful to show explicitly the importance of requiring covariance in order to restrict the charge ambiguity, in situations like this where the reference solution is not enough or not relevant.

As a closing remark, we point out that despite not being able to construct a covariant local corner charge algebra, a covariant charge algebra can on the other hand be achieved if one defines the surface charges only as relative quantities with respect to a given reference solution in the distance future. For instance if we assume that spacetime settles down to Minkowski spacetime at $u \rightarrow + \infty$, then we can define the charge at a cross section $S$ of $\scri^+$ as the integral of the current between $\scri^+_+$ and $S$. By doing this, the charge vanishes automatically at any cross section $S$ in Minkowski spacetime, since the current vanishes in this case, and the charge algebra is free of cocycles. 

The values we found for the cocycle of the charges rest on the assumption of \cite{Barnich:2011mi} that one can discard boundary terms when integrating by parts. If boundary terms do contribute, then the value of the cocycle will be affected.\footnote{It seems however difficult to imagine that such terms could alter the cocycle to the point of removing it. If the boundary terms are contact terms localized at the poles of the symmetry vector fields, one could consider doing a conformal transformation that trivializes in the vicinity of all poles. The cocycle we found would change since it is not conformally invariant, but the contribution from the boundary terms would not.}
The covariance of the symplectic potential and of the Noether currents, hence the lack of cocycles in their algebra, is on the other hand independent of the assumption, because no integration by parts is needed.

\section{Conclusions}

We wrote this paper with two goals in mind. First, to bring to the forefront some aspects of the physics and geometry of $\scri$ that although well explained in the existing reviews, may not have received the attention they deserve in order to explain and relate different results in the literature.
Notably the fact that the covariant shear constructed in \cite{Compere:2018ylh} can be derived in a coordinate-independent way from the radiative phase space of \cite{Ashtekar:1981bq,Ashtekar:1981sf}, hence explaining that the super-translation field aka super-translation Goldstone field is not a new degree of freedom, 
and it is present in Ashtekar-Streubel phase space. It can be thought of as a bad cut (final) temporal condition acquire a status like an edge mode when it is used as a temporal boundary condition to enlarge the radiative phase space with a temporal boundary condition. We also explained that the seemingly contradictory statements that the time derivative of the shear is not conformally invariant  \cite{Dray:1984rfa,Barnich:2016lyg,Compere:2018ylh}, and that it is \cite{Grant:2021sxk}, are due to the alternative options of preserving or not a shear associated to a Lie-dragged auxiliary vector field. The first option results in simpler transformations laws and simpler flux formulas, and that are actually the ones used by the community working in Bondi coordinates.
We provided a general formula for the news tensor that is not often found, and resurfaced Dray's argument explaining why it is convenient not to restrict attention to only Bondi frames when studying flux-balance laws between arbitrary cross sections. 
By presenting all results both in covariant language as well as in Bondi coordinates, we hope to have contributed in helping communication between the two communities. 
A further technical result of our paper is to show that covariance and conformal invariance can be studied in a practical and economical way using finite BMSW transformations, similarly to what done in \cite{Barnich:2016lyg}, seen as an auxiliary transformations and not as an extension of the symmetry. We also hope to have convinced the reader that 
the anomaly operator is a very convenient tool, that makes checking covariance as well as conformal invariance straightforward and simultaneous, and most practically, doable in a fixed coordinate system.

The second goal of the paper was to present the details of the results announced in \cite{Rignon-Bret:2024wlu}.
We have shown that the 2-cocycle found in \cite{Barnich:2011mi} is a result of a non-covariant split in the definition of the charges, and that it is possible to find a Wald-Zoupas potential that removes it. For BMS the right potential was already known, and we have checked that both currents and charges realize the algebra under the Barnich-Troessaert bracket without central extensions. For eBMS it is a new result, and generalizes previous formulas appeared in \cite{Campiglia:2020qvc,Donnay:2022hkf}.
There is an interesting similarity between the two constructions, in the sense that in both cases the key role is played by Geroch tensor, with the crucial difference that it is universal in the first case, non-universal and rather an edge mode in the second. The similarities however stop here. In the eBMS case, covariance requires to modify the symplectic 2-form by a corner term, and can only be achieved for the fluxes and a subset of the charges. The full charge algebra presents a time-independent cocycle determined by the boundary fields $u_0$ and $\r_{ab}$ and independent of radiative degrees of freedom such as the shear.

Having found a covariant symplectic potential for eBMS was quite a surprise to us. It follows from $(i)$ the subtle role that the generalization of Geroch's tensor plays in restoring covariance even if it is no longer universal, in fact \emph{precisely} because it is no longer universal and it transforms inhomogeneously, and $(ii)$
using an appropriate symplectic 2-form with the corner improvement $\d(u_0N_{AB})\d\r^{AB}\eps_\scri$.
Our results suggest that case studies in which field-dependent cocycles appear in the literature
(see e.g. \cite{Afshar:2015wjm,Barnich:2017ubf, Distler:2018rwu, Barnich:2019vzx,Compere:2020lrt, Campiglia:2021bap,Geiller:2024amx})
are probably afflicted by loss of covariance in their construction, at one level or another. 

The fact that the charge covariance requires additional input than flux covariance is in line with the results of \cite{Ashtekar:1981bq,Ashtekar:2024mme,Ashtekar:2024stm} showing that the fluxes have also a more consistent interpretation as canonical generators on the radiative phase space, than the charges on the partial Cauchy slice phase space.

Applications of our method to the generalizations of the BMS symmetry to larger groups will appear elsewhere \cite{AS2}.
Our results are likely to be relevant also relevant for the current ongoing research on $w_{1+\infty }$\cite{Strominger:2021mtt,Freidel:2021ytz,Compere:2022zdz,Blanchet:2023pce,Geiller:2024bgf}. For other recent work on asymptotic gravitational symmetries, see e.g. 
\cite{
Ciambelli:2019lap, Wieland:2020gno, Wieland:2021eth,Kolanowski:2021hwo,Adami:2023wbe,Ciambelli:2024kre,Freidel:2024tpl,Barnich:2024aln}.

\subsubsection*{Acknowledgements}
We are very grateful to Glenn Barnich for help, sharing calculations and insights into eBMS, and pushing us to get to the bottom of things. 
We also would like to thank Abhay Ashtekar, Adrien Fiorucci and Marc Geiller for many discussions, and an anonymous referee for valuable comments. Simone
is grateful for the hospitality of Perimeter Institute where part of this work was carried out. Research at Perimeter Institute is supported in part by the Government of Canada through the Department of Innovation, Science and Economic Development and by the Province of Ontario through the Ministry of Colleges and Universities.

\appendix

\section{Shear transformation}\label{AppdxiS}

In this Appendix we review how \eqref{dxiAbhay} reduces to \eqref{dxiC} for a Lie dragged $l$.
We start from the identity
\be
[\pounds_\xi, D_{\la a}]l_{b\ra} = l^c(\xi^d\hat R_{d\la ab\ra c} -D_{\la a}D_{b\ra}\xi_c).
\ee
For the vertical part of the symmetry vector field, $\xi=f n$, we have
\be
l^c n^d\hat R_{d\la ab\ra c} = \f12\hat S_{\la ab\ra} 
\ee
because the conformal Weyl tensor vanishes at $\scri$, and
\be
-l^c D_{\la a}D_{b\ra} (fn_c) = D_{\la a}D_{b\ra}f = \Dd_{\la a}\Dd_{b\ra} f + \s_{ab}\dot f + 2l_{\la a}D_{b\ra}\dot f
\ee
where we used the divergence-free frame condition and the first identity in footnote~\ref{doublederivative}.
For the horizontal part, we have
\be
 l^c Y^d\hat R_{d\la ab\ra c} =  Y^c[D_c,D_{\la a}]l_{b\ra} = Y^cD_c\s_{ab} - Y^c D_{\la b} \s_{a\ra c}
\ee
where we already set $\t_a=0$, and
\be
- l^c D_aD_bY_c = \pounds_Y\s_{ab} - Y^cD_c\s_{ab} +Y^cD_{\la a}\s_{b\ra c}.
\ee
Adding up and using \eqref{SchoutenBondi}, we recover
\be
\d_{\xi} \s_{ab}= (f\p_u+\pounds_Y-\dot{f}) \s_{ab} + {\Dd}_{\langle a} \Dd_{b\rangle} f = -\f12 \d_{\xi} C_{ab}.
\ee

\subsection{On the flux-balance laws}\label{AppDray}

In this appendix we investigate two different methods for computing the flux of the BMS charges between two arbitrary cross sections of $\scri^+$ using the covariance properties that we studied in the main text. Imagine for example that we want to compute the flux between the cross section $S_1$ defined at $u_1 = 0$ and $S_2$ defined at $u_2 = \alpha(x^A)$. The two cross sections are related to a supertranslation $u_2 = u_1 + \alpha$, and so the coordinates $u' = u - \alpha, x^{A'} = x^A$ define a nother Bondi coordinate system where $S_2$ is located at $u' = 0$. If we adapt the background structure to the first coordinate system $(u, x^A)$ by usiong an auxiliary vector $l = -du$ tangent to the foliation induced by the first Bondi coordinate system, we can compute the charges at $u = 0$ which are given by 
\be
	Q_\xi^{\sscr BMS} [S_1] = \frac{1}{16 \pi} \oint_{S_1} (4 M_\rho T + 2 Y^A J_A) \eps_S
\ee
where $M$ and $J^A$ are computed using $l = -du$. However, on the cross section $S_2$, the charge is not given in general by
\be \label{first try} 
	Q_\xi^{\sscr BMS} [S_2] = \frac{1}{16 \pi} \oint_{S_2} (4 M_\rho T + 2 Y^A J_A) \eps_S
\ee
since $M$ and $J^A$ are not adapted to the foliation to which $S_2$ belongs to. However, we can compute the BMS charge on $S_2$ by adapting the background structure made of $l$ and $\xi$ to the cross section $S_2$, such that
\be \label{goodcharges2}
	Q_{\xi'}^{\sscr BMS} [S_2] = \frac{1}{16 \pi} \oint_{S_2} (4 M_\rho^{'} {T'} + 2 {Y'}^{A} J'_{A}) \eps_S
\ee
where the coordinate system $A'$ labels the cross section $S_2$, $M'_\rho$ and $J'_{A}$ are computed by transforming $l = -du$ into $l' = -d(u - \alpha)$ adapted to the foliations of supertranslated Bondi coordinate system, which corresponds to the application of the (finite) anomaly operator associated to the super-translation $\alpha$. Furthermore, we have to adapt the vector field $\xi$ to the new coordinate system, such that
\be
	\xi = T \p_u + Y^A \p_A = T' \p_{u'} + Y'{}^{A} \p_{A}',
\ee
and $\p_{A}'$ being tangent to $S_2$, and which components $(T', Y'{}^{A})$ are obtained by computing the bracket $[\alpha, \xi]$. It is how the charges have been computed in \cite{Flanagan:2015pxa,Chen:2022fbu} for instance, without referring to the anomaly operator. Furthermore, we know that the relation between \eqref{first try} and \eqref{goodcharges2} is given by the integration of the infinitesimal relation \eqref{dxiQam} to a finite supe-rtranslation parameter $\alpha$. Indeed, this relation tells us that at linearized order the anomaly of the charge is equivalent to computing the anomaly of the vector fields, i.e their bracket. In particular, for any super-translation $T$, since $[\alpha, T] = 0$, \eqref{dxiQam} can easily be integrated and we have that 
\be
	Q_\xi [S] = \frac{1}{4 \pi} \int_S (TM_{\rho} +\Dd_A V^A_{(C+u\r,T)}) \eps_S =  \frac{1}{4 \pi} \int_S (TM_{\rho}' + \Dd_{A}' V^{A}_{(C',T)}) \eps_S
\ee
which outlines the invariance of the super-momentum with respect to the choice of $l$. Nevertheless, the super-momentum can be written as the integral of $T M_\rho$ on $S$ only if $l$ is adapted to the cross section $S$, since $\Dd_A$ depends on the choice of $l$. However, the angular momentum does not commute with an arbitrary super-translation $\alpha$, and so the angular momentum is dependent on the choice of $l$, but in a covariant manner, since \eqref{first try} and \eqref{goodcharges2} are related by \eqref{dxiQam}. Therefore, if we want to compute the angular momentum charge using an auxiliary vector $l$ which is not adapted to the cross-section, we have to integrate \eqref{dxiQam} to a finite super-translation parameter $\alpha$. 

\vspace{0.3 cm}

However, translating between two different Bondi coordinate systems can be painful and so we outline here another elegant way of computing the flux of angular momentum which has been worked out by Dray \cite{Dray:1984gz}. The Dray-Straubel angular momentum charge is computed using an auxiliary vector $l$ that is tangent to the cross section $S$ on which we evaluate the angular momentum charge. However in the example above we do not have an auxiliary vector $l$ that is tangent to both cross sections at the same time, for the same reason as the vector $\xi$ cannot be tangent to both cross sections, since for a general $\alpha$ it will have a vertical component on $S_2$ if it does not have one on $S_1$. Nevertheless, the Ashtekar-Streubel flux is conformally invariant, so we can compute the flux and the charges in any conformal frame. Under conformal transformations (followed by a super-translation $T$), we remind that the affine coordinate $u$ transforms into 
\be
	u \rightarrow u' = \omega (u + T) + O(\Omega)
\ee 
which is another affine coordinate. If we choose now $\omega = \frac{1}{u_2 - u_1}$ and $T = -\frac{u_1}{u_2 - u_1}$, then the cross sections $S_1$ and $S_2$ belong the same foliation, and are located respectively at $u' = 0$ and $u' = 1$. In other words, for any pair of cross sections $(S_1, S_2)$ there exists a privileged conformal frame in which the auxiliary field $l' = -du'$ adapted to $S_1$ is also adapted to $S_2$. Therefore in this conformal frame the two cross sections belong to the same foliation and we can compute the flux and the charges by neglecting the contribution of the total divergences $\Dd_A' X^A \eps_\scri$ all along (where the two dimensional derivative operator $\Dd_A'$ is associated with $l' = -du'$).

\section{Explicit calculation of the 2-cocycle's removal}\label{AppProofs}

In this Appendix we prove that the Wald-Zoupas charges satisfy the algebra without 2-cocycle with an explicit calculation and without reference to the anomaly operator. 
The cocycle is removed if 
\be\label{tobe}
\d_\xi (q_\chi^{\sscr BT} - q_\chi^{\sscr BMS}) - i_\xi I_\chi (\theta^{\sscr BT}-\th^{\sscr BMS}) = 
- K^{\sscr BT}_{(\xi,\chi)} - (q_{[\xi,\chi]}^{\sscr BT} - q_{[\xi,\chi]}^{\sscr BMS}).
\ee
where
\begin{align}
& \ell^{\sscr G} = -\f12 \r_{AB}C^{AB}\eps_\scri, \qquad \theta^{\sscr BT}-\th^{\sscr BMS} = -\d\ell^{\sscr G} = \f1{32\pi} \r_{AB}\d C^{AB}\eps_\scri, \\
& q_\chi^{\sscr BT} - q_\chi^{\sscr BMS} = -i_\chi\ell^{\sscr G} = \f {f_\chi}{32\pi}\rho_{AB} C^{AB}\eps_S, \qquad i_\chi \eps_\scri = f_\chi\eps_S.
\end{align}
From the latter we have
\be
q_{[\xi,\chi]}^{\sscr BT} - q_{[\xi,\chi]}^{\sscr BMS} = \f {f_{[\xi,\chi]}}{32\pi}\rho_{AB} C^{AB}\eps_S = 
\f 1{32\pi}\left(f_\xi\dot f_\chi +Y_\xi [f_\chi] -  (\xi\leftrightarrow\chi) \right)\rho_{AB} C^{AB} \eps_S,
\ee
hence the right-hand side of \eqref{tobe} is equal to
\be\label{tobe2}
- \f 1{32\pi}\left(f_\xi(\p^Af_\chi \p_A \cR+ 2C^{AB}\Dd_A \Dd_B\dot f_\chi) + (f_\xi\dot f_\chi +Y_\xi [f_\chi])\r_{AB} C^{AB} - (\xi\leftrightarrow \chi)\right)\eps_S,
\ee
where we used \eqref{KBT}. On the left hand side we have
\begin{align}\label{tobe3}
& \d_\xi (q_\chi^{\sscr BT} - q_\chi^{\sscr BMS}) - i_\xi I_\chi (\theta^{\sscr BT}-\th^{\sscr BMS}) = 
\f {f_\chi}{32\pi}\rho_{AB} \, \d_\xi C^{AB} \, \eps_S -  (\xi\leftrightarrow\chi) 
\nn\\
&\qquad\qquad = \f {1}{32\pi} f_\chi \rho_{AB} \left( (\pounds_{Y_\xi} + 3\dot f_\xi ) C^{AB} -2 \Dd^{\la A}\Dd^{B\ra}f_\xi\right) \eps_S - (\xi\leftrightarrow\chi),
\end{align}
where we used \eqref{DxiCinv}.
In the first term we integrate by parts:
\begin{align}
& {f_\chi}\rho_{AB} \pounds_{Y_\xi} C^{AB} \eps_S = - C^{AB} \pounds_{Y_\xi}\left( {f_\chi}\rho_{AB} \eps_S\right)
= - C^{AB}( \r_{AB} Y_\xi[f_\chi] + {f_\chi} \pounds_{Y_\xi}\rho_{AB} + 2{f_\chi} \rho_{AB} \dot f_\xi ) \eps_S, 
\end{align}
using $d\eps_S=0$ and $ d (i_{Y_\xi}\eps_S) = DY_\xi\eps_S=2\dot f_\xi\eps_S$. 
The crucial ingredient at this point is the transformation law \eqref{lierho} of Geroch's tensor, which allows us to rewrite \eqref{tobe3} as
\be
-\f1{32\pi}\left(C^{AB}Y_\xi[f_\chi]  - {f_\chi} \rho_{AB} \dot f_\xi C^{AB} + 2f_\chi \big(C_{AB} \Dd^{\la A}\Dd^{B\ra}\dot f_\xi +\r_{\la AB\ra } \Dd^A\Dd^B f_\xi\big)\right) \eps_S - (\xi\leftrightarrow\chi).
\ee
The first three terms match the first three terms of \eqref{tobe2}, including the anti-symmetrization.
The last term we integrate by parts, finding
\be
-f_\chi \r_{\la AB\ra } \Dd^A\Dd^B f_\xi = \f12f_\chi \p_A{\cal R} \p^A f_\xi + \r_{\la AB\ra } \Dd_A f_\chi \Dd^B f_\xi.
\ee
The second term vanishes from the anti-symmetrization, and the derivation of \eqref{tobe} is complete. 

This derivation can be applied to the eBMS case if we replace  \eqref{lierho} by \eqref{dxirho}. In this case we also have an extra contribution $- \frac{1}{32 \pi} C_{AB} \d_\xi \rho^{AB} \eps_S f_\chi - (\xi \leftrightarrow \chi)$ to $\d_\xi q_\chi^{\sscr BMS} - i_\xi I_\chi \theta^{\sscr BMS}$ and thus we understand that it is in fact the combination $\pounds_\xi \rho_{AB} - \d_\xi \rho_{AB} = - \Delta_\xi \rho_{AB}$ that appears in the more general derivation. However this quantity is the same for both BMS and eBMS symmetries, and the rest of the proof follows the same steps. 

\section{Conformal stress-energy tensor and super-rotations}\label{AppSET}

The equivalent of Geroch's tensor in the context or eBMS is still defined by  \eqref{rhoCond}, but one can now look for a reference solution on a flat conformal frame, with $\cR=0$. This means that it is trace-less and divergence-free, like the stress energy tensor of a conformal field theory. If we use complex coordinates $(z,\bar z)$,
the only non-vanishing component of the flat metric is $q_{z\bar z}=1$, and the trace-free and divergence-free conditions read
\be
\rho_{z\bar{z}} = 0, \qquad \p_{\bar{z}} \rho_{zz} = 0 \qquad  \p_z \rho_{\bar{z} \bar{z}} = 0.
\ee
The solution to these equations is $\rho_{zz} = \r_{zz}(z)$ an arbitrary meromorphic function while $\rho_{\bar{z}\bar{z}} = {\r}_{\bar z\bar z}(\bar{z})$ is an arbitrary anti-meromorphic function. 
 We thus have an infinite number of solutions, parametrizable in terms of conformal field theories.\footnote{One can in fact show that being divergence-free and trace-less implies the third condition in \eqref{rhoCond}.} Therefore the generalization of Geroch's tensor is not unique nor universal,
 which was to be expected since the sphere's topology was crucial to establish those properties.
Let us now look at its transformation rules. Under a (meromorphic)  coordinate transformation, both the induced metric and $\r_{ab}$ transform as a tensor, hence in complex coordinates,
\be
q_{z\bar z}' = \left| \f{\p z}{\p z'} \right|^2,  \qquad \rho_{zz}' = \left( \f{\p z}{\p z'} \right)^2 \rho_{zz}(z').
\ee
The first equation coincides with a conformal transformation of the metric. 
Therefore just like for BMS vector fields, preservation of the background metric is achieved if we perform simultaneously an asymptotic diffeomorphism and a conformal transformation with factor $\om= \left| \f{\p z}{\p z'} \right|^{-1}$, 
\be
q_{z\bar z}' = \left| \f{\p z}{\p z'} \right|^2 \left| \f{\p z}{\p z'} \right|^{-2}q_{z\bar z}=q_{z\bar z}.
\ee
This establishes invariance of the background metric for finite eBMS transformations, 
namely $\d_\xi q_{ab}=0$ at the infinitesimal level.
On the other hand, the behaviour of $\r_{ab}$ under conformal transformation is still given by \eqref{rhoconf}, consistently with the defining conditions.
As a consequence, the combined action of cross-section meromorphism plus (inverse) conformal rescaling defining a eBMS symmetry  gives
\be \label{rhotransformation}
\rho_{zz}' = \left( \f{\p z}{\p z'} \right)^2 \rho_{zz}(z') + {\rm Schw}(z'),
\ee
where ${\rm Schw}(f):=\f{f'''}{f'}-\f32 \left(\f{f''}{f'}\right)^2$ is the Schwarzian derivative.
The infinitesimal version of this combined transformation is 
\be\label{dxirhoApp}
\d_\xi \r_{AB} = \pounds_\xi \r_{AB} + 2D_A D_B\dot f. 
\ee
The remarkable property \eqref{lierho} is now lost, because it was due to the topological properties of the sphere, and it is 
 replaced by \eqref{dxirhoApp}.
The inhomogeneous term means that an eBMS transformation can introduce a non-trivial $\r_{ab}$ even if one initially has $\r_{ab}=0$, and this plays a key role in the study of eBMS covariance.

\subsection{Recovering the universality of Geroch's tensor}\label{AppG1}

In this Appendix we prove that without assuming sphere topology for the cross sections, it is still possible to show that there is a choice of Geroch tensor such that $\d_\xi \r_{AB}=0$ for globally defined CKVs, in any frame. To do so, we consider first the flat metric. There $\p_{\la A}\p_{B\ra}\dot f=0$, which in complex coordinates is the statement that $\p_z^3 Y^z = 0$ for the Mobius transformations $Y^z(z) = a + bz + cz^2$. Then $\d_\xi \r_{AB}=\pounds_\xi\r_{AB}=0$ is solved trivially by $\r_{AB}=0$.
Starting from this special solution on the plane, the correspondent one in an arbitrary frame $q_{AB}=\om^2\d_{AB}$ is
\begin{align}
	\mathring{\rho}_{AB} = -2\om^{-1}\Dd_A \Dd_B \om +4\om^{-2} \Dd_A\om \Dd_B\om -\om^{-2} g_{AB} \Dd^C\om \Dd_C\om.
	\label{conformalrho}
\end{align}
The right-hand side is manifestly universal, hence this solution satisfies $\d_\xi \mathring{\rho}_{AB}=0$ in any frame. 
Conversely, had we started from a different solution than the trivial one, the resulting transformation law $\d_\xi \r_{AB}$ may not be zero for globally defined CKVs.

\providecommand{\href}[2]{#2}\begingroup\raggedright\endgroup

\end{document}